\documentclass[prd,twocolumn,superscriptaddress,preprintnumbers,nofootinbib]{revtex4-1}

\usepackage{graphicx}
\usepackage{epsfig}
\usepackage{units}
\usepackage{color}
\usepackage{mathrsfs}
\usepackage{enumitem}
\usepackage[export]{adjustbox}% for valign
\begin{document}
\title{{PBR Sensitivity to Boosted Dark Matter Explanation for the KM3-230213A Event}}
\author{M. H. Reno}
\affiliation{Department of Physics and Astronomy, University of Iowa, Iowa City, IA, United States}

\author{T. Venters}
\affiliation{NASA Goddard Space Flight Center, Greenbelt, MD, United States}

\author{M. Abrate}
\affiliation{Sezione di Torino, Istituto Nazionale di Fisica Nucleare, Torino, Italy}
\affiliation{Università Torino, Torino, Italy}

\author{J. Adams}
\affiliation{CSPAR, University of Alabama Huntsville, Huntsville, AL, United States}

\author{J. Alfaro}
\affiliation{Departments of Physics \& Astronomy and Astrophysics, Institute for Gravitation and the Cosmos, Pennsylvania State University, University Park, PA, United States}

\author{D. Allard}
\affiliation{Laboratoire APC, Université Paris Cité, Paris, France}

\author{P. Alldredge}
\affiliation{RSESC, University of Alabama Huntsville, Huntsville, AL, United States}

\author{R. Aloisio}
\affiliation{Laboratori Nazionali del Gran Sasso, Istituto Nazionale di Fisica Nucleare, L'Aquila, Italy}
\affiliation{Gran Sasso Science Institute, L'Aquila, Italy}

\author{R. Ammendola}
\affiliation{Sezione di Roma Tor Vergata, Istituto Nazionale di Fisica Nucleare, Rome, Italy}

\author{A. Anastasio}
\affiliation{Sezione di Napoli, Istituto Nazionale di Fisica Nucleare, Naples, Italy}

% ...

\begin{abstract} \noindent {POEMMA-Balloon with Radio (PBR) is a hybrid fluorescence and optical Cherenkov telescope that is a pathfinder instrument for a future Probe Of Extreme Multi-Messenger Astrophysics (POEMMA), together with a radio instrument,  to be flown on a super pressure balloon in Spring 2028. One of its science objectives is follow up of astrophysical event alerts via signals of tau decays in the atmosphere, sourced by Earth-skimming tau neutrinos. The recent report of the KM3NeT Collaboration's muon event with $ E_{\mu}=\unit[120^{+110}_{-60}]{PeV}$ sparked searches for beyond the standard model (BSM) physics explanations. Using a specific, boosted dark matter scenario  which together with the KM3NeT observation, sets a scale for the fluence of boosted dark matter in this energy range, we assess the capabilities of PBR to constrain or discover such a source for a boosted dark matter energy spectrum above $\sim 10^8$ GeV.  PBR has discovery potential for dark matter energies above $10^9$ GeV, and depending on BSM input parameters, to dark matter energies as low as $1.5\times 10^8$ GeV. While the analysis is focused on one BSM example, the methodology is applicable to any BSM physics that affects tau production in the Earth.}
\end{abstract}

\maketitle

\section{Introduction}
\label{sec:intro}

Very high energy neutrino astronomy, 
as part of a multimessenger program to understand the most powerful accelerators in the Universe, has made significant advances in recent years \cite{Ackermann:2022rqc, Guepin:2022qpl}. With the discovery of the diffuse astrophysical neutrino flux \cite{IceCube:2013low} and more recent associations of neutrino events with astrophysical transients
\cite{IceCube:2018dnn,IceCube:2018cha,Stein:2020xhk,Reusch:2021ztx,IceCube:2022der}, neutrino telescopes are pushing the frontiers of high-energy and ultrahigh-energy neutrino detection. Until recently, it has been the IceCube Neutrino Observatory  \cite{IceCube:2016zyt} that detected the highest energy neutrinos via through-going muons expected to originate from muon neutrinos in the $1-10$ PeV energy range \cite{Abbasi:2021qfz} and a shower with visible energy $6.05\pm 0.72$ PeV that likely originates from an incident $\bar\nu_e$ of similar energy \cite{IceCube:2021rpz}. The observation on February 13, 2023 of a nearly horizontal muon track, with a reconstructed energy of
$ E_{\mu}=\unit[120^{+110}_{-60}]{PeV}$ \cite{KM3NeT:2025npi} by KM3NeT/ARCA
detector~\cite{KM3Net:2016zxf, KM3NeT:2022pnv}, designated as KM3-230213A, may have been sourced by a muon neutrino of even higher energy, in which case it represents the highest energy neutrino event in a neutrino telescope. However, diffuse flux limits from other experiments and the absence of a clear association with a transient source \cite{KM3NeT:2025npi} make the interpretation of KM3-230213A challenging. Instruments with pointing capabilities that are sensitive to neutrino sources in this energy range can facilitate the study such extreme energy events and potentially reveal their origins. Part of the Joint Exploratory Missions for
an Extreme Universe Observatory (JEM-EUSO) program \cite{ICRC2025Plebaniak}, the upcoming POEMMA-Balloon with Radio (PBR) mission \cite{ICRC2025Eser,PBRTeam:2026wxs} will be poised to make these observations or establish new limits on transient neutrino sources.

The PBR mission is a stratospheric balloon-based pathfinder for future satellite-based instruments such as the Probe of Multi-Messenger Astrophysics (POEMMA) \cite{POEMMA:2020ykm} and M-EUSO \cite{ICRC2025Plebaniak}. Scheduled to fly on a NASA Super Pressure Ballon (SPB), PBR will be launched from Wanaka, New Zealand in Spring 2028 and circle the Southern Ocean. A mission time of at least 20 days is anticipated. During the flight, science goals will be to observe high-altitude horizontal air showers \cite{Cummings:2021bhg}, make first measurements of fluorescence emission from ultra-high-energy cosmic ray (UHECR) air showers from above \cite{Battisti:2024rfv}, and most important to this discussion, search for Target-of-Opportunity (ToO) neutrino sources \cite{Heibges:2025yvw}. Multi-messenger event alerts will direct the schedule of instrument pointing for ToO searches \cite{Heibges:2025llt}.

PBR will have a hybrid focal surface comprising a 1.1 m Schmidt telescope with a Fluorescence Camera (FC) and Cherenkov Camera (CC) as a pathfinder instrument for POEMMA. Overall, the field of view (FoV) will be 36$^{\circ}$ (vertical) by 24$^{\circ}$
(horizontal). The CC uses silicon photomultipliers with a pixel size of 3 mm $\times$ 3 mm and pixel instantaneous FoV of $0.2^\circ$ by $0.2^\circ$. With 2048 pixels, the CC has a total FoV of $6^\circ$ (vertical) by $12^\circ$ (horizontal), 10\,ns integration time and a wavelength range of $320-900$ nm. The CC can capture ultra-fast pulses from optical Cherenkov light from up-going extensive air showers (EAS) sourced by Earth-skimming neutrinos \cite{Venters:2019xwi} with estimated neutrino energy thresholds of $\gtrsim 10$ PeV \cite{Cummings:2020ycz,PBRTeam:2026wxs}. With a NASA-supplied azimuthal rotator and a custom-designed tilt system, the instrument will be pointed to optimize ToO searches.
Typical astrophysical ToO events remain in the CC FoV for more than 20 minutes without active tracking \cite{Heibges:2025llt}.  The FC and CC designs for PBR build on experience from prior missions: EUSO-SPB1 \cite{JEM-EUSO:2023ypf} and EUSO-SPB2 \cite{Adams:2026ltd,Bagheri:2024byu,Adams:2024gsj}, and from the POEMMA design study \cite{POEMMA:2020ykm}.

New on PBR will be a dedicated Radio Instrument (RI) that benefits from the extensive development of the design of the Low Frequency instrument for the Payload for Ultra-High Energy (PUEO) experiment \cite{Abarr_2021}. The RI will be comprised of sinusoidal radio antennas with a broadband gain between $50-500$\,MHz. They will be mounted below the hybrid FC/CC telescope. The RI FoV will cover the entire optical FoV with 5~dB points of $\pm30^{\circ}$ from the boresight. The RI will be externally triggered by the CC, lowering its threshold to a few 10 PeV. {With coincident Cherenkov and radio detection, more direct information about the energy that initiates the shower may be extracted.}

Combined, the instruments on PBR are designed to detect neutrino signatures above $\sim 10$ PeV energies. PBR instruments will also be sensitive to events in this energy range from beyond-standard-model (BSM) physics, including 
models that may have  signatures like the one associated with the KM3-230213A event. Here, we calculate the PBR CC sensitivity to KM3-230213A-like
events in a BSM scenario with very energetic non-thermal dark matter (boosted DM) from an astrophysical source and determine the PBR discovery reach. A preliminary discussion of PBR CC capabilities for KM3-230213A-like
events was presented in ref. \cite{ICRC2025Olinto}. Here, we expand the analysis to include a  BSM scenario by Farzan and Hostert  \cite{Farzan:2025ydi} and present updated results.  In Sec.~\ref{sec:2} we review the characteristics of KM3-230213A event and its relation to the atmospheric muon background.   In Sec.~\ref{sec:3}
we discuss how PBR can investigate potential interpretations of KM3-230213A; in Sec.~\ref{sec:FH_model} we reformulate the  boosted dark matter scenario with the CC. In Sec.~\ref{sec:5} we evaluate the PBR discovery reach, and finally, we conclude in Sec.~\ref{sec:6}.

\section{The KM3-230212A event and the atmospheric muon background}
\label{sec:2}

\begin{figure}[tb]
  %\centering
 \includegraphics[width=0.48\textwidth]{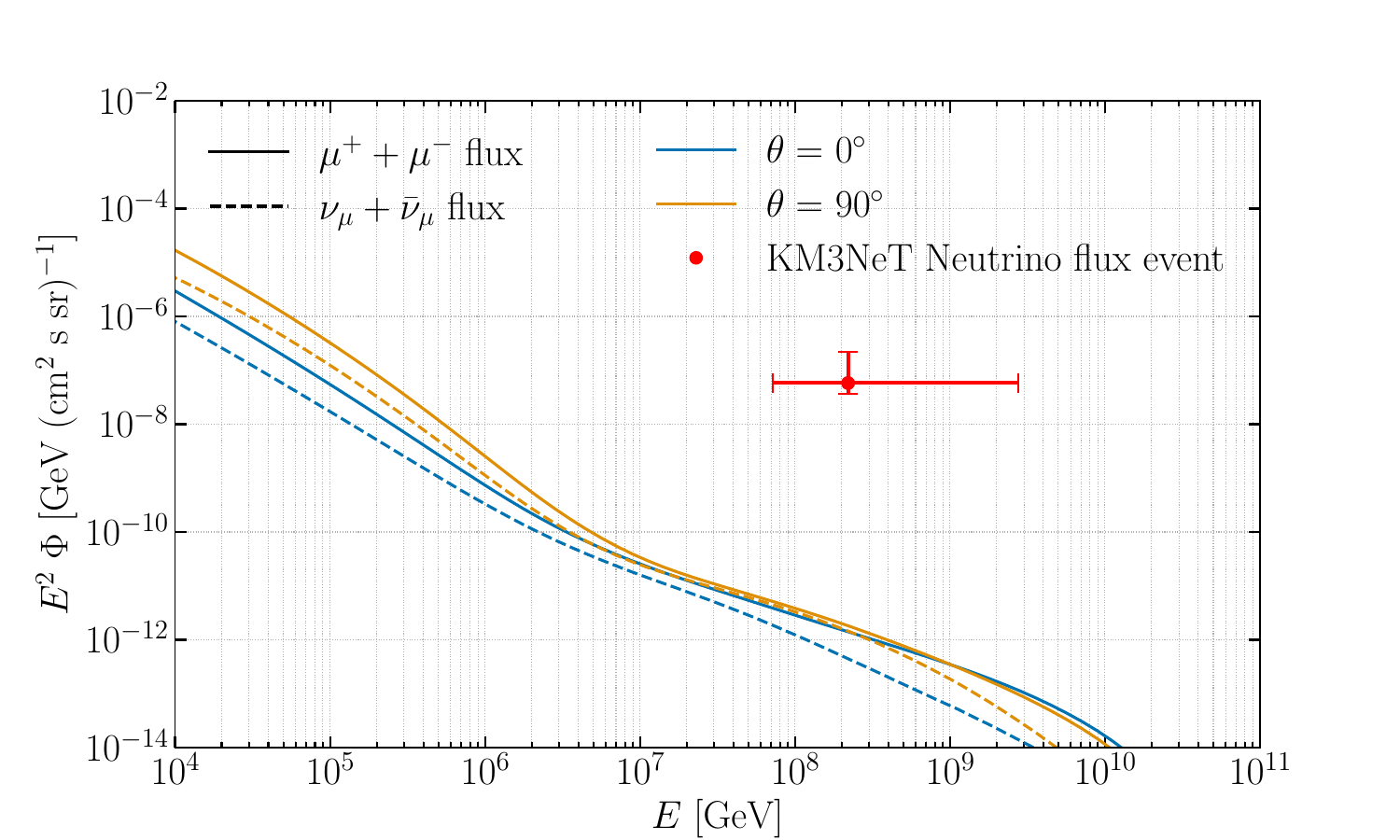} 
  \caption{Fluxes of atmospheric muons (solid lines, scaled by $E_\mu^2$) and atmospheric muon neutrinos (dashed lines, scaled by $E_\nu^2$) at sea level for zenith angles $0^\circ$ and $90^\circ$ evaluated with  MCEq~\cite{Fedynitch:2018cbl} using Sibyll 2.3c \cite{Fedynitch:2018cbl}
  and Gaisser-Hillas H3p cosmic ray spectrum \cite{Gaisser:2011klf}. The red data point with error bars shows the KM3NeT scaled isotropic per flavor neutrino flux~\cite{KM3NeT:2025npi}. Figure from ref. \cite{ICRC2025Olinto}.}
  \label{fig:atm-muons}
\end{figure}

With a 21 string partial implementation of the KM3NeT/ARCA detector and 284.7 days of observation, the KM3-230212A event was reported with a reconstructed energy $E_\mu = 120^{+110}_{-60}$\,PeV and 90\% CL energy range of $35-380$\,PeV~\cite{KM3NeT:2025npi}. The muon track was reconstructed as slightly down-going, with a nominal incident direction of $0.6^\circ$ above the horizon and an azimuth of $259.8^\circ$. As discussed below, the incident angle above the horizon is important for the interpretation of the event. The directional uncertainty as quoted in ref.~\cite{KM3NeT:2025npi} is $1.5^\circ$ (68\% CL). In principle, detection of Cherekov light in water by multiple detectors will permit an improved angular resolution, at the level of $0.12^\circ$, once detector positioning is better determined. Translated to celestial coordinates, the muon trajectory points back to ${\rm RA}=94.3^\circ$, ${\rm DEC}=-7.8^\circ$ with MJD $=59988.0533299$. Given KM3NeT/ARCA's effective aperture and exposure, and 
assuming the event originates from a steady isotropic neutrino flux, the scaled per flavor flux is $E_\nu^2 \Phi(E_\nu) = 5.8^{+10.1}_{-3.7} \times 10^{-8}$ GeV/(cm$^2$ s sr)~\cite{KM3NeT:2025npi}. This is shown with the red data point and error bars in fig.~\ref{fig:atm-muons}.

A steady isotropic $\nu_\mu$ flux at this level in the energy range of 72 PeV -- 2.6 EeV \cite{KM3NeT:2025npi} is in tension with limits from IceCube \cite{IceCube:2018fhm} and the Pierre Auger Observatory \cite{PierreAuger:2023pjg}.
Atmospheric muons are a potential source of background events to through-going muon events like the KM3-230212A event, however, the KM3NeT collaboration has ruled out this source. We review the argument here, starting with atmospheric muon flux on the surface, at sea level.

Using the Matrix Cascade Equation (MCEq) simulation software~\cite{Fedynitch:2018cbl} with the 
Sibyll 2.3c hadronic interaction model
\cite{Fedynitch:2018cbl} and the Gaisser--Hillas H3p cosmic ray spectrum and composition~\cite{Gaisser:2011klf}, the upper and lower solid lines in fig. \ref{fig:atm-muons} show the horizontal (zenith angle $\theta=90^\circ$) and vertical (zenith angle $\theta=0^\circ$) atmospheric $\mu^++\mu^-$ flux scaled by the muon energy $E_\mu^2$. The nominal incident muon zenith angle for the event is $\theta=89.4^\circ$, nearly horizontal. Above $E_\mu\sim 10$ PeV, prompt decays of charm hadrons and light unflavored mesons are the dominant sources of muons~\cite{Illana:2009qv, Illana:2010gh}, so the atmospheric muon flux is nearly isotropic in this energy regime. For reference, fig.~\ref{fig:atm-muons} also shows the horizontal and vertical atmospheric $\nu_\mu+\bar\nu_\mu$ fluxes.

Already at sea level, the flux of very high energy muons is small for $E_\mu>35$\,PeV, the lower end of the KM3-230212A energy range. The horizontal atmospheric muon flux follows an approximate energy scaling behavior: 
$E_\mu^2\phi_\mu \simeq  3.6\times 10^{-12}$(100 PeV/$E)^{0.9}$\,GeV/(cm$^2$s\,sr) for $E>10$ PeV. For $E>35$\,PeV, this corresponds to an integrated flux of $\sim 1.4\times 10^{-19}$ muons/(cm$^2$ s sr), thus $\sim 3.5\times 10^{-12}$ muons/(cm$^2$ sr) at sea level given the 287.4 days of data taking, assuming a 100\% duty cycle. Using a solid angle corresponding to a cone apex angle of $3^\circ = 2\times 1.5^\circ$ and a detector area of $\sim 0.4$\,km$^2$, the expected number of muons with $E_\mu  > 35$\,PeV at sea level that could be a background to the KM3NeT event is $\sim 2\times 10^{-3}$. At higher energies of $E_\mu > 120$\,PeV, this number decreases by roughly an order of magnitude.

While muons, with their long lifetimes, can penetrate to great depths, they also lose energy through electromagnetic interactions as they transit rock and water, so the number of atmospheric muons at the detector with energies above 35\,PeV in the direction of KM3-230112A is further reduced. To first approximation, muon energy loss depends only on the column depth $X$ traversed by the muon.  The average energy loss approximation for very high energy muons is $\langle dE_\mu/dX\rangle \simeq - b_\mu E_\mu$ with 
$b_\mu=3.5\times 10^{-6}$cm$^2$/g~\cite{Groom:2001kq,mu-eloss-web}. With the column depth of $X\sim 300$ kmwe between sea level and the detector for $\theta=89.4^\circ$, the $\langle dE_\mu/dX\rangle$ formula for the initial muon energy, $E_\mu^i\simeq E_\mu^f\exp(b_\mu X)$ gives an extremely high initial muon energy if $E_\mu^f=35$ PeV. Even for a shorter column depth of $X=60$ kmwe that corresponds to a trajectory $2.6^\circ$ above the horizon, $\theta=87.4^\circ$, $E_\mu^i\simeq 4.6\times 10^{10}$ PeV, again, an extremely high muon energy at the surface.

\begin{figure}[htb]
    \centering
    \includegraphics[width=0.45\textwidth]{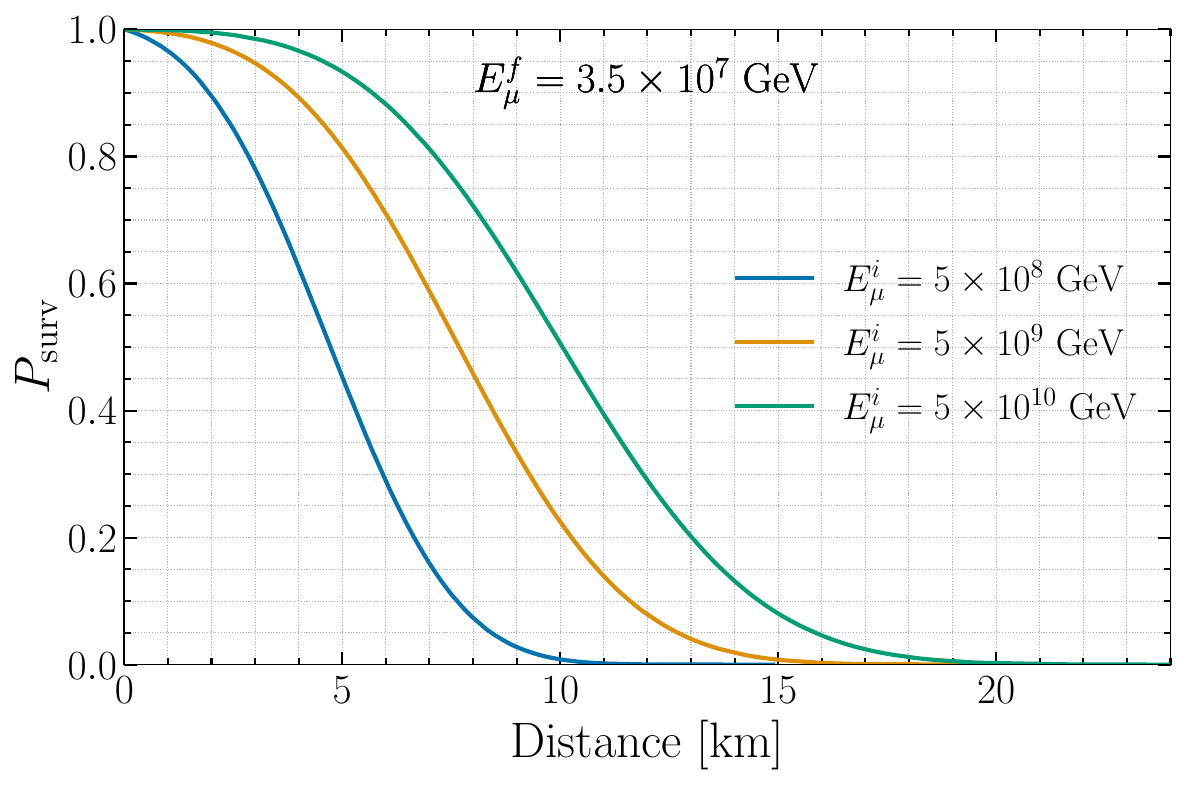}
    \includegraphics[width=0.46\textwidth]{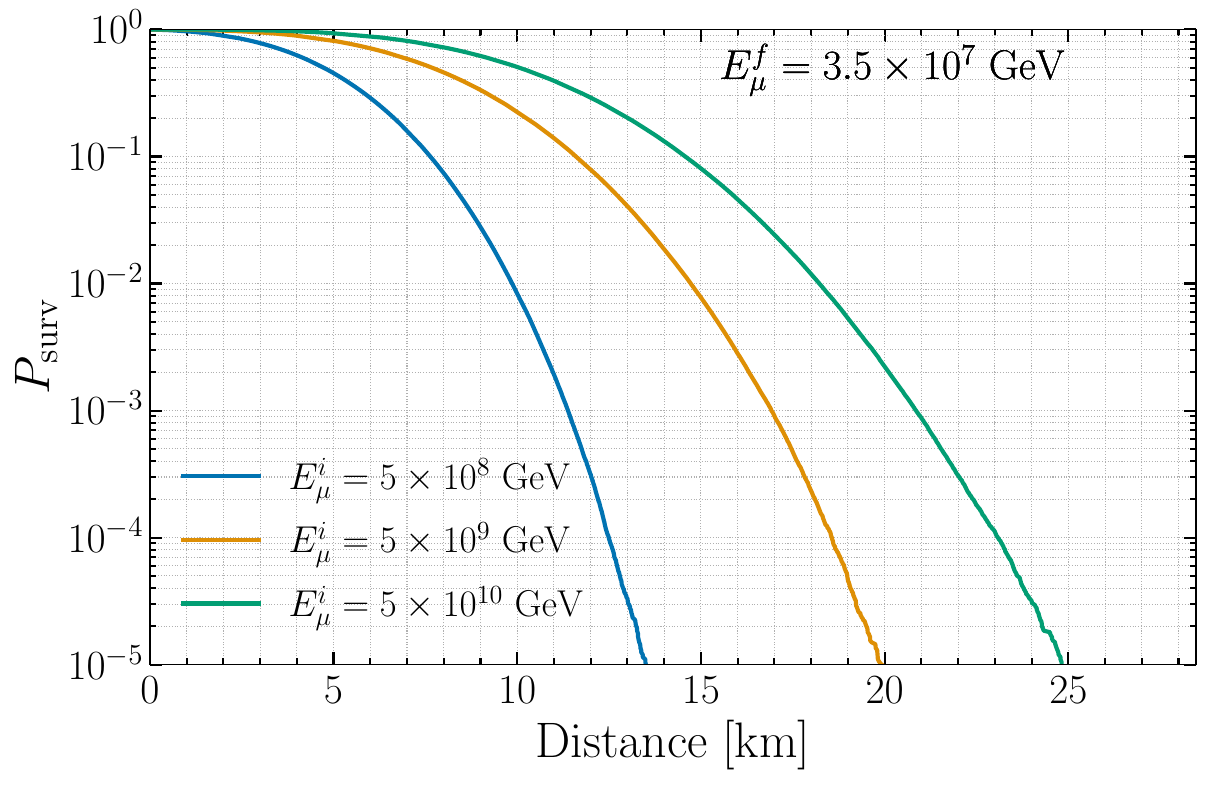}
    \caption{Muon survival probability as a function of distance traveled in water ($\rho=1$\,g/cm$^3$) for three initial energies, with a final muon energy of $E_\mu^f = 35$\,PeV. The lower panel shows the same results on log-linear axes. Figure from ref. \cite{ICRC2025Olinto}.}
    \label{fig:psurv-muons}
    \vspace{-2ex}
\end{figure}

Fluctuations in energy loss of muons that transit from the surface to the KM3NeT detector are not captured by the $\langle dE_\mu/dX\rangle$ formula \cite{Lipari:1991ut}. 
We modeled muon stochastic energy loss \cite{Dutta:2000hh} accounting for bremsstrahlung, pair production pair production contributions~\cite{Lohmann:1985qg} and photonuclear energy loss \cite{Abramowicz:1997ms,Abramowicz:1991xz}. We evaluated the muon survival probability as a function of distance traveled in water \cite{Lipari:1991ut, Dutta:2000hh} in the lowest energy in the 90\%\ CL energy range of the KM3-230213A event, $E_\mu^f= 35$\,PeV. The results for incident muons at the surface with $E_\mu^i=500,\ 5\,000$ and $50\,000$\,PeV and $E_\mu^f=35$\,PeV are shown in the upper (linear survival probability) and lower (log survival probability) panels of fig.~\ref{fig:psurv-muons}. 
Even for $E_\mu^i = 50\,000$\,PeV, the survival probability given $E_\mu^f>35$\,PeV falls below $10^{-5}$ after  $X\simeq 25$\,kmwe. It is negligible for $X=60$\,kmwe. Lower energy muons have an even smaller probabilities of surviving a distance of $60$\,kmwe for $E_\mu^f>35$\,PeV. As emphasized by the KM3NeT Collaboration, this rules out atmospheric muons as the origin of KM3-230213A event, even under the most conservative assumptions and at the lower end of the 90\% confidence interval for the reconstructed energy.

\section{Possible neutrino origins of KM3-230213A}
\label{sec:3}

% sec 3
With atmospheric muons ruled out as the source of the KM3-230213A event, we turn to investigate possible neutrino sources. Astrophysical neutrinos are expected to be produced as {\it smoking gun} byproducts when UHECRs (protons and heavy nuclei) interact with ambient matter and radiation~\cite{Gaisser:1994yf,Halzen:2002pg,Kotera:2011cp,Anchordoqui:2018qom,Coleman:2022abf}. These interactions occur either inside the extreme cosmic accelerators (like active galactic nuclei~\cite{Stecker:1991vm}, gamma-ray bursts~\cite{Waxman:1997ti}, and starburst galaxies~\cite{Loeb:2006tw}) or through the Greisen-Zatsepin-Kuzmin (GZK) effect~\cite{Greisen:1966jv,Zatsepin:1966jv} as cosmic rays propagate through intergalactic space~\cite{Berezinsky:1969erk}.

While the KM3NeT event could be attributed to a diffuse flux of cosmogenic neutrinos originating in GZK interactions~\cite{KM3NeT:2025vut,Muzio:2025gbr,Kuznetsov:2025ehl,Alhebsi:2026bdk}, reconciling this hypothesis with other observatories is problematic. Despite operating for considerably longer times and with greater effective areas, neither IceCube~\cite{IceCubeCollaborationSS:2025jbi} nor the Pierre Auger Observatory~\cite{PierreAuger:2023pjg} has detected neutrinos with $E_\nu \gtrsim 
40~{\rm PeV}$. %10^{7.6}~{\rm GeV}$. 
This null result underscores a $3.5\sigma$ tension between KM3NeT/ARCA and IceCube data for the cosmogenic hypothesis~\cite{Li:2025tqf,KM3NeT:2025ccp}. The tension between KM3NeT/ARCA and IceCube measurements is somewhat smaller if the neutrino flux originates from a transient source, at the level of $\sim 2.6\sigma$. Actually, assuming specific conditions for the transient model, a transient source is the least disfavored scenario and could reduce the tension to $2\sigma$~\cite{Yuan:2025isv}.

Separate constraints come from noting that gamma rays accompany ultra-high energy neutrinos in transparent astrophysical sources. The absence of a gamma-ray signal constrains both the source redshift and the intergalactic magnetic field strength between the source and Earth~\cite{Fang:2025nzg,Crnogorcevic:2025vou}. The gamma-ray constraints suggest that potential flaring sources are located at cosmological distances~\cite{Neronov:2025jfj,Yuan:2025zwe}. Despite these limitations, several potential associations with conventional astrophysical sources have been investigated. For example, a time delay between the gamma-ray and neutrino signals, attributed to in-vacuo dispersion, has been proposed for the gamma-ray burst GRB~090401B~\citep{Amelino-Camelia:2025lqn}. Additionally, multiple blazars have been found in spatial coincidence with KM3-230213A, with PMNJ0606-0724 exhibiting a notable radio flare 5 days before the neutrino event (chance coincidence probability $p = 0.26\%$)~\citep{KM3NeT:2025bxl}. This study also highlights theoretical work addressing neutrino production in both flaring and non-flaring gamma-ray states~\citep{Padovani:2019xcv,Petropoulou:2019zqp}. Furthermore, ref.~\cite{Sherman:2025gir} explores potential synchrotron pair-echo emissions if KM3-230213A originates within the large-scale structure surrounding a source of ultra-high-energy cosmic rays. Finally, intense pair production inside the source (through $\gamma\gamma \rightarrow e^{+}e^{-}$) can significantly suppress the gamma-ray emission while still allowing ultra-high-energy neutrinos to be produced~\citep{Kivokurtseva:2025sui}.

The difficulties in pinpointing conventional astrophysical sources have motivated the exploration of BSM physics scenarios that can produce extreme-energy neutrinos. For example, one scenario attributes the origin of the KM3NeT neutrino to the decay of super-heavy dark matter (SHDM) clustered in the Galactic halo~\cite{Borah:2025igh, Kohri:2025bsn, Jho:2025gaf, Barman:2025hoz, Murase:2025uwv, Khan:2025gxs,Aloisio:2025nts,Barger:2026zkd}. However, there are two major objections to interpreting the KM3NeT event as a signature of SHDM. First, the clustering of decaying particles in the halo precludes the necessary absorption of the gamma-ray flux. Second, the observed arrival direction is incompatible with the standard SHDM prediction of a dominant signal from the Galactic Center. Finally, a recent analysis by the KM3NeT Collaboration investigated the event's potential SHDM origin~\cite{KM3NeT:2026flu}. They determined that, across all scenarios, the most likely dark matter mass exceeds $\approx 100\text{ PeV}$ (at a 95\% confidence level), with a best-fit lifetime between $10^{26}\text{ and }10^{27}\text{ seconds}$. These parameter ranges are in direct conflict with current limits from gamma-ray observations and other neutrino telescopes~\cite{KM3NeT:2026flu}. However, scenarios with neutrinos that come from super-heavy long-lived relic decays in the early Universe evade gamma-ray constraints since the gamma-ray emission is reprocessed \cite{GrimbaumYamamoto:2025nai,GrimbaumYamamoto:2026kam}.

A somewhat different but related theoretical framework models dark matter as primordial black holes (PBHs). A recent study~\cite{Klipfel:2025jql} explores whether the KM3NeT-ARCA event originated from a transient, localized PBH explosion inside the Solar System at the end of the object's evaporation cycle. Indeed, realistic PBH mass distributions yield an 8\% likelihood of a primordial black hole explosion occurring within $10^3~{\rm AU}$ of Earth every 14 years.\footnote{Following ref.~\cite{Klipfel:2025jql}, we define the Solar System with its boundary encompassing the Kuiper Belt and Oort Cloud, reaching estimated distances of $10^{3}~{\rm AU}$ and $10^{5}~{\rm AU}$.} However, just as in the SHDM scenario, this hypothesis is challenged by the lack of an observed photon counterpart~\cite{Airoldi:2025opo}. A counterargument to this claim was recently presented in ref.~\cite{Klipfel:2026jrx}, showing that
if KM3-230213A originated in a PBH
explosion within the Oort Cloud the electromagnetic signal might have gone undetected given its large distance from Earth and the temporary outage of a major gamma-ray detector. Although future PBH explosions at similar ranges could be measured (both in $\gamma$-rays and neutrinos), success will depend heavily on the fortuitous alignment of the explosion locations with the active fields of view of our instrumentation at that moment. In addition, in models with large extra dimensions, PBHs may avoid $\gamma$-ray constraints because their Hawking radiation is dominated by bulk modes (such as sterile neutrinos)~\cite{Anchordoqui:2025xug,Anchordoqui:2025opy}. However, in the  strong-gravity regime, the branching ratios of sterile neutrinos favor competitive decay pathways into $l^\pm W^\mp$, $\nu
Z$, and $\nu h$, meaning an accompanying photon signal cannot be ruled out~\cite{Anchordoqui:2026ret}. Furthermore, similar to distant flaring sources~\cite{Neronov:2025jfj, Yuan:2025zwe}, attributing the KM3NeT detection to PBH models requires the observed event to be a statistical upward fluctuation to remain compatible with IceCube flux limits~\cite{Klipfel:2025jql,Anchordoqui:2025xug}. Consequently, both cosmological-scale flaring sources and nearby PBHs lie beyond the observational reach of PBR.

Yet another model for the KM3NeT neutrino event suggests that sterile-to-active neutrino oscillations, amplified by Earth's matter potential, could explain why the signal appeared at KM3NeT-ARCA but not at IceCube~\cite{Brdar:2025azm}. The varying path lengths and matter densities traversed by the neutrinos allow for this directional discrepancy. Two similar models, exploiting propagation distance differences have been proposed in~\cite{Farzan:2025ydi,Dev:2025czz}. Rather than sterile neutrinos, these models rely on interaction chains  inside the Earth initiated by ultra-high-energy  dark sector particles. 
Under this class of models, the KM3NeT detection is treated as an observation of a high energy source, not an observation of a statistical fluctuation capable of accommodating IceCube data, thereby implying such sources could fall within the reach of PBR. We examine one such model comprehensively in the following section.

\section{BSM Boosted Dark Matter:  Farzan-Hostert Scenario}\label{sec:FH_model}

As discussed in the previous section, a potential explanation of the very high energy event, KM3-230213A, is a series of interactions that begin with a source of very high energy dark sector particles denoted here by $\chi$ and generically as dark matter. As outlined by Farzan and Hostert in ref.~\cite{Farzan:2025ydi}, one scenario that could account for the KM3NeT observation of a muon with a 90\%\ confidence interval energy of $35\,{\rm PeV}<E_\mu<380\,{\rm PeV}$ is a series of interactions and decays that lead to a pair of muons that are nearly collinear and mimic a single high energy muon. Here, we take the same scenario, however, with a tau pair in the final state rather than a muon pair.

Following ref.~\cite{Farzan:2025ydi}, we consider a flux of energetic $\chi$ that scatters coherently off a nucleus in the Earth with mass number $A$ to produce $\chi'$, $\chi A\to \chi' A$. The $\chi'$ promptly decays to a 2-body final state which includes a new meta-stable particle $\phi$: $\chi'\to \chi + \phi$. The long lifetime of the $\phi$, followed by $\phi\to \mu\mu$, is one of the keys to this scenario's  account of the KM3NeT observation in the absence of an IceCube observation. Here, we expand the possible final states of the $\phi$ decay to $\phi\to \ell\ell$ for $\ell = \mu$ or $\ell = \tau$ and reference the branching fractions for $\phi$ decays to lepton pairs as $B_{\mu\mu}$ and $B_{\tau\tau}$, respectively. All of the discussion below follows this scenario \cite{Farzan:2025ydi}: an astrophysical source with a fluence of high-energy  $\chi$ which arrive at Earth, interact, and through $\chi\to\phi\to\ell$, produce observable signals of the leptons.

Relevant distance scales are the $\chi$ interaction distance which depends on the inverse of the $\chi A$ cross section, $\sigma_{\chi A}$, and the decay length of the $\phi$ ($\ell_{d,\phi}$). We assume that the decay length of the $\chi'$ is very short compared to the interaction distance and the $\phi$ decay length. In terms of column depth, the $\chi$ interaction length and $\phi$ decay length are written,
\begin{eqnarray}
   \lambda_\chi &=& (\sigma_{\chi A}N_A/A)^{-1}\\ 
   \lambda_{d,\phi}&=&\ell_{d,\phi}\rho\,
\end{eqnarray}
where $N_A$ is Avogadro's number, $A$ is the target atomic mass and $\rho$ is the target density.
Guided by ref. \cite{Farzan:2025ydi} to explain the KM3NeT event, we write the coherent cross section in terms of the $\chi-$proton cross section $\sigma_0$, set $A=22$, and assume that the $\phi$ decay length for $E_\phi=100$ PeV is 100 km, 
\begin{eqnarray}
\label{eq:sigma-xiA}
    \sigma_{\chi N} &=& A^2\sigma_0 \\
\label{eq:ell-dphi}
     \ell_{d,\phi} &=& 100\ {\rm km}\ \Bigl(\frac{E_\phi}{ 100\ {\rm PeV}}\Bigr)\,.    
\end{eqnarray}
As in ref. \cite{Farzan:2025ydi}, we assume that $\sigma_0$ is independent of energy. 
We also simplify the cross section and decay kinematics in the same manner as in ref. \cite{Farzan:2025ydi}.
In this scenario, the average energies are related to the initial energy of the $\chi$ by
\begin{equation}
\label{eq:energy-relations}
    E_\chi = 2E_\phi = 4E_\ell^i \,,
\end{equation}
for $\ell = \mu$ or $\ell = \tau$ for $\phi\to \mu\mu$ or $\phi\to \tau\tau$. The superscript $i$ in $E_\ell^i$ allows us to account for electromagnetic energy loss of the produced $\ell$ pair as they transit from the decay point of the $\phi$ to the detector, or to emerge from the Earth to produce a signal in PBR. We use the average energies of eq.~(\ref{eq:energy-relations}) to evaluate the sensitivity of PBR to this scenario for energetic tau-pairs in the final state.

\begin{figure}[htb]
    \centering
\includegraphics[width=0.45\textwidth]{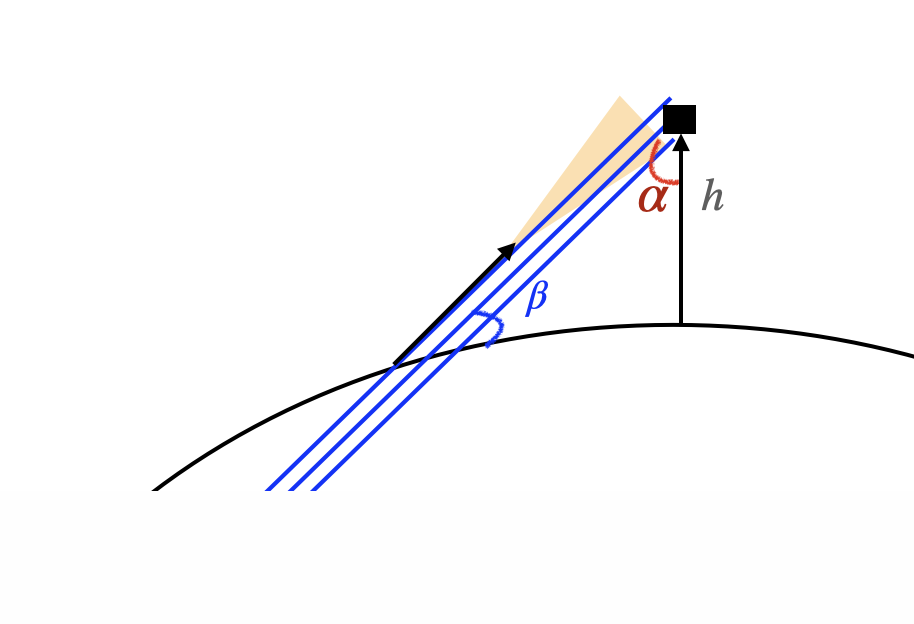}
\caption{Geometry of distant $\chi$ source, which through $\chi\to \phi\to \tau$ generates EAS from $\tau$-lepton decays in the atmosphere. To illustrate, an exaggerated Cherenkov cone is shown with the orange triangle. The Earth-emergence angle is labeled by $\beta$. Not to scale. \label{fig:geometry}} 
\end{figure}

To determine the sensitivity of PBR to the final state taus, we evaluate the probability that a $\chi$ with energy $E_\chi$ incident on the Earth at angle $\beta$ produces a tau that exits the Earth, $P_{\rm exit}(E_\chi,\beta)$. Here, $\beta$ is relative to the tangent to the Earth's surface: small $\beta$ means Earth-skimming trajectories, as shown in fig. \ref{fig:geometry}. For PBR on a balloon payload at $h=33$ km, the Earth's limb is $5.8^\circ$ below the payload's horizon. With the zenith coverage of $6^\circ$ of the Cherenkov telescope \cite{PBRTeam:2026wxs}, if the camera is pointed to cover the limb of the Earth and below, $\beta\simeq 0^\circ-10^\circ$ \cite{Reno:2021veo}. We also evaluate the energy distribution of the outgoing taus for incident $E_\chi$ and $\beta$.

The tau exit probabilities account for attenuation of $\chi$ in the Earth, the $\phi$ decay length, and the tau electromagnetic energy loss and its decay probability. Again we can use the approximate lepton electromagnetic energy loss formula for column depth $X$, now for high energy tau-leptons, $\langle dE_
\tau/dX\rangle\simeq -b_\tau E_\tau$, 
with $b_\tau\simeq 5\times 10^{-7}$\,cm$^2$/g (see, e.g., ref. \cite{Dutta:2000hh}). 
The approximate energy loss formula and energy independent $b_\tau$ greatly simplify the discussion \cite{Dutta:2005yt}, since tau energy and column depth traveled are related. 
$E_\tau^i$ is the tau energy at the $\phi$ decay point. The electromagnetic energy loss relates the energy $E_\tau$ at a column depth $\Delta X$ from its production point by
\begin{equation}
\label{eq:etau-z}
    E_\tau(\Delta X) = E_\tau^i\exp(-b_\tau \Delta X)\,.
\end{equation}
This means that for a tau to emerge with energy $E_\tau$ at the end of the trajectory with a total column depth $X_\beta$, the energy loss formula and eq. (\ref{eq:energy-relations}) prescribe the relation between $E_\tau^i$, $E_\tau$ and where the tau is produced ($X$) according to
\begin{equation}
    E_\tau = \frac{E_\chi}{4} \exp \left[-b_\tau(X_\beta - X)\right]\,.
\end{equation}
We use this relation between $X$ and $E_\tau$
\begin{equation}
\label{eq:z-beta}
    X = X_\beta -\frac{1}{b_\tau}\ln\left(\frac{E_\chi}{4E_\tau}\right)
\end{equation}
in the equations below, noting that $X$ depends on $\beta$ through $X_\beta$.

We use the Preliminary Earth Reference Model \cite{PREM} to calculate the column depth $X_\beta$ along the trajectory described by $\beta$. For our evaluation of the tau exit probability $P_{\rm exit}$ and the tau energy distribution as it emerges from the Earth, we approximate the density along the trajectory by the average density $\rho = \rho(\beta)$. With these approximations, we write the $\phi$ and $\tau$ survival probabilities \cite{Dutta:2005yt}:
\begin{eqnarray}
\label{eq:psurv-phi}
    P_{\phi,\rm surv}(E_\phi,X) &=& \exp \Biggl(-\frac{X}{\lambda_{d,\phi}(E_\phi)}\Biggr)\,,\\
    \label{eq:psurv-tau}
     P_{\tau,\rm surv}(E_\tau,E_\tau^i) &=& \exp \left[-\frac{m_\tau}{c\,\tau_\tau\,\rho \,b_\tau}\left(\frac{1}{E_\tau}-\frac{1}{E_\tau^i}\right)\right]  
\end{eqnarray}
where $E_\tau = E_\tau(X)$.

\begin{figure}
    \centering
    \includegraphics[width=0.45\textwidth]{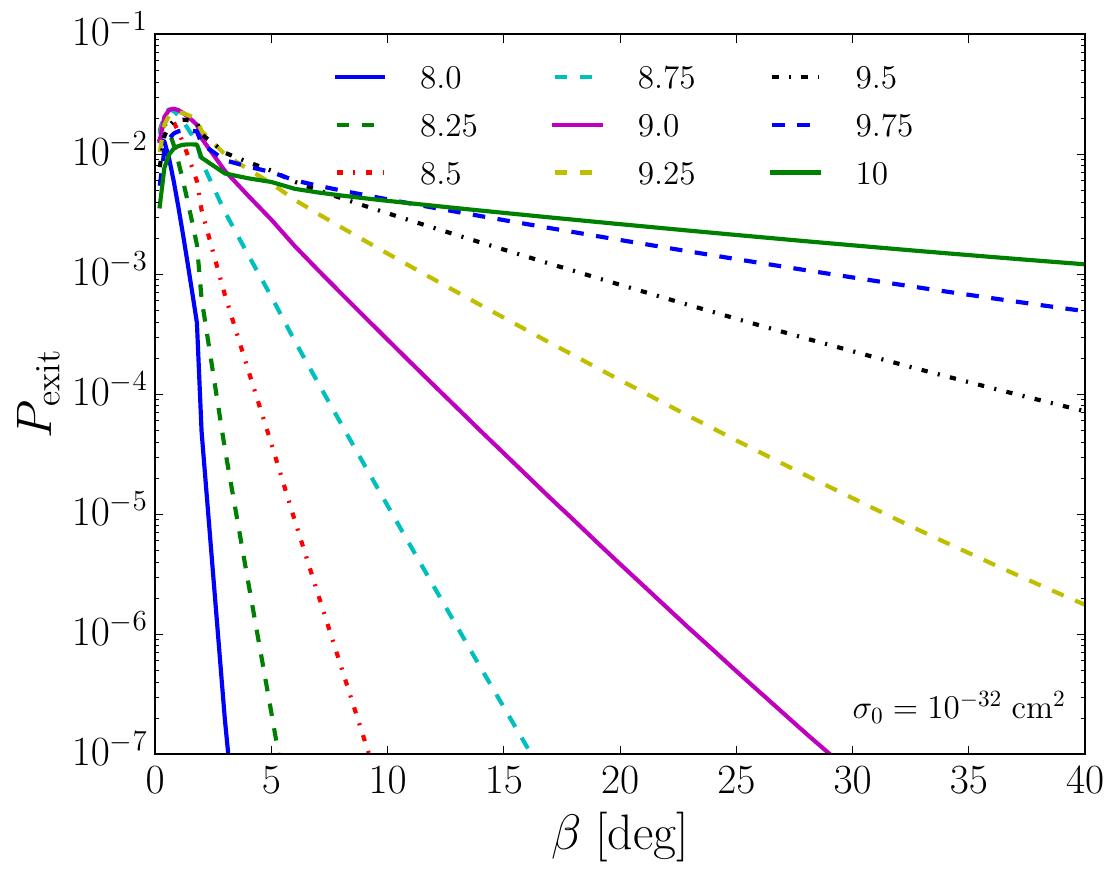}  
    \includegraphics[width=0.45\textwidth]{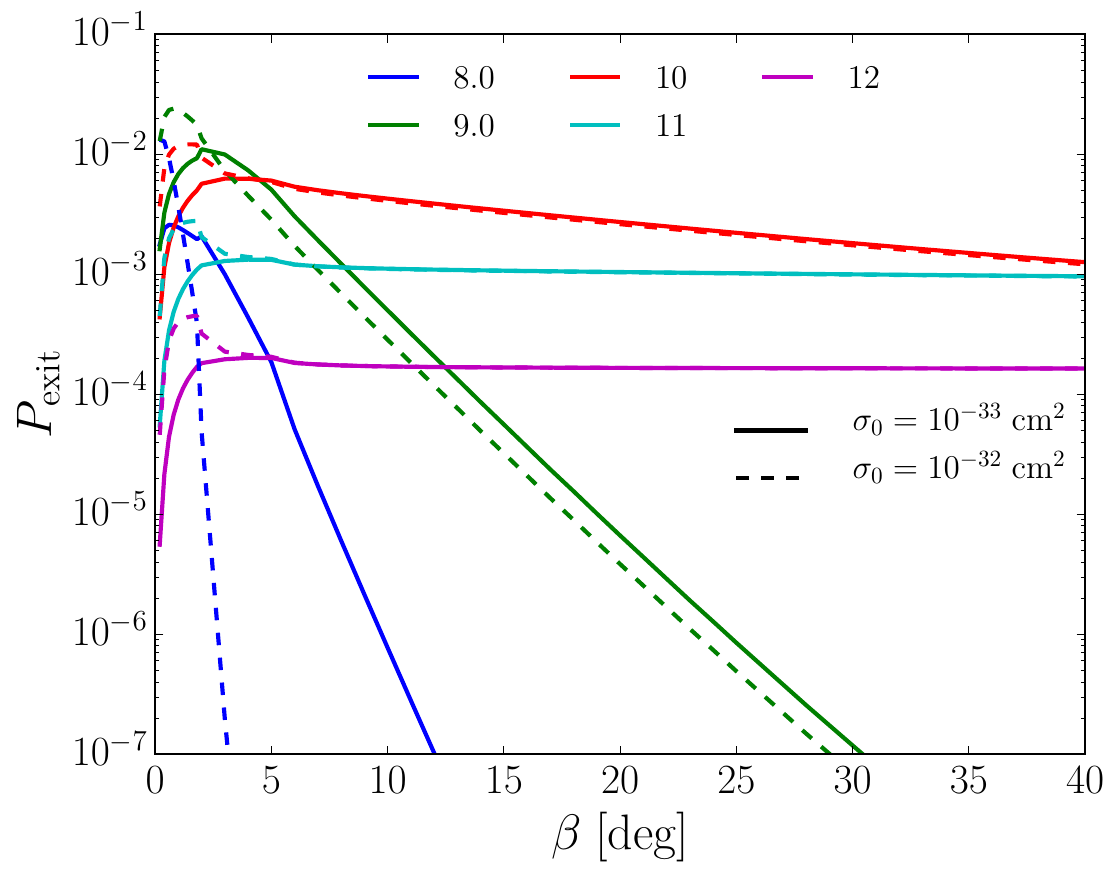}
    \caption{Upper: The tau exit probabilities as a function of Earth emergence angle $\beta$ for $\sigma_0=10^{-32}$ cm$^2$, for $E_\chi=10^n$ GeV with $n=8-10$ as indicated in the figure labels. Lower: The tau exit probabilities as a function of $\beta$ for 
    $\sigma_0=10^{-33}$ cm$^2$ (solid) and $\sigma_0=10^{-32}$ cm$^2$ (dashed) for selected $E_\chi$ values.
    In both panels, the $\phi\to \tau\tau$ branching fraction $B_{\tau\tau}=1$ is assumed, and a factor of 2 accounts for the 2 tau-leptons in the final state.}
    \label{fig:pexit}
\end{figure}

Subsequently, the energy distribution of emerging taus can be evaluated numerically, so that the tau exit probability, accounting for 2 taus in the final state, is
\begin{eqnarray}
P_{\rm exit}(E_\chi,\beta) &=&
\int\, dE_\tau 
        \frac{dN_\tau(E_\tau,\beta)}{dE_\tau}\\
        \nonumber
         \frac{dN_\tau(E_\tau,\beta)}{dE_\tau}
        &=& 
    \frac{2 B_{\tau\tau}}{E_\tau \,b_\tau} \exp\left[ -\frac{m_\tau}{c\,\tau_\tau \,b_\tau\,\rho(\beta)} \left(\frac{1}{E_\tau} - \frac{1}{ E_\chi/4}\right)\right]\\
    \nonumber
    &\times& \left[\exp\left(-\frac{z}{\lambda_{d,\phi} (E_\chi/2)}\right)- \exp\left(-\frac{z}{\lambda_{\chi} ( E_\chi)}\right) \right]\\
    &\times & \left[ \lambda_{d,\phi}(E_\chi/2)
    -{\lambda_\chi(E_\chi)} \right]^{-1}\,.
\end{eqnarray}
We will also use the notation
$d_\tau(E)\equiv (E_\tau/m_\tau)c\tau_\tau$ and $\lambda_{d,\tau}=d_\tau \rho$. 
For $\sigma_0=10^{-32}$ cm$^2$, independent of energy, and $B_{\tau\tau}=1$, the results for the tau exit probabilities for several values of $E_\chi$ are shown as a function of $\beta$ in the upper panel of fig.~\ref{fig:pexit}. In the lower panel, the solid lines show the tau exit probabilities for $\sigma_0=10^{-33}$ cm$^2$, while the dashed lines are for $\sigma_0=10^{-32}$ cm$^2$ for a range of $E_\chi$.

The main dependence of the tau exit probabilities on $\beta$ is through $X$  from eq.~(\ref{eq:z-beta}). 
The qualitative behavior of $P_{\rm exit}(E_\chi,\beta)$ follows from a simplified picture. At least for some values of $\sigma_0$, we can approximate  $\chi$ interactions as occurring immediately inside the Earth to produce a $\phi$ with $E_\phi =E_\chi/2$. For a non-zero exit probability, the $\phi$ must survive almost the whole distance $d_\beta\equiv X_\beta/\rho(\beta)$ along the chord length for a given angle $\beta$. 
By requiring the $\phi$ to decay within  the range of the tau such that an energetic tau emerges, approximately $d^{\rm eff}_\tau(E_\chi/4)\equiv\min (d_\tau(E_\tau^i), 20$ km) \cite{Dutta:2000hh}, as long as $d^{\rm eff}_\tau\ll d_\beta$, the exit probability for the $\tau$ is roughly
\begin{equation}
\label{eq:pexit-approx}
    P_{\rm exit}^{\rm approx}
    \sim \exp\Biggl[-\frac{d_\beta}{\ell_{d,\phi}(E_\phi=E_\chi/2)}\Biggr]\frac{d^{\rm eff}_\tau(E_\chi/4)}{\ell_{d,\phi}(E_\chi/2)}\, 2 B_{\tau\tau}\,.
\end{equation}
This approximate equation, which does not depend on $\lambda_\chi$,  represents the exit probability to within a factor of $\sim 1/2-2$ for most energies for $\beta\gtrsim 5^\circ$ as long as $X_\beta\gg \lambda_\chi$. 
The overlap of the exit probabilities for the two values of $\sigma_0$ for $E_\chi=10^{10}-10^{12}$ GeV and $\beta\gtrsim 5^\circ$ in the lower panel fig.~\ref{fig:pexit} illustrates this effect.
When $X_\beta$ is smaller than or comparable to $\lambda_\chi$, the rough approximation  in eq.~(\ref{eq:pexit-approx}) breaks down and the shapes of the exit probability curves change. 
At lower energies, and at smaller angles, the dashed and solid curves in the lower panel of fig.~\ref{fig:pexit} do not overlap.

For $\sigma_0=10^{-33}$ cm$^2$, $\lambda_\chi\simeq 800$ kmwe. For $\beta\lesssim 3^\circ$, $X_\beta<\lambda_\chi$. In this case, fewer $\chi$ interact than when $\lambda_\chi\simeq 80$ kmwe ($\sigma_0=10^{-32}$ cm$^2$), so exit probabilities are lower, regardless of $E_\chi$.
In the lower energy range, there are additional effects at larger angles. For example, for $\beta=5^\circ$, $X_\beta\simeq 3000$ kmwe. For $E_\chi=10^8$ GeV, the relevant $\phi$ and $\tau$ decay lengths times density are $\lambda_{d,\phi}\simeq 130$ kmwe and $\lambda_{d,\tau}\simeq 3$ kmwe. With $\lambda_\chi\simeq 800$ kmwe, the probability that a $\phi$ survives to within 3 kmwe of the Earth's surface is several orders of magnitude larger than for $\lambda_\chi\simeq 80$ kmwe. On the other hand, for $E_\chi=10^9$\,GeV, $\lambda_{d,\phi}\simeq 1300$ kmwe, so a much larger fraction of $\phi$ traverse $X_\beta$, regardless of $\lambda_\chi$. The solid and dashed curves in the lower panel of fig.~\ref{fig:pexit} differ by less than a factor of two for $E_\chi=10^9$\,GeV. In the evaluations of the projected PBR sensitivities, the results will show better sensitivities for $\sigma_0=10^{-33}$\,cm$^2$ than for $\sigma_0=10^{-32}$\,cm$^2$ for $E_\chi\lesssim 3\times 10^9$\,GeV.

\begin{figure}
    \centering
\includegraphics[width=0.45\textwidth]{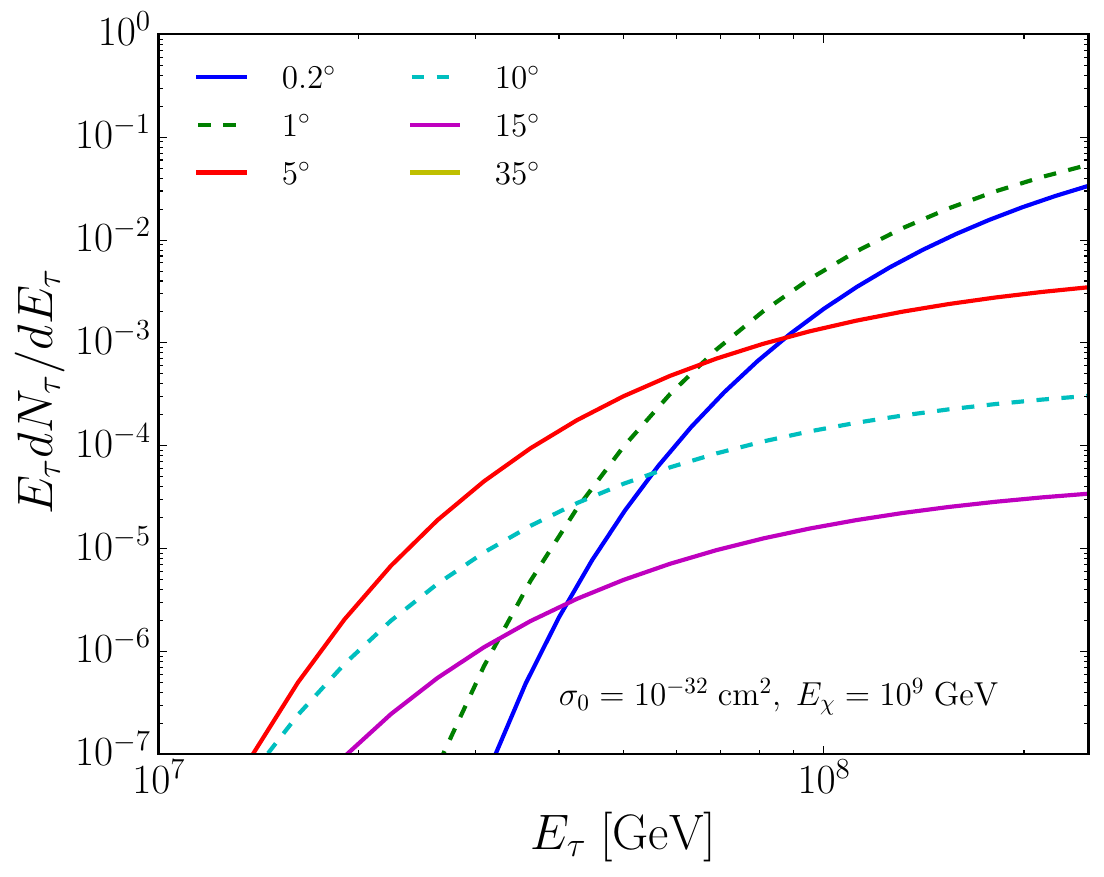}  
\includegraphics[width=0.45\textwidth]{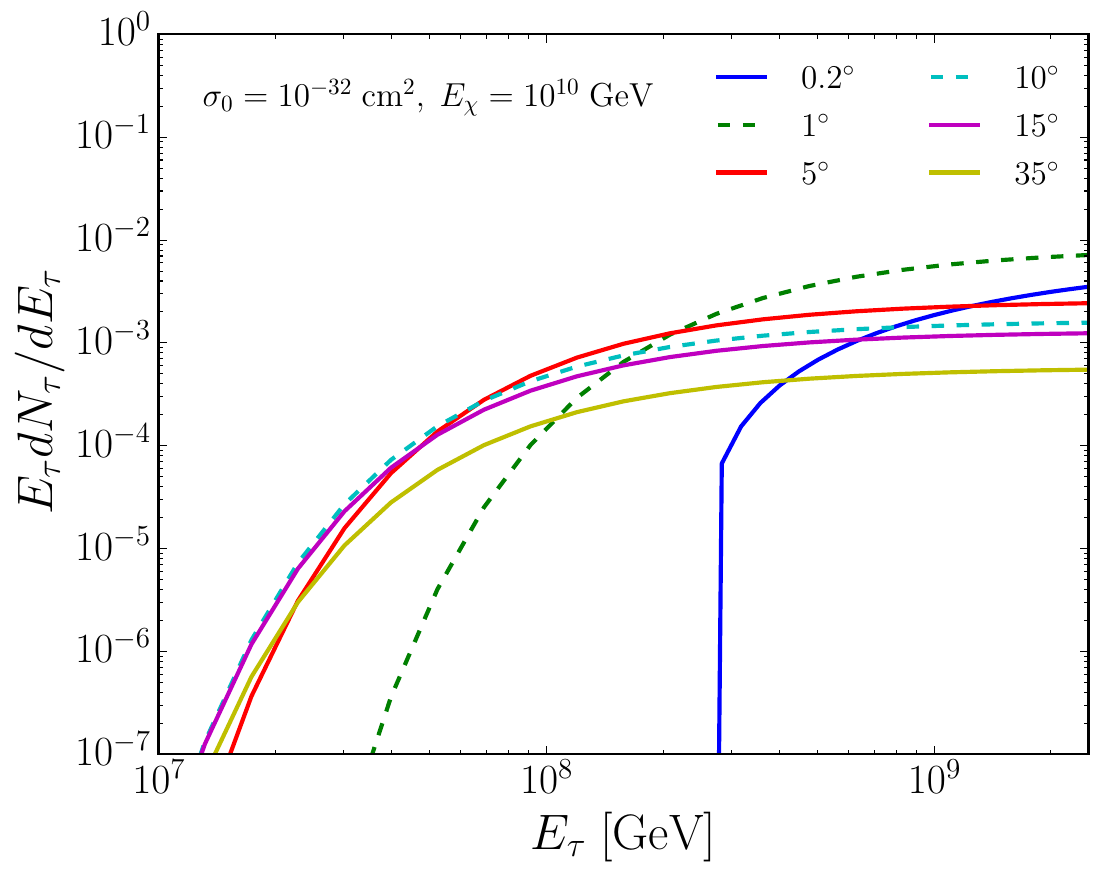}  
\caption{Scaled by tau energy, the differential energy distribution of taus that exit the Earth for $\beta=0.2^\circ - 35^\circ$ as labeled in the figure for $E_\chi = 10^9$ GeV (upper) and $10^{10}$ GeV (lower), assuming  $B_{\tau\tau}=1$ and $\sigma_0=10^{-32}$ cm$^2$.\label{fig:endne}}
\end{figure}

The energy distributions of the exiting taus are shown  for some select angles in fig.~\ref{fig:endne} for $E_\chi=10^9$\,GeV and $10^{10}$\,GeV for $\sigma_0=10^{-32}$\,cm$^2$. To evaluate PBR's acceptance, the differential energy distributions are used to generate cumulative distribution functions which are sampled as a function of $E_\chi$ and $\beta$ to get emerging tau energies.

Figures \ref{fig:pexit} and \ref{fig:endne} are shown with $B_{\tau\tau}=1$. Below, we will show the projected sensitivities for the $\chi$ fluence times $B_{\ell\ell}$ for the cases where $B_{\mu\mu}=B_{\tau\tau}$ and for a flavor dependent relation $B_{\tau\tau}=B_{\mu\mu}\times (m_\tau/m_\mu)^2$. Appendix \ref{App:A} shows that if $\phi$ is a pseudoscalar that couples to leptons axially, the enhancement factor of $(m_\tau/m_\mu)^2$ follows in the limit that $m_\tau/m_\phi$ is small.

\section{PBR discovery reach}
\label{sec:5}

\begin{figure}[tb]
    \centering
\includegraphics[width=0.45\textwidth]{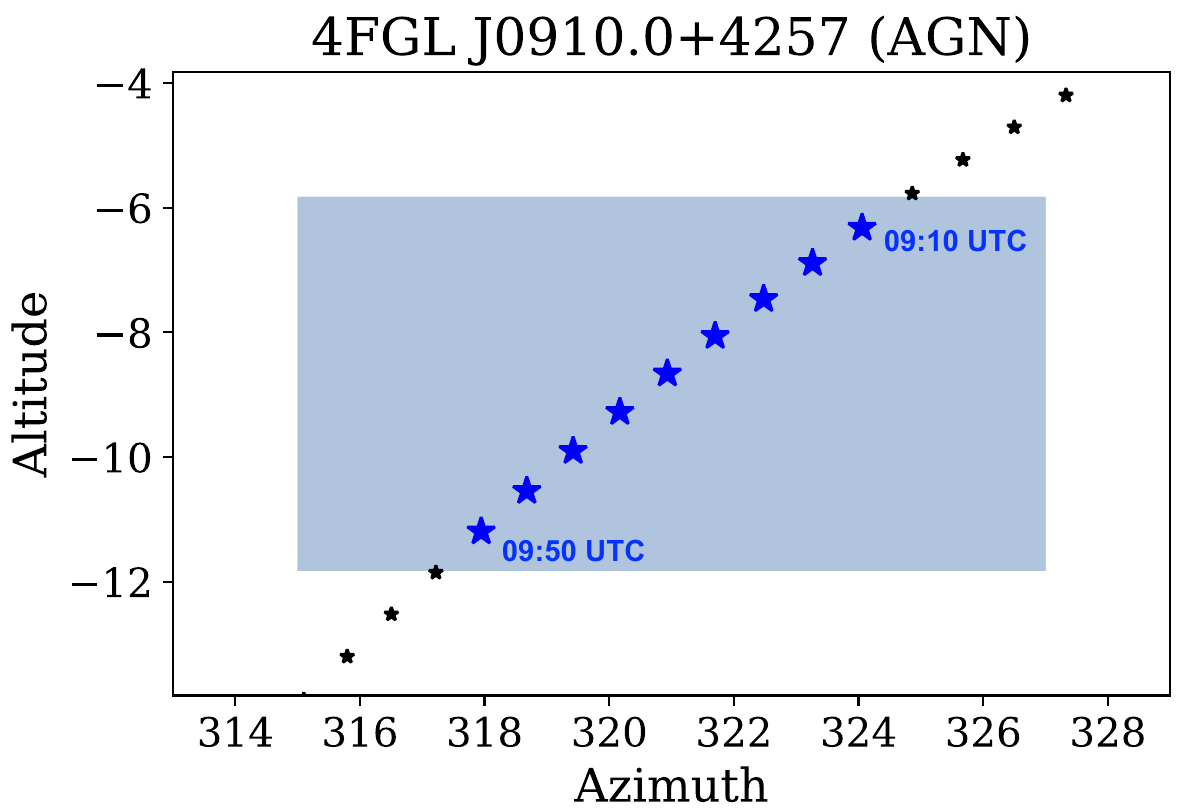}
\caption{Sample source that passes through the PBR Cherenkov camera field of view. The camera's zenith field of view begins at the Earth's limb and spans $6^\circ$, and azimuth $\Delta \phi=12^\circ$. 
\label{fig:source-fov}} 
\end{figure}

For standard model signals, the principal means of PBR to observe sources of very high energy neutrinos is through the Earth-skimming technique with $\nu_\tau\to \tau$ conversions in the Earth followed by $\tau$-leptons exiting the Earth and initiating EASs in the atmosphere. The EASs generate Cherenkov radiation that can be detected by the PBR CC. At sea level, the optical Cherenkov angle is approximately $\sim 1.5^\circ$ \cite[see, e.g., ref.][for more details]{Reno:2019jtr}. Standard model interactions determine the tau exit probabilities and tau energy distributions \cite{Reno:2025tpw,Garg:2022ugd,Alvarez-Muniz:2018owm}, which in turn are used to evaluate the instrument's $\nu_{\tau}$ acceptance and ultimately, its sensitivity \cite{Venters:2019xwi}. In the BSM Farzan-Hostert scenario with $\phi\to\tau\tau$, the $\phi$ decays in the Earth produce $\tau$-leptons that emerge to initiate the EASs. Therefore, using the BSM $P_{\rm exit}$ and $dN_\tau/dE_\tau$ distributions from the previous section, the sensitivity evaluation for $\chi$ sources is similar to that of standard model $\nu_{\tau}$ sources.

Regardless of how the $\tau$-leptons that emerge from the Earth are produced, the PBR exposure must account for the motion of an astrophysical source through its FoV. In the scenario considered here, it is the source of boosted $\chi$s that moves through the PBR CC's FoV. An example of a typical source trajectory (an active galactic nucleus in this example) passing through the FoV is shown in fig.~\ref{fig:source-fov} \cite{Heibges:2025nuts}.
Then, assuming the KM3-230313A observation was in fact linked to a $\chi$ event, we can compare our $\chi$ fluence sensitivity to the $\chi$ fluence nominally detected by KM3NeT, as estimated in ref.~\cite{Farzan:2025ydi}. 
As noted above, we focus our discussion on the capabilities of PBR's CC for observing optical Cherenkov signals. The RI will be triggered by the CC. Future improvements can be made by combining the RI and CC measurements. 

The acceptance is limited by observing conditions. An essential condition is dark skies: astronomical night and little to no light from the Moon. In the discussion below, $f(t)$ represents an efficiency factor that accounts for observational dead periods due to the Sun and Moon \cite{Venters:2019xwi}. In our simulations, we assume PBR's trajectory as a function of time follows historical wind patterns corresponding to a 20-day flight launching roughly in mid-late April from Wanaka, New Zealand. At every instant in time, we determine $f\left(t\right)$ by calculating the elevation angles of the Sun and the Moon and the illumination of the Moon using the UTC time and the balloon's locations as inputs. If the Sun's elevation angle as viewed by the balloon is less than 18 degrees below the Earth's limb, $f\left(t\right) = 0$ at that instant in time. The calculation for the Moon is more complex as periods in which only a small fraction of the Moon is illuminated may permit observations. To determine when such periods occur, we calculate the average intensity of moonlight, including direct and scattered components, which we rescale from radiative transfer simulations of sunlight performed using libRadtan~\cite{2016GMD.....9.1647E}. The moonlight intensity is modulated by the phase of the Moon, and for our calculations, we determine the critical phase angle at which the calculated moonlight intensity is $\gtrsim  10$\% of the estimated intensity of the night sky airglow ($\sim 18800$ phot m$^{-2}$ ns$^{-1}$ sr$^{-1}$ integrated from $300$--$1000$ nm to cover the wavelength range of the CC). At a given instant in time, observation is possible ($f\left(t\right) = 1$) if the Moon's phase angle is greater than the critical phase angle.\footnote{Note that the phase angle ranges from $0^{\circ}$--$180^{\circ}$, where $0^{\circ}$ corresponds to a full moon and $180^{\circ}$ corresponds to a new moon.} For phase angles less than or equal to the critical phase angle, observation will only be possible if the Moon is below the Earth's limb. For more details on the Sun and Moon calculations, see ref.~\cite{Venters:2021xos}.

Not all  sources of $\chi$ will be observationally accessible to PBR on a given night, even with dark skies. A $\chi$ source, located at equatorial coordinate right ascension $\phi$ and  declination $\delta$, must pass below the Earth's limb as seen by PBR to take advantage of the Earth skimming technique. Again, the location of the balloon with PBR is relevant.
As fig. \ref{fig:source-fov} indicates, typical transit times through the  CC FoV are of order 20 min \cite{Heibges:2025llt}.
Even so, because of the 33 km float altitude of the balloon, the instantaneous acceptance for VHE neutrinos is expected to be
comparable to, or better than, ground-based detectors \cite{Venters:2021RV,Reno:2021veo,Heibges:2023yhn}. 

The PBR sensitivity to the $\chi$ fluence from a point source depends on the time average of the source and balloon location-specific instantaneous acceptance. The instantaneous acceptance 
starts with the area of the detector FoV projected on to the surface of the Earth and  
incorporates $P_{\rm exit}$ and $dN_\tau/dE_\tau$ 
of an emerging $\tau$-leptons, which depend on the location at which the $\tau$-lepton emerges from the Earth, given in terms of Earth latitude, $\lambda_{\oplus}$, and longitude, $\phi_{\oplus}$.
Light propagation to the detector  
and detector light thresholds are also components of the acceptance.

The instantaneous acceptance for a given energy $E_\chi$,
$A_{\tau}(\beta(t,\lambda_{\oplus},\phi_{\oplus}),E_{\chi},\phi,\delta)$,
 is averaged over the observation time $T_{\rm obs}$ and weighted by the time dependent observation efficiency $f(t)$.
Because the $\phi\to \tau\tau$ branching fraction ($B_{\tau\tau}$) is unknown, we 
define $\langle A_{\tau}(E_\chi, \phi,\delta)\rangle_{T_{\rm obs}}$ with $B_{\tau\tau}=1$ and explicitly include the branching fraction as a factor.
The time-averaged acceptance can be written
\begin{eqnarray}
\label{eq:avga}
B_{\tau\tau}\langle A_{\tau}(E_{\chi},\phi,\delta)\rangle_{T_{\rm obs}}
&=& \frac{1}{T_{\rm obs}}\int _{t_0}^{t_0+T_{\rm obs}}\, dt \ f(t)\\
\nonumber
 &\times&
B_{\tau\tau} A_{\tau}(\beta(t,\lambda_{\oplus},\phi_{\oplus}),E_{\chi},\phi,\delta)\,.
\end{eqnarray}
Some of the variables on which $A_{\tau}$ depends are suppressed for simplicity.

We performed Monte Carlo simulations to sample the region of interest and to calculate the detection probabilities of sampled events, accounting for the features described above. We have set the minimum photon density at the detector to be 200 $\gamma$/m$^2$ and used an average photon detection efficiency of 20\%. Simulations that model the PBR CC instrument with the EUSO-OffLine software \cite{JEM-EUSO:2023fyg}, including a wavelength dependent photon detection efficiency, indicate that these adopted values are conservative.

The sensitivity for $E_\chi^2$ times the $\chi$ energy-differential fluence ${\cal F}_\chi$ over observation time $T_{\rm obs}$ per energy decade for PBR can be approximated by  
\begin{equation}\label{eq:sensitivity}
\mbox{Sensitivity}\times B_{\tau\tau} = \frac{2.44}{\ln(10)}
\times\frac{E_{\chi}}{ \langle A_{\tau} (E_{\chi})\rangle_{T_{\rm obs}}} \ ,
\end{equation}
where $\ln(10)$ accounts for the per-decade sensitivity~\cite{Anchordoqui:2002vb}. The factor of 2.44 comes from the Feldman-Cousins $90\%$ unified confidence level assuming no signal and no background \cite{Feldman:1997qc}. 

\begin{figure}
    \centering
    \includegraphics[width=0.85\linewidth, trim = 11cm 1cm 7cm 1cm, clip]{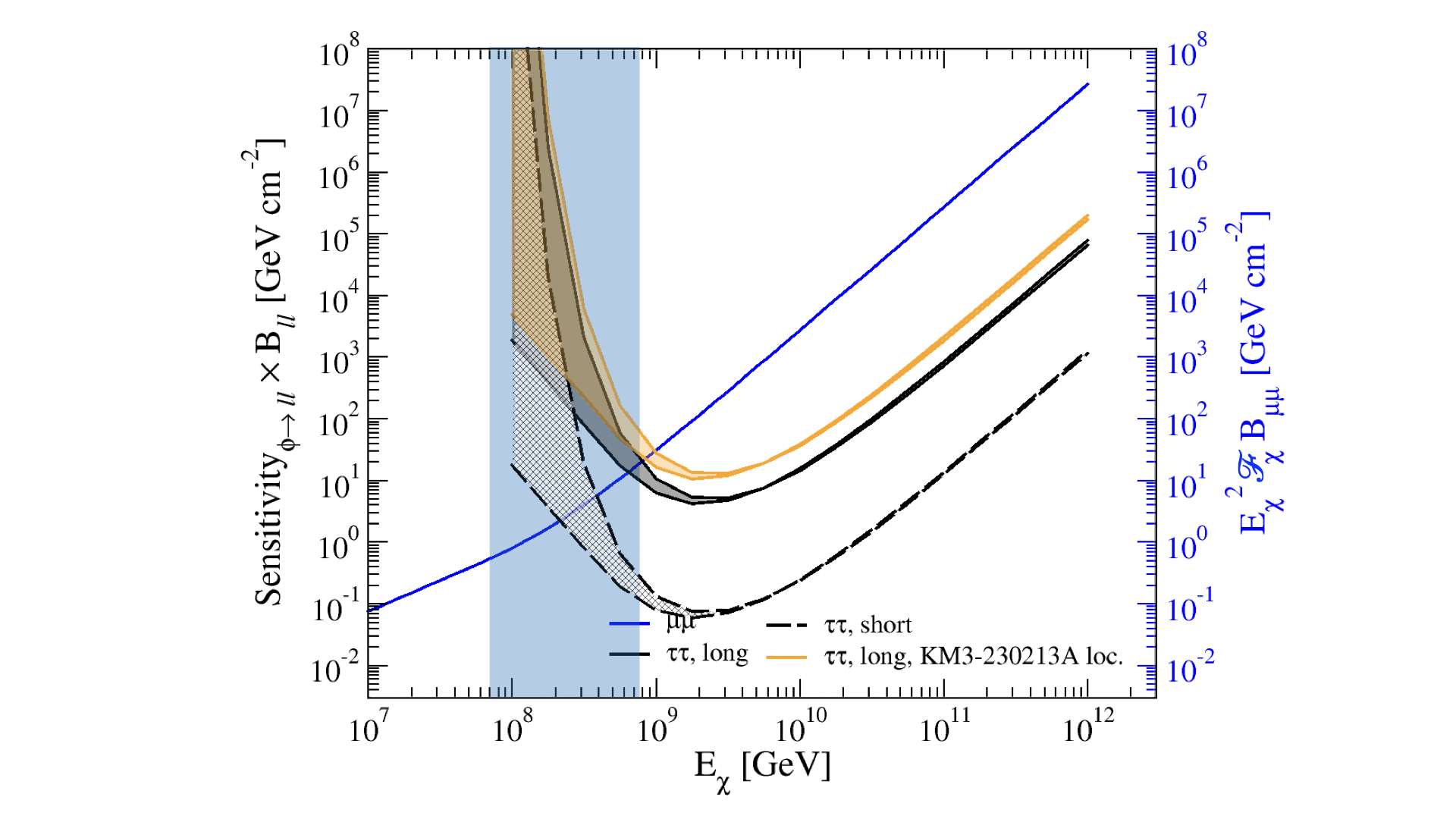}
    \caption{The projected PBR sensitivity multiplied by $B_{\tau\tau}$ as a function of $E_\chi$ to EASs generated by $\chi\to\phi\to \tau$ in Earth, 
    analogous to a scenario of $\chi\to \phi\to \mu$ in ref. \cite{Farzan:2025ydi}. The black and orange bands represent the ranges in sensitivity corresponding to 
      $\sigma_0 = 10^{-33}-10^{-32}$ cm$^{2}$ for the lower and upper range of PBR sensitivity bands. Black curves and corresponding bands represent the best achievable sensitivities 
    for a $\chi$ source in an optimal position on the sky. The orange curves and band represent sensitivities at the nominal sky location of KM3-230213A. Solid curves are for a transient event lasting over the duration of the flight (taken to be $20$ days). Dashed curves are for a burst lasting $T_{\rm obs}=1000$ s. The solid blue curve is the approximate sensitivity of KM3Net to $\mu\mu$ pairs. The blue shaded region shows the energy range $E_\chi=70-760$ PeV, assuming the energy of KM3-230213A is the sum of the energies of the two muons.}
    \label{fig:PBRSensBtautau}
\end{figure}

\begin{figure}
    \centering
    \includegraphics[width=0.85\linewidth, trim = 11cm 1cm 7cm 1cm, clip]{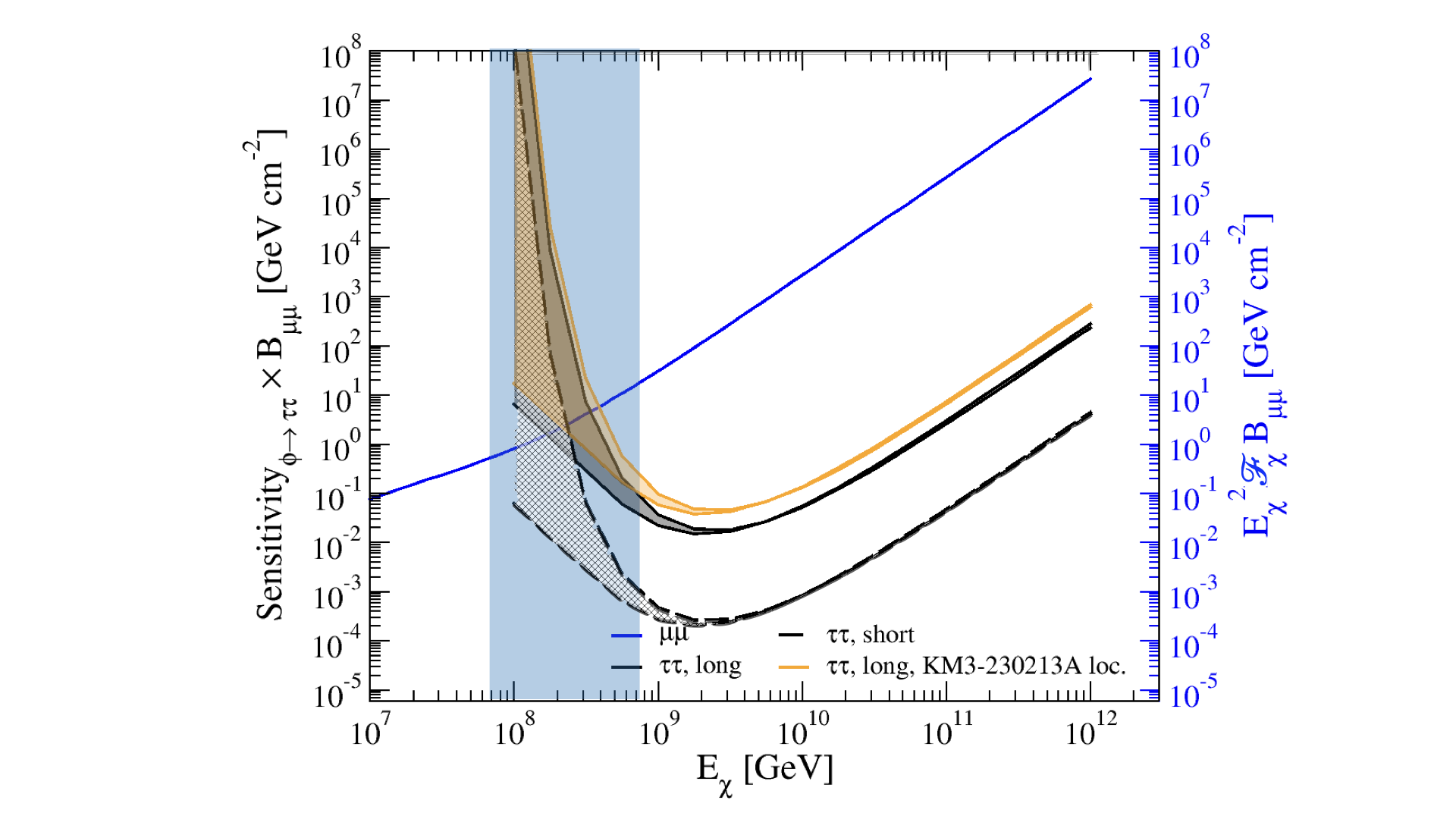}
    \caption{Same as in Fig.~\ref{fig:PBRSensBtautau} except that the PBR sensitivity is multiplied by $B_{\mu\mu}$ with the $\tau$-lepton branching fraction scaling with the square of the $\tau$-lepton mass ($B_{\tau\tau}=B_{\mu\mu}\times (m_\tau/m_\mu)^2$). 
    Curves and bands are the same as in Fig.~\ref{fig:PBRSensBtautau}.}
    \label{fig:PBRSensBmumu}
\end{figure}

We consider two observation scenarios. For the first scenario, we consider a short, 1000-s burst of a transient $\chi$ source. For this scenario, we assume ideal observing conditions and that the source is in the optimal position to be observed at the start of the burst. This means that the skies are dark ($f\left(t\right)=1$) and there is no cloud coverage in the direction of the source. Also, the telescope is pointing at the Earth's limb in the direction of the source just as it dips below the horizon at the burst's initial time $t_0$. For such a short burst, the balloon motion is negligible. The second scenario assumes $\chi$ emission lasting longer than $\sim 20-50$ days during which time PBR observes the source. The choice of 20 days is motivated by typical balloon mission durations. 
For the long observation periods, we use a balloon trajectory following historical wind patterns, after an assumed PBR launch date of April 25, 2028. Results for PBR's sensitivities to high energy $\chi$ sources are shown in figs. \ref{fig:PBRSensBtautau} and \ref{fig:PBRSensBmumu}. Black curves and corresponding bands are calculated for a source in an optimal position on the sky.

The dashed black curves in fig. \ref{fig:PBRSensBtautau} show the projected PBR sensitivity of the energy-scaled quasi-differential $\chi$ fluence times the $
\phi$ branching fraction to $\tau$-lepton pairs $(\ell\ell=\tau
\tau$) as a function of energy for $T_{\rm obs}=1000$ s. The upper black  dashed curve is evaluated using a $\chi N$ cross section of $\sigma_0=10^{-32}$ cm$^2$, while the lower dashed curve is for $\sigma_0=10^{-33}$ cm$^2$. Taking $\sigma_0=10^{-31}$ cm$^2$ yields approximately the same sensitivity as for $\sigma_0=10^{-32}$ cm$^2$.

The solid black curves show the PBR sensitivity to the energy-scaled fluence times $B_{\tau\tau}$ with optimal observing conditions for $T_{\rm obs}=20$ days, while the orange curves  show the 20-day sensitivity for a $\chi$ source at the position of KM3-230213A.  The spread in the solid black curves and orange curves again shows the range of $\sigma_0 = 10^{-33}-10^{-32}$ cm$^{2}$.

PBR's burst sensitivity to $\tau\tau$ pairs from $\chi\to\phi\to \tau$ is complementary to the  $\mu\mu$ pair fluence that follows from KM3-230213A as discussed in ref. \cite{Farzan:2025ydi}. The $\chi$ fluence times $B_{\mu\mu}$, extrapolated to higher energies following the discussion in ref. \cite{Farzan:2025ydi}, is shown here with the blue curve and $y$-axis labels in blue using 
\begin{eqnarray}
\label{eq:aeffmu}
        \langle A_{\rm eff}^\mu(E_\chi)\rangle &\simeq  &
    A_0\frac{1-\exp(-L_{km}/\ell_{d,\phi}(E_\phi))}{1-\exp(-L_{km}/100\ {\rm km})}\\
    \label{eq:fhsensitivity}
    E_\chi^2{\cal F}_\chi (E_\chi)B_{\mu\mu} &\simeq & \frac{E_\chi}{\langle A_{\rm eff}^\mu(E_\chi)\rangle } 
   \,,
\end{eqnarray}   
where $L_{km}=140$ km is the path length distance in the Earth of the trajectory of KM3-230213A, and $A_0\simeq  10^{-2}$ km$^2$ is the effective area for KM3NeT for $\ell_{d,\phi}(E_\phi=100\ {\rm PeV})= 100$ km and $\sigma_0=10^{-32}$ cm$^2$
(see fig. 4 of ref. \cite{Farzan:2025ydi}). The estimated value of $A_0$ is a factor of $\sim 0.2$ smaller for $\sigma_0=10^{-33}$ cm$^2$, meaning that $E_\chi^2{\cal F}_\chi (E_\chi)B_{\mu\mu}$ would be a factor of $\sim 5$ larger in this case.
In ref. \cite{Farzan:2025ydi}, the $\mu\mu$ pair with a very small opening angle carries the energy of the $\phi$, responsible for the muon signal in KM3NeT in the range of $E_\mu^{\rm tot}= 35-380$ PeV. In this scenario, $E_\chi=70-760$ PeV. The energy range of the $\chi$ for KM3-230213A is shown with the blue band in fig. \ref{fig:PBRSensBtautau}. 

To compare the PBR sensitivity to the nominal KM3NeT $\chi$ fluence, an assumption must be made about the relation between $B_{\tau\tau}$ and $B_{\mu\mu}$. For the relevant $\chi$ energy range and assuming $B_{\tau\tau}=B_{\mu\mu}$, the crossover of the PBR sensitivity to a 1000 s $\chi$ burst and the KM3NeT fluence represented by the blue line is at  $E_\chi\simeq 4\times 10^8$ GeV for $\sigma_0=10^{-32}$ cm$^2$. 
For $\sigma_0=10^{-33}$ cm$^2$, the estimated value in eq. (\ref{eq:fhsensitivity}) scales with $A_0^{-1}$, and with $A_0 \simeq 2\times 10^{-3}$ km$^2$, the KM3NeT fluence of $\chi$ increases by a factor of $\sim 5$. The crossover energy in this case would be $E_\chi\simeq 1.5\times 10^8$ GeV. The 
PBR sensitivity is considerably lower than the fluence of eq. (\ref{eq:fhsensitivity}) at higher energies, for example, by a factor of $10^{-3}$ at $E_\chi=10^9$ GeV for $\sigma_0=10^{-32}$ cm$^2$.

For long bursts, with a 20-day observing period, with $\sigma_0=10^{-32}$ cm$^2$, PBR's sensitivity crosses the blue curve in fig. \ref{fig:PBRSensBtautau} at $E_\chi\sim 8\times 10^8$ GeV, just outside the blue band. For $\sigma_0=10^{-33}$ cm$^2$, the crossover occurs at {$E_\chi\simeq 5\times 10^8$ GeV}.
The figure shows that the nominal sky location of KM3-230213A is close to optimally located. While not as sensitive in the KM3NeT energy range,  PBR will be more than an order of magnitude more sensitive than the $\chi$ fluence represented by the blue line for $E_\chi$ above $\sim 5\times 10^9$ GeV.

A reasonable alternative to equal $\phi$ branching fractions to muons and taus is for $B_{\tau\tau}=B_{\mu\mu}\times (m_\tau/m_\mu)^2$. This assumption relies on the premise that $\phi$ has a pseudoscalar coupling to lepton pairs. Conservation of angular momentum requires the final-state leptons from a spin-$0$ particle decay to have identical helicities. However, the chiral nature of the axial interaction prefers opposite helicities. Thus, to conserve both energy and spin, one of the leptons must undergo a spin-flip. The probability of this spin-flip is proportional to the mass of the lepton $m_{\ell}$. Consequently, the decay rate is heavily suppressed for lighter leptons, leading to a branching fraction that inherently depends on $m_{\ell}^{2}$. For details, see Appendix~\ref{App:A}. The results with this assumption for $B_{\tau\tau}$ are shown in fig. \ref{fig:PBRSensBmumu}, where for both vertical axes, the PBR sensitivity to $\phi\to \tau\tau$  and the $\chi$ fluence from $\phi\to\mu\mu$, are multiplied by the same unknown $B_{\mu\mu}$. This assumption for the branching fraction for $\tau$-lepton pairs improves the sensitivity by a factor of $\sim 20$. For $\sigma_0=10^{-33}$ cm$^2$, the evaluated PBR sensitivity is stronger than the approximate KM3NeT sensitivity represented by eqs. (\ref{eq:aeffmu}-\ref{eq:fhsensitivity})  with $A_0=2\times 10^{-3}$ km$^2$ for the whole $E_\chi$ energy range in the blue band.

\begin{figure}[htb]
    \centering
    \begin{minipage}[t]{0.99\linewidth}
        \centering
        \includegraphics[width=0.92\linewidth, trim = 0.05mm 0.05mm 3mm 3mm, clip]{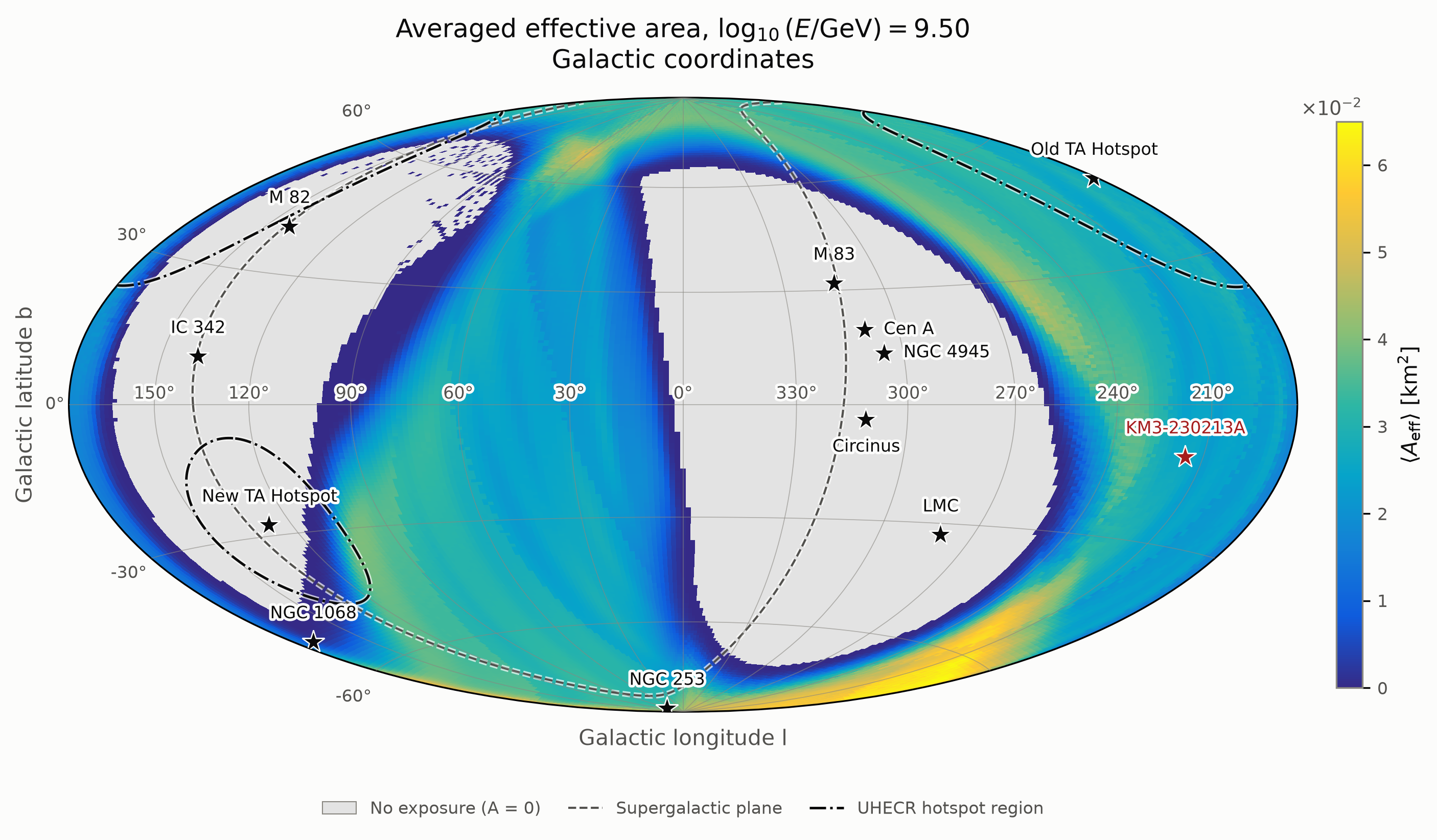}
        \caption{PBR sky coverage in Galactic coordinates assuming a 20-day flight duration for a balloon trajectory that follows historical wind patterns. The launch date was set to April 25, 2028 to coincide with the new moon. The $\chi$ acceptances  were calculated for $\langle A_\tau (E_\chi)\rangle$ with $E_\chi=10^{9.5}~{\rm GeV}$ and assuming $B_{\tau\tau} = 1$. For reference, the black stars indicate selected nearby astrophysical sources and reported hotspots from the Telescope Array (TA) experiment, whereas the red star indicates the arrival direction of KM3-230213A. 
        \label{fig:skycoverage}} 
    \end{minipage}
\hfill
\vspace{-2ex}
\end{figure}

As already noted, PBR's acceptance depends on the sky location of the source of high energy $\chi$ particles. Not all regions of the sky are accessible during a 20-day flight, and not all observable locations are close to optimal. Figure \ref{fig:skycoverage} shows the PBR sky coverage in Galactic coordinates for a 20-day flight after an April 25, 2028 launch. The sky map confirms that the KM3-230213A location is close to optimal for a flight with the stated assumptions about launch and duration. A longer than nominal flight duration will permit observations of a greater portion of the sky.

On a final note, we discuss the impact of clouds on the calculations presented above. In the BSM Farzan-Hostert scenario, as in the Earth-skimming $\nu_{\tau}$ scenario, $\tau$-leptons emerge from the Earth and then decay at some altitude to initiate upward-going EASs. For $\tau$-leptons that decay at altitudes below the height of the clouds, the Cherenkov light from their EASs will be obscured. We do not account for this effect in the calculations and results presented above since we are showcasing the best achievable sensitivities for PBR and because the presence of clouds is sporadic. However, we can provide a conservative estimate of their impact on the PBR sensitivity. Based on studies of clouds using the Modern-Era Retrospective Analysis for Research and Applications, Version 2~\cite[MERRA-2; ][]{2017JCli...30.5419G} database, the average cloud coverage for the ocean at New Zealand latitudes is $\sim 60$\%--$70$\% and ranges in altitude up to $\sim 2$ km. To study the impact of clouds, we performed Monte Carlo calculations of the $\chi$ acceptance as above but cutting simulated $\tau$-leptons that decay at altitudes below $2$ km. In these calculations, we find that the greatest impact is at lower energies with the energy threshold increasing from $\sim 2 \times 10^{8}$ GeV to $\sim 3 \times 10^{8}$ GeV and the acceptance at $10^{9.5}$ GeV worsening by about a factor of 3. At higher energies, the effect of clouds becomes increasingly marginal as high-energy $\tau$-leptons are more likely to decay at higher altitudes. We emphasize that these estimates are conservative. Cloud coverage is not expected to be $100$\% for all times as we are assuming here. Also, clouds are not expected to be completely opaque, and EASs that start below clouds may still be detectable if they are sufficiently bright. Nonetheless, even in this conservative approach, we see that PBR will still have sensitivity to $\chi$ sources over a large range of energies.

\vspace{-2ex}

\section{Conclusions}
\label{sec:6}

Among the observing goals of the PBR mission is the search for up-going EASs initiated by Earth-skimming tau neutrinos. During these observation periods where the CC is pointed below the Earth's limb, the instrument will be sensitive to other, more exotic signals that may originate from BSM physics. The KM3NeT/ARCA observation of a very high energy nearly horizontal muon event has sparked searches for BSM physics explanations, introducing new particles and/or new interactions. Non-observation of similar events by IceCube and the Pierre Auger Observatory dictate how the particles that produced the muon in the event transit the Earth. This is analogous to requirements that the unusual up-going ANITA \cite{ANITA:2016vrp,ANITA:2018sgj} events put on BSM physics at higher energies and larger angles (see, e.g. ref. \cite{Chipman:2019vjm}). PBR's capabilities to detect up-going EASs via their optical Cherenkov signal from a balloon-based platform will allow to probe properties of the BSM physics associated with particle transit in the Earth.

In this paper, we have selected one scenario \cite{Farzan:2025ydi} with boosted DM $\chi$ to illustrate the scale of PBR's sensitivity reach and energy range. The scenario has a particle interaction and decay chain that takes $\chi\to\phi\to\ell$ ($\ell =\mu,\,\tau$) in the Earth. Taking the high energy muon event as a signal of a transient $\chi$ source, the KM3-230212A observation sets the scale of the fluence of boosted DM from a source in the direction identified by KM3NeT. 

Our results are summarized in figs. \ref{fig:PBRSensBtautau} and \ref{fig:PBRSensBmumu}, where comparisons between the expected fluence and the PBR sensitivity are made assuming $B_{\tau\tau}=B_{\mu\mu}$ and $B_{\tau\tau}=B_{\mu\mu}\times (m_\tau/m_\mu)^2$, respectively. The blue energy band shows the range of $E_\chi$ that would be associated with the KM3NeT event, but in principle, a source of boosted DM would have a spectrum of energies. PBR has discovery capabilities above $E_\chi \sim 1.8-8\times 10^8$ GeV, depending on parameters such as the the $\chi N$ cross section $\sigma_0$, $B_{\tau\tau}$ and other parameters that are inputs to the evaluation of $P_{\rm exit}$ and $dN_\tau/dE_\tau$.

Under ideal observing conditions, when $B_{\tau\tau}=B_{\mu\mu}$ and $\sigma_0=10^{-32}$ cm$^2$, PBR will have a sensitivity to short (1000 s) $\chi$ bursts for $E_\chi\gtrsim 4\times 10^8$ GeV, namely, to the more energetic portion of the KM3NeT range of $E_\chi\simeq 70-760$ PeV. For long bursts, with a 20-day observing period, PBR's optimal sensitivity to the $\chi$ fluence begins at higher energy, for energies above the blue band in fig. \ref{fig:PBRSensBtautau}. For both long and short bursts, EAS detection by PBR improves until $E_\chi\sim 2\times 10^9$ GeV. Over the whole high energy range above the blue band, PBR will be sensitive to a $\chi$ fluence determined by the KM3-230213A event. For $\sigma_0=10^{-33}$ cm$^2$, PBR will be optimally sensitive to $E_\chi\gtrsim 1.5\times 10^8$ GeV for short bursts and $\gtrsim 5\times 10^8$ GeV for long bursts. For $B_{\tau\tau}=B_{\mu\mu}\times (m_\tau/m_\mu)^2$, the crossovers between the PBR sensitivity and the $\chi$ fluence occur at lower energies and include almost the whole KM3NeT  energy band for $E_\chi$. At high energies, the PBR sensitivity is dramatically lower than the projected flux from eq. (\ref{eq:fhsensitivity}).

PBR is well-positioned to observe nearby bursting $\chi$ sources.
While the PBR sensitivity results presented here are for a single boosted DM scenario \cite{Farzan:2025ydi} designed to address the KM3NeT observation of KM3-230213A, the overall methodology applies equally well to other scenarios with BSM effects that modify $P_{\rm exit}$ and $dN_\tau/dE_\tau$ relative to their standard model values; each BSM model that produces up-going EASs from tau decays will have a model specific $P_{\rm exit}$ and $dN_\tau/dE_\tau$. Thus, PBR observations with its Cherenkov camera pointed below the Earth's limb will extend its science reach to BSM physics.

\section*{Acknowledgments}
\noindent
\small{ The authors would like to acknowledge the support by NASA
  awards RTOP 21-APRA21-0071, 80NSSC22K1488, and 80NSSC24K1780, by the French space agency
  CNES and the Italian Space agency ASI.  OP
  JAC financed by ESIF and the MEYS
  CZ.02.01.01/00/22\_008/0004596 also supported this work. The work of L.A.A. and K.P.C. is
  supported by the US National Science Foundation (grant
  PHY-2412679). M.H.R.'s work was performed in part at the Aspen Center for Physics, which is supported by National Science Foundation grant PHY-2210452. We thank M. Hostert for discussions.
We gratefully acknowledge the collaboration and expert
  advice provided by the PUEO collaboration. We also acknowledge the
  invaluable contributions of the administrative and technical staffs
  at our home institutions.}

\appendix

\section{Flavor-dependent dark matter branching fractions to leptons}
\label{App:A}

Hot dark matter can decay into Standard Model leptons with branching
fractions that depend directly on the lepton masses under specific
theoretical frameworks. This mass-dependent decay structure is heavily
dictated by the spin and underlying portal mechanism of the dark
matter particle. In this Appendix we show that if $\phi$ is a
pseudoscalar that couples to leptons  axially, a mass dependency on the
branching fraction arises due to helicity suppression.

Consider a dimension-five operator coupling a derivative acting on $\phi$ with a leptonic axial current
\begin{equation}
\mathscr{L} \supset \frac{\partial _{\mu }\phi
}{2f_{a}} \left(\bar{\ell}_i \ \gamma ^{\mu } \ \gamma^{5} \ \ell_j \right) \,,
\label{Lagrangian}
\end{equation}
where $\ell_i$ denotes a charged
lepton $(i = e, \mu, \tau)$ and the axion decay constant $f_a$ is a scale related to the spontaneous breaking of the a
global $U(1)$ symmetry the axion is associated to~\cite{Marsh:2015xka}. We can apply integration by parts to shift the derivative from the
axion field to the leptonic axial current. Specifically, 
 using the standard calculus product rule for
derivatives it follows that
\begin{equation}
\partial _{\mu }\left(\phi \, \bar{\ell}_i \, \gamma ^{\mu } \, \gamma
  ^{5} \, \ell_j \right)=(\partial _{\mu }\phi) \, \bar{\ell }_i \, \gamma
^{\mu } \, \gamma ^{5} \, \ell_j + \phi \, \partial _{\mu
}\left(\bar{\ell}_i \, \gamma ^{\mu } \, \gamma ^{5} \, \ell_j \right) \, .
\label{pmu}
\end{equation}
Rearranging (\ref{pmu}) to isolate the original term in (\ref{Lagrangian}) gives:
\begin{equation}
(\partial _{\mu } \phi )\bar{\ell}_i \, \gamma ^{\mu } \, \gamma ^{5}
\, \ell_j =\partial _{\mu }\left(\phi \, \bar{\ell}_i \, \gamma ^{\mu }
  \, \gamma ^{5} \, \ell_j \right)- \phi \, \partial _{\mu
}\left(\bar{\ell}_i \, \gamma ^{\mu } \, \gamma ^{5} \, \ell_j \right) \, .
\label{Lrearrangement}
\end{equation}
When integrated over all space and time, the first term on the right
of (\ref{Lrearrangement}) is a total derivative, and so by Gauss'
theorem, it turns into a boundary term at infinity where fields drop
to zero, so it vanishes. The interaction Lagrangian can then be
rewritten as
\begin{equation}
\mathscr{L} \supset -\frac{\phi }{2f_{a}} \, \partial _{\mu
}\left(\bar{\ell}_i \, \gamma ^{\mu } \, \gamma^{5} \, \ell_j \right) \, .
\label{L2}
\end{equation}
The problem is now reduced to evaluating the derivative of the
axial-vector current. Using the product rule again it follows that
\begin{equation}
\partial _{\mu }\left(\bar{\ell}_i \, \gamma ^{\mu } \, \gamma ^{5} \,
  \ell_j
\right)  =  (\partial _{\mu }\bar{\ell}_i) \, \gamma ^{\mu } \, \gamma
^{5} \, \ell_j
+\bar{\ell}_i \, \gamma ^{\mu } \, \gamma ^{5} \, (\partial _{\mu }\ell_j ) \,,
\end{equation}
or equivalently, 
\begin{equation}
\partial _{\mu }\left(\bar{\ell}_i \, \gamma ^{\mu } \, \gamma ^{5} \,
  \ell_j
\right)=\left[(\partial _{\mu }\bar{\ell}_i) \, \gamma ^{\mu
  }\right] \, \gamma ^{5} \, \ell_j - \bar{\ell}_i \, \gamma ^{5} \,
\left[\gamma ^{\mu } \, (\partial _{\mu }\ell_j )\right] \, .
\end{equation}
because the $\gamma ^{5}$ matrix  anti-commutes with all other gamma matrices ($\gamma^\mu \gamma^5 = -\gamma^5 \gamma^\mu$).

Using the Standard Model Dirac equation and its adjoint (conjugate) equation for on-shell leptons ($\partial_\mu \gamma^\mu \ell = -i m_\ell \ell$ and $\partial_\mu \bar{\ell} \gamma^\mu = i m_\ell \bar{\ell}$), the divergence of the axial-vector current yields
\begin{equation}
\partial _{\mu }\left(\bar{\ell }_{i} \, \gamma ^{\mu } \, \gamma^{5}
  \, \ell_{j}\right)= i (m_{i}+m_{j}) \, \bar{\ell
}_{i}\gamma^{5}\ell_{j} \, .
\end{equation}
Substituting this back into (\ref{L2}) reveals an effective pseudo-scalar Yukawa
coupling
\begin{equation}
\mathscr{L} \supset -i\frac{(m_{i}+m_{j})}{2f_{a}} \, \phi \, 
\left(\bar{\ell }_{i} \, \gamma^{5} \, \ell _{j}\right) \, .
\end{equation}
Because the effective coupling constant $g_{\phi \ell \ell} \propto
(m_i + m_j) / f_a$, any matrix element $\mathscr{M}$ calculated from this
vertex will pick up a factor of the
lepton mass.

Now, since axions are spin-$0$ pseudoscalars, for a flavor-conserving
decay ($i = j$) such as $\phi \rightarrow \ell^+ \ell^-$, the
final-state leptons must be produced with matching helicities (e.g.,
both right-handed or both left-handed) to conserve the zero total
angular momentum along the decay axis. However, the weak/chiral nature of the axial current dictates that a massless scalar would naturally produce particles of opposite helicity. To force them into matching helicity states to satisfy angular momentum conservation, a helicity flip is required. The probability of a helicity flip is directly proportional to the mass of the lepton $m_{\ell }$.

Squaring the matrix element to get the decay rate yields:
\begin{equation}
\sum |\mathscr{M}|{}^{2}\propto |g_{\phi \ell \ell }|{}^{2}\propto
m_{\ell }^{2} \, .
\end{equation}
For flavor-conserving decays, the full analytical width translates to~\cite{Calibbi:2022izs}
\begin{equation}
\Gamma (\phi \rightarrow \ell ^{+}\ell ^{-})=\frac{m_{\phi }}{8\pi
}\left(\frac{m_{\ell }}{f_{a}}\right)^{2}\sqrt{1-\frac{4m_{\ell
    }^{2}}{m_{\phi }^{2}}} \, .
\end{equation}
When the axion is much heavier than the lepton ($m_\phi \gg m_\ell$),
the phase-space square root factor approaches 1, leaving the width (and its resulting branching fraction) strictly proportional to the
square of the lepton mass, $m_{\ell }^{2}$~\cite{Nomura:2008ru}.

\vspace{-2ex}

%\bibliography{references}

\begin{thebibliography}{114}%
\makeatletter
\providecommand \@ifxundefined [1]{%
 \@ifx{#1\undefined}
}%
\providecommand \@ifnum [1]{%
 \ifnum #1\expandafter \@firstoftwo
 \else \expandafter \@secondoftwo
 \fi
}%
\providecommand \@ifx [1]{%
 \ifx #1\expandafter \@firstoftwo
 \else \expandafter \@secondoftwo
 \fi
}%
\providecommand \natexlab [1]{#1}%
\providecommand \enquote  [1]{``#1''}%
\providecommand \bibnamefont  [1]{#1}%
\providecommand \bibfnamefont [1]{#1}%
\providecommand \citenamefont [1]{#1}%
\providecommand \href@noop [0]{\@secondoftwo}%
\providecommand \href [0]{\begingroup \@sanitize@url \@href}%
\providecommand \@href[1]{\@@startlink{#1}\@@href}%
\providecommand \@@href[1]{\endgroup#1\@@endlink}%
\providecommand \@sanitize@url [0]{\catcode `\\12\catcode `\$12\catcode
  `\&12\catcode `\#12\catcode `\^12\catcode `\_12\catcode `\%12\relax}%
\providecommand \@@startlink[1]{}%
\providecommand \@@endlink[0]{}%
\providecommand \url  [0]{\begingroup\@sanitize@url \@url }%
\providecommand \@url [1]{\endgroup\@href {#1}{\urlprefix }}%
\providecommand \urlprefix  [0]{URL }%
\providecommand \Eprint [0]{\href }%
\providecommand \doibase [0]{http://dx.doi.org/}%
\providecommand \selectlanguage [0]{\@gobble}%
\providecommand \bibinfo  [0]{\@secondoftwo}%
\providecommand \bibfield  [0]{\@secondoftwo}%
\providecommand \translation [1]{[#1]}%
\providecommand \BibitemOpen [0]{}%
\providecommand \bibitemStop [0]{}%
\providecommand \bibitemNoStop [0]{.\EOS\space}%
\providecommand \EOS [0]{\spacefactor3000\relax}%
\providecommand \BibitemShut  [1]{\csname bibitem#1\endcsname}%
\let\auto@bib@innerbib\@empty
%</preamble>
\bibitem [{\citenamefont {Ackermann}\ \emph {et~al.}(2022)\citenamefont
  {Ackermann} \emph {et~al.}}]{Ackermann:2022rqc}%
  \BibitemOpen
  \bibfield  {author} {\bibinfo {author} {\bibfnamefont {M.}~\bibnamefont
  {Ackermann}} \emph {et~al.},\ }\href {\doibase 10.1016/j.jheap.2022.08.001}
  {\bibfield  {journal} {\bibinfo  {journal} {JHEAp}\ }\textbf {\bibinfo
  {volume} {36}},\ \bibinfo {pages} {55} (\bibinfo {year} {2022})},\ \Eprint
  {http://arxiv.org/abs/2203.08096} {arXiv:2203.08096 [hep-ph]} \BibitemShut
  {NoStop}%
\bibitem [{\citenamefont {Gu{\'e}pin}\ \emph {et~al.}(2022)\citenamefont
  {Gu{\'e}pin}, \citenamefont {Kotera},\ and\ \citenamefont
  {Oikonomou}}]{Guepin:2022qpl}%
  \BibitemOpen
  \bibfield  {author} {\bibinfo {author} {\bibfnamefont {C.}~\bibnamefont
  {Gu{\'e}pin}}, \bibinfo {author} {\bibfnamefont {K.}~\bibnamefont {Kotera}},
  \ and\ \bibinfo {author} {\bibfnamefont {F.}~\bibnamefont {Oikonomou}},\
  }\href {\doibase 10.1038/s42254-024-00801-5} {\bibfield  {journal} {\bibinfo
  {journal} {Nature Rev. Phys.}\ }\textbf {\bibinfo {volume} {4}},\ \bibinfo
  {pages} {697} (\bibinfo {year} {2022})},\ \Eprint
  {http://arxiv.org/abs/2207.12205} {arXiv:2207.12205 [astro-ph.HE]}
  \BibitemShut {NoStop}%
\bibitem [{\citenamefont {Aartsen}\ \emph {et~al.}(2013)\citenamefont {Aartsen}
  \emph {et~al.}}]{IceCube:2013low}%
  \BibitemOpen
  \bibfield  {author} {\bibinfo {author} {\bibfnamefont {M.~G.}\ \bibnamefont
  {Aartsen}} \emph {et~al.} (\bibinfo {collaboration} {IceCube}),\ }\href
  {\doibase 10.1126/science.1242856} {\bibfield  {journal} {\bibinfo  {journal}
  {Science}\ }\textbf {\bibinfo {volume} {342}},\ \bibinfo {pages} {1242856}
  (\bibinfo {year} {2013})},\ \Eprint {http://arxiv.org/abs/1311.5238}
  {arXiv:1311.5238 [astro-ph.HE]} \BibitemShut {NoStop}%
\bibitem [{\citenamefont {Aartsen}\ \emph
  {et~al.}(2018{\natexlab{a}})\citenamefont {Aartsen} \emph
  {et~al.}}]{IceCube:2018dnn}%
  \BibitemOpen
  \bibfield  {author} {\bibinfo {author} {\bibfnamefont {M.~G.}\ \bibnamefont
  {Aartsen}} \emph {et~al.} (\bibinfo {collaboration} {IceCube, Fermi-LAT,
  MAGIC, AGILE, ASAS-SN, HAWC, H.E.S.S., INTEGRAL, Kanata, Kiso, Kapteyn,
  Liverpool Telescope, Subaru, Swift NuSTAR, VERITAS, VLA/17B-403}),\ }\href
  {\doibase 10.1126/science.aat1378} {\bibfield  {journal} {\bibinfo  {journal}
  {Science}\ }\textbf {\bibinfo {volume} {361}},\ \bibinfo {pages} {eaat1378}
  (\bibinfo {year} {2018}{\natexlab{a}})},\ \Eprint
  {http://arxiv.org/abs/1807.08816} {arXiv:1807.08816 [astro-ph.HE]}
  \BibitemShut {NoStop}%
\bibitem [{\citenamefont {Aartsen}\ \emph
  {et~al.}(2018{\natexlab{b}})\citenamefont {Aartsen} \emph
  {et~al.}}]{IceCube:2018cha}%
  \BibitemOpen
  \bibfield  {author} {\bibinfo {author} {\bibfnamefont {M.~G.}\ \bibnamefont
  {Aartsen}} \emph {et~al.} (\bibinfo {collaboration} {IceCube}),\ }\href
  {\doibase 10.1126/science.aat2890} {\bibfield  {journal} {\bibinfo  {journal}
  {Science}\ }\textbf {\bibinfo {volume} {361}},\ \bibinfo {pages} {147}
  (\bibinfo {year} {2018}{\natexlab{b}})},\ \Eprint
  {http://arxiv.org/abs/1807.08794} {arXiv:1807.08794 [astro-ph.HE]}
  \BibitemShut {NoStop}%
\bibitem [{\citenamefont {Stein}\ \emph {et~al.}(2021)\citenamefont {Stein}
  \emph {et~al.}}]{Stein:2020xhk}%
  \BibitemOpen
  \bibfield  {author} {\bibinfo {author} {\bibfnamefont {R.}~\bibnamefont
  {Stein}} \emph {et~al.},\ }\href {\doibase 10.1038/s41550-020-01295-8}
  {\bibfield  {journal} {\bibinfo  {journal} {Nature Astron.}\ }\textbf
  {\bibinfo {volume} {5}},\ \bibinfo {pages} {510} (\bibinfo {year} {2021})},\
  \Eprint {http://arxiv.org/abs/2005.05340} {arXiv:2005.05340 [astro-ph.HE]}
  \BibitemShut {NoStop}%
\bibitem [{\citenamefont {Reusch}\ \emph {et~al.}(2022)\citenamefont {Reusch}
  \emph {et~al.}}]{Reusch:2021ztx}%
  \BibitemOpen
  \bibfield  {author} {\bibinfo {author} {\bibfnamefont {S.}~\bibnamefont
  {Reusch}} \emph {et~al.},\ }\href {\doibase 10.1103/PhysRevLett.128.221101}
  {\bibfield  {journal} {\bibinfo  {journal} {Phys. Rev. Lett.}\ }\textbf
  {\bibinfo {volume} {128}},\ \bibinfo {pages} {221101} (\bibinfo {year}
  {2022})},\ \Eprint {http://arxiv.org/abs/2111.09390} {arXiv:2111.09390
  [astro-ph.HE]} \BibitemShut {NoStop}%
\bibitem [{\citenamefont {Abbasi}\ \emph
  {et~al.}(2022{\natexlab{a}})\citenamefont {Abbasi} \emph
  {et~al.}}]{IceCube:2022der}%
  \BibitemOpen
  \bibfield  {author} {\bibinfo {author} {\bibfnamefont {R.}~\bibnamefont
  {Abbasi}} \emph {et~al.} (\bibinfo {collaboration} {IceCube}),\ }\href
  {\doibase 10.1126/science.abg3395} {\bibfield  {journal} {\bibinfo  {journal}
  {Science}\ }\textbf {\bibinfo {volume} {378}},\ \bibinfo {pages} {538}
  (\bibinfo {year} {2022}{\natexlab{a}})},\ \Eprint
  {http://arxiv.org/abs/2211.09972} {arXiv:2211.09972 [astro-ph.HE]}
  \BibitemShut {NoStop}%
\bibitem [{\citenamefont {Aartsen}\ \emph {et~al.}(2017)\citenamefont {Aartsen}
  \emph {et~al.}}]{IceCube:2016zyt}%
  \BibitemOpen
  \bibfield  {author} {\bibinfo {author} {\bibfnamefont {M.~G.}\ \bibnamefont
  {Aartsen}} \emph {et~al.} (\bibinfo {collaboration} {IceCube}),\ }\href
  {\doibase 10.1088/1748-0221/12/03/P03012} {\bibfield  {journal} {\bibinfo
  {journal} {JINST}\ }\textbf {\bibinfo {volume} {12}},\ \bibinfo {pages}
  {P03012} (\bibinfo {year} {2017})},\ \bibinfo {note} {[Erratum: JINST 19,
  E05001 (2024)]},\ \Eprint {http://arxiv.org/abs/1612.05093} {arXiv:1612.05093
  [astro-ph.IM]} \BibitemShut {NoStop}%
\bibitem [{\citenamefont {Abbasi}\ \emph
  {et~al.}(2022{\natexlab{b}})\citenamefont {Abbasi} \emph
  {et~al.}}]{Abbasi:2021qfz}%
  \BibitemOpen
  \bibfield  {author} {\bibinfo {author} {\bibfnamefont {R.}~\bibnamefont
  {Abbasi}} \emph {et~al.},\ }\href {\doibase 10.3847/1538-4357/ac4d29}
  {\bibfield  {journal} {\bibinfo  {journal} {Astrophys. J.}\ }\textbf
  {\bibinfo {volume} {928}},\ \bibinfo {pages} {50} (\bibinfo {year}
  {2022}{\natexlab{b}})},\ \Eprint {http://arxiv.org/abs/2111.10299}
  {arXiv:2111.10299 [astro-ph.HE]} \BibitemShut {NoStop}%
\bibitem [{\citenamefont {Aartsen}\ \emph {et~al.}(2021)\citenamefont {Aartsen}
  \emph {et~al.}}]{IceCube:2021rpz}%
  \BibitemOpen
  \bibfield  {author} {\bibinfo {author} {\bibfnamefont {M.~G.}\ \bibnamefont
  {Aartsen}} \emph {et~al.} (\bibinfo {collaboration} {IceCube}),\ }\href
  {\doibase 10.1038/s41586-021-03256-1} {\bibfield  {journal} {\bibinfo
  {journal} {Nature}\ }\textbf {\bibinfo {volume} {591}},\ \bibinfo {pages}
  {220} (\bibinfo {year} {2021})},\ \bibinfo {note} {[Erratum: Nature 592, E11
  (2021)]},\ \Eprint {http://arxiv.org/abs/2110.15051} {arXiv:2110.15051
  [hep-ex]} \BibitemShut {NoStop}%
\bibitem [{\citenamefont {Aiello}\ \emph {et~al.}(2025)\citenamefont {Aiello}
  \emph {et~al.}}]{KM3NeT:2025npi}%
  \BibitemOpen
  \bibfield  {author} {\bibinfo {author} {\bibfnamefont {S.}~\bibnamefont
  {Aiello}} \emph {et~al.} (\bibinfo {collaboration} {KM3NeT}),\ }\href
  {\doibase 10.1038/s41586-024-08543-1} {\bibfield  {journal} {\bibinfo
  {journal} {Nature}\ }\textbf {\bibinfo {volume} {638}},\ \bibinfo {pages}
  {376} (\bibinfo {year} {2025})}\BibitemShut {NoStop}%
\bibitem [{\citenamefont {Adrian-Martinez}\ \emph {et~al.}(2016)\citenamefont
  {Adrian-Martinez} \emph {et~al.}}]{KM3Net:2016zxf}%
  \BibitemOpen
  \bibfield  {author} {\bibinfo {author} {\bibfnamefont {S.}~\bibnamefont
  {Adrian-Martinez}} \emph {et~al.} (\bibinfo {collaboration} {KM3Net}),\
  }\href {\doibase 10.1088/0954-3899/43/8/084001} {\bibfield  {journal}
  {\bibinfo  {journal} {J. Phys. G}\ }\textbf {\bibinfo {volume} {43}},\
  \bibinfo {pages} {084001} (\bibinfo {year} {2016})},\ \Eprint
  {http://arxiv.org/abs/1601.07459} {arXiv:1601.07459 [astro-ph.IM]}
  \BibitemShut {NoStop}%
\bibitem [{\citenamefont {Aiello}\ \emph {et~al.}(2022)\citenamefont {Aiello}
  \emph {et~al.}}]{KM3NeT:2022pnv}%
  \BibitemOpen
  \bibfield  {author} {\bibinfo {author} {\bibfnamefont {S.}~\bibnamefont
  {Aiello}} \emph {et~al.} (\bibinfo {collaboration} {KM3NeT}),\ }\href
  {\doibase 10.1088/1748-0221/17/07/P07038} {\bibfield  {journal} {\bibinfo
  {journal} {JINST}\ }\textbf {\bibinfo {volume} {17}},\ \bibinfo {pages}
  {P07038} (\bibinfo {year} {2022})},\ \Eprint
  {http://arxiv.org/abs/2203.10048} {arXiv:2203.10048 [astro-ph.IM]}
  \BibitemShut {NoStop}%
\bibitem [{\citenamefont {Plebaniak}(2025)}]{ICRC2025Plebaniak}%
  \BibitemOpen
  \bibfield  {author} {\bibinfo {author} {\bibfnamefont {Z.}~\bibnamefont
  {Plebaniak}},\ }\href@noop {} {\bibfield  {journal} {\bibinfo  {journal}
  {PoS}\ }\textbf {\bibinfo {volume} {ICRC2025}},\ \bibinfo {pages} {360}
  (\bibinfo {year} {2025})}\BibitemShut {NoStop}%
\bibitem [{\citenamefont {Eser}\ \emph {et~al.}(2025)\citenamefont {Eser},
  \citenamefont {Olinto}, \citenamefont {Osteria} \emph
  {et~al.}}]{ICRC2025Eser}%
  \BibitemOpen
  \bibfield  {author} {\bibinfo {author} {\bibfnamefont {J.}~\bibnamefont
  {Eser}}, \bibinfo {author} {\bibfnamefont {A.~V.}\ \bibnamefont {Olinto}},
  \bibinfo {author} {\bibfnamefont {G.}~\bibnamefont {Osteria}},  \emph
  {et~al.},\ }\href@noop {} {\bibfield  {journal} {\bibinfo  {journal} {PoS}\
  }\textbf {\bibinfo {volume} {ICRC2025}},\ \bibinfo {pages} {249} (\bibinfo
  {year} {2025})}\BibitemShut {NoStop}%
\bibitem [{\citenamefont {Adams}\ \emph
  {et~al.}(2026{\natexlab{a}})\citenamefont {Adams} \emph
  {et~al.}}]{PBRTeam:2026wxs}%
  \BibitemOpen
  \bibfield  {author} {\bibinfo {author} {\bibfnamefont {J.}~\bibnamefont
  {Adams}} \emph {et~al.} (\bibinfo {collaboration} {PBR Team}),\ }\href
  {\doibase 10.1088/1475-7516/2026/07/001} {\bibfield  {journal} {\bibinfo
  {journal} {JCAP}\ }\textbf {\bibinfo {volume} {07}},\ \bibinfo {pages} {001}
  (\bibinfo {year} {2026}{\natexlab{a}})},\ \Eprint
  {http://arxiv.org/abs/2601.19997} {arXiv:2601.19997 [astro-ph.IM]}
  \BibitemShut {NoStop}%
\bibitem [{\citenamefont {Olinto}\ \emph {et~al.}(2021)\citenamefont {Olinto}
  \emph {et~al.}}]{POEMMA:2020ykm}%
  \BibitemOpen
  \bibfield  {author} {\bibinfo {author} {\bibfnamefont {A.~V.}\ \bibnamefont
  {Olinto}} \emph {et~al.} (\bibinfo {collaboration} {POEMMA}),\ }\href
  {\doibase 10.1088/1475-7516/2021/06/007} {\bibfield  {journal} {\bibinfo
  {journal} {JCAP}\ }\textbf {\bibinfo {volume} {06}},\ \bibinfo {pages} {007}
  (\bibinfo {year} {2021})},\ \Eprint {http://arxiv.org/abs/2012.07945}
  {arXiv:2012.07945 [astro-ph.IM]} \BibitemShut {NoStop}%
\bibitem [{\citenamefont {Cummings}\ \emph
  {et~al.}(2021{\natexlab{a}})\citenamefont {Cummings}, \citenamefont
  {Aloisio}, \citenamefont {Eser},\ and\ \citenamefont
  {Krizmanic}}]{Cummings:2021bhg}%
  \BibitemOpen
  \bibfield  {author} {\bibinfo {author} {\bibfnamefont {A.}~\bibnamefont
  {Cummings}}, \bibinfo {author} {\bibfnamefont {R.}~\bibnamefont {Aloisio}},
  \bibinfo {author} {\bibfnamefont {J.}~\bibnamefont {Eser}}, \ and\ \bibinfo
  {author} {\bibfnamefont {J.}~\bibnamefont {Krizmanic}},\ }\href {\doibase
  10.1103/PhysRevD.104.063029} {\bibfield  {journal} {\bibinfo  {journal}
  {Phys. Rev. D}\ }\textbf {\bibinfo {volume} {104}},\ \bibinfo {pages}
  {063029} (\bibinfo {year} {2021}{\natexlab{a}})},\ \Eprint
  {http://arxiv.org/abs/2105.03255} {arXiv:2105.03255 [astro-ph.IM]}
  \BibitemShut {NoStop}%
\bibitem [{\citenamefont {Battisti}\ \emph {et~al.}(2024)\citenamefont
  {Battisti}, \citenamefont {Eser}, \citenamefont {Filippatos}, \citenamefont
  {Olinto}, \citenamefont {Osteria},\ and\ \citenamefont
  {Parizot}}]{Battisti:2024rfv}%
  \BibitemOpen
  \bibfield  {author} {\bibinfo {author} {\bibfnamefont {M.}~\bibnamefont
  {Battisti}}, \bibinfo {author} {\bibfnamefont {J.}~\bibnamefont {Eser}},
  \bibinfo {author} {\bibfnamefont {G.}~\bibnamefont {Filippatos}}, \bibinfo
  {author} {\bibfnamefont {A.}~\bibnamefont {Olinto}}, \bibinfo {author}
  {\bibfnamefont {G.}~\bibnamefont {Osteria}}, \ and\ \bibinfo {author}
  {\bibfnamefont {E.}~\bibnamefont {Parizot}} (\bibinfo {collaboration}
  {JEM-EUSO}),\ }\href {\doibase 10.1016/j.nima.2024.169727} {\bibfield
  {journal} {\bibinfo  {journal} {Nucl. Instrum. Meth. A}\ }\textbf {\bibinfo
  {volume} {1068}},\ \bibinfo {pages} {169727} (\bibinfo {year} {2024})},\
  \Eprint {http://arxiv.org/abs/2408.14867} {arXiv:2408.14867 [astro-ph.IM]}
  \BibitemShut {NoStop}%
\bibitem [{\citenamefont {Heibges}\ \emph {et~al.}(2025)\citenamefont
  {Heibges}, \citenamefont {Gu{\'e}pin-Detrigne}, \citenamefont {Garg},
  \citenamefont {Kupari}, \citenamefont {Reno}, \citenamefont {Venters},\ and\
  \citenamefont {Wiencke}}]{Heibges:2025yvw}%
  \BibitemOpen
  \bibfield  {author} {\bibinfo {author} {\bibfnamefont {T.}~\bibnamefont
  {Heibges}}, \bibinfo {author} {\bibfnamefont {C.}~\bibnamefont
  {Gu{\'e}pin-Detrigne}}, \bibinfo {author} {\bibfnamefont {D.}~\bibnamefont
  {Garg}}, \bibinfo {author} {\bibfnamefont {L.}~\bibnamefont {Kupari}},
  \bibinfo {author} {\bibfnamefont {M.~H.}\ \bibnamefont {Reno}}, \bibinfo
  {author} {\bibfnamefont {T.~M.}\ \bibnamefont {Venters}}, \ and\ \bibinfo
  {author} {\bibfnamefont {L.}~\bibnamefont {Wiencke}} (\bibinfo
  {collaboration} {JEM-EUSO}),\ }\href {\doibase 10.22323/1.501.1051}
  {\bibfield  {journal} {\bibinfo  {journal} {PoS}\ }\textbf {\bibinfo {volume}
  {ICRC2025}},\ \bibinfo {pages} {1051} (\bibinfo {year} {2025})},\ \Eprint
  {http://arxiv.org/abs/2511.10332} {arXiv:2511.10332 [astro-ph.HE]}
  \BibitemShut {NoStop}%
\bibitem [{\citenamefont {Heibges}\ \emph
  {et~al.}(2026{\natexlab{a}})\citenamefont {Heibges}, \citenamefont
  {Gu{\'e}pin}, \citenamefont {Kupari}, \citenamefont {Wistrand}, \citenamefont
  {Lahm}, \citenamefont {Eser}, \citenamefont {Reno}, \citenamefont {Venters},\
  and\ \citenamefont {Wiencke}}]{Heibges:2025llt}%
  \BibitemOpen
  \bibfield  {author} {\bibinfo {author} {\bibfnamefont {T.}~\bibnamefont
  {Heibges}}, \bibinfo {author} {\bibfnamefont {C.}~\bibnamefont {Gu{\'e}pin}},
  \bibinfo {author} {\bibfnamefont {L.}~\bibnamefont {Kupari}}, \bibinfo
  {author} {\bibfnamefont {H.}~\bibnamefont {Wistrand}}, \bibinfo {author}
  {\bibfnamefont {R.}~\bibnamefont {Lahm}}, \bibinfo {author} {\bibfnamefont
  {J.}~\bibnamefont {Eser}}, \bibinfo {author} {\bibfnamefont {M.~H.}\
  \bibnamefont {Reno}}, \bibinfo {author} {\bibfnamefont {T.~M.}\ \bibnamefont
  {Venters}}, \ and\ \bibinfo {author} {\bibfnamefont {L.}~\bibnamefont
  {Wiencke}},\ }\href {\doibase 10.1016/j.cpc.2026.110310} {\bibfield
  {journal} {\bibinfo  {journal} {Comput. Phys. Commun.}\ }\textbf {\bibinfo
  {volume} {328}},\ \bibinfo {pages} {110310} (\bibinfo {year}
  {2026}{\natexlab{a}})},\ \Eprint {http://arxiv.org/abs/2509.13844}
  {arXiv:2509.13844 [astro-ph.IM]} \BibitemShut {NoStop}%
\bibitem [{\citenamefont {Venters}\ \emph {et~al.}(2020)\citenamefont
  {Venters}, \citenamefont {Reno}, \citenamefont {Krizmanic}, \citenamefont
  {Anchordoqui}, \citenamefont {Gu\'epin},\ and\ \citenamefont
  {Olinto}}]{Venters:2019xwi}%
  \BibitemOpen
  \bibfield  {author} {\bibinfo {author} {\bibfnamefont {T.~M.}\ \bibnamefont
  {Venters}}, \bibinfo {author} {\bibfnamefont {M.~H.}\ \bibnamefont {Reno}},
  \bibinfo {author} {\bibfnamefont {J.~F.}\ \bibnamefont {Krizmanic}}, \bibinfo
  {author} {\bibfnamefont {L.~A.}\ \bibnamefont {Anchordoqui}}, \bibinfo
  {author} {\bibfnamefont {C.}~\bibnamefont {Gu\'epin}}, \ and\ \bibinfo
  {author} {\bibfnamefont {A.~V.}\ \bibnamefont {Olinto}},\ }\href {\doibase
  10.1103/PhysRevD.102.123013} {\bibfield  {journal} {\bibinfo  {journal}
  {Phys. Rev. D}\ }\textbf {\bibinfo {volume} {102}},\ \bibinfo {pages}
  {123013} (\bibinfo {year} {2020})},\ \Eprint
  {http://arxiv.org/abs/1906.07209} {arXiv:1906.07209 [astro-ph.HE]}
  \BibitemShut {NoStop}%
\bibitem [{\citenamefont {Cummings}\ \emph
  {et~al.}(2021{\natexlab{b}})\citenamefont {Cummings}, \citenamefont
  {Aloisio},\ and\ \citenamefont {Krizmanic}}]{Cummings:2020ycz}%
  \BibitemOpen
  \bibfield  {author} {\bibinfo {author} {\bibfnamefont {A.~L.}\ \bibnamefont
  {Cummings}}, \bibinfo {author} {\bibfnamefont {R.}~\bibnamefont {Aloisio}}, \
  and\ \bibinfo {author} {\bibfnamefont {J.~F.}\ \bibnamefont {Krizmanic}},\
  }\href {\doibase 10.1103/PhysRevD.103.043017} {\bibfield  {journal} {\bibinfo
   {journal} {Phys. Rev. D}\ }\textbf {\bibinfo {volume} {103}},\ \bibinfo
  {pages} {043017} (\bibinfo {year} {2021}{\natexlab{b}})},\ \Eprint
  {http://arxiv.org/abs/2011.09869} {arXiv:2011.09869 [astro-ph.HE]}
  \BibitemShut {NoStop}%
\bibitem [{\citenamefont {Abdellaoui}\ \emph {et~al.}(2024)\citenamefont
  {Abdellaoui} \emph {et~al.}}]{JEM-EUSO:2023ypf}%
  \BibitemOpen
  \bibfield  {author} {\bibinfo {author} {\bibfnamefont {G.}~\bibnamefont
  {Abdellaoui}} \emph {et~al.} (\bibinfo {collaboration} {JEM-EUSO}),\ }\href
  {\doibase 10.1016/j.astropartphys.2023.102891} {\bibfield  {journal}
  {\bibinfo  {journal} {Astropart. Phys.}\ }\textbf {\bibinfo {volume} {154}},\
  \bibinfo {pages} {102891} (\bibinfo {year} {2024})},\ \Eprint
  {http://arxiv.org/abs/2401.06525} {arXiv:2401.06525 [astro-ph.IM]}
  \BibitemShut {NoStop}%
\bibitem [{\citenamefont {Adams}\ \emph
  {et~al.}(2026{\natexlab{b}})\citenamefont {Adams} \emph
  {et~al.}}]{Adams:2026ltd}%
  \BibitemOpen
  \bibfield  {author} {\bibinfo {author} {\bibfnamefont {J.~H.}\ \bibnamefont
  {Adams}, \bibfnamefont {Jr.}} \emph {et~al.},\ }\href {\doibase
  10.1016/j.astropartphys.2026.103263} {\bibfield  {journal} {\bibinfo
  {journal} {Astropart. Phys.}\ }\textbf {\bibinfo {volume} {182}},\ \bibinfo
  {pages} {103263} (\bibinfo {year} {2026}{\natexlab{b}})}\BibitemShut
  {NoStop}%
\bibitem [{\citenamefont {Bagheri}\ \emph {et~al.}(2025)\citenamefont {Bagheri}
  \emph {et~al.}}]{Bagheri:2024byu}%
  \BibitemOpen
  \bibfield  {author} {\bibinfo {author} {\bibfnamefont {M.}~\bibnamefont
  {Bagheri}} \emph {et~al.},\ }\href {\doibase 10.1016/j.nima.2024.169999}
  {\bibfield  {journal} {\bibinfo  {journal} {Nucl. Instrum. Meth. A}\ }\textbf
  {\bibinfo {volume} {1070}},\ \bibinfo {pages} {169999} (\bibinfo {year}
  {2025})},\ \Eprint {http://arxiv.org/abs/2406.08274} {arXiv:2406.08274
  [astro-ph.IM]} \BibitemShut {NoStop}%
\bibitem [{\citenamefont {Adams}\ \emph {et~al.}(2025)\citenamefont {Adams}
  \emph {et~al.}}]{Adams:2024gsj}%
  \BibitemOpen
  \bibfield  {author} {\bibinfo {author} {\bibfnamefont {J.~H.}\ \bibnamefont
  {Adams}, \bibfnamefont {Jr.}} \emph {et~al.},\ }\href {\doibase
  10.1016/j.astropartphys.2024.103046} {\bibfield  {journal} {\bibinfo
  {journal} {Astropart. Phys.}\ }\textbf {\bibinfo {volume} {165}},\ \bibinfo
  {pages} {103046} (\bibinfo {year} {2025})},\ \Eprint
  {http://arxiv.org/abs/2406.13673} {arXiv:2406.13673 [astro-ph.IM]}
  \BibitemShut {NoStop}%
\bibitem [{\citenamefont {{PUEO Collaboration}}(2021)}]{Abarr_2021}%
  \BibitemOpen
  \bibfield  {author} {\bibinfo {author} {\bibnamefont {{PUEO
  Collaboration}}},\ }\href {\doibase 10.1088/1748-0221/16/08/p08035}
  {\bibfield  {journal} {\bibinfo  {journal} {Journal of Instrumentation}\
  }\textbf {\bibinfo {volume} {16}},\ \bibinfo {pages} {P08035} (\bibinfo
  {year} {2021})}\BibitemShut {NoStop}%
\bibitem [{\citenamefont {Olinto}\ \emph {et~al.}(2025)\citenamefont {Olinto}
  \emph {et~al.}}]{ICRC2025Olinto}%
  \BibitemOpen
  \bibfield  {author} {\bibinfo {author} {\bibfnamefont {A.~V.}\ \bibnamefont
  {Olinto}} \emph {et~al.},\ }\href@noop {} {\bibfield  {journal} {\bibinfo
  {journal} {PoS}\ }\textbf {\bibinfo {volume} {ICRC2025}},\ \bibinfo {pages}
  {980} (\bibinfo {year} {2025})},\ \Eprint {http://arxiv.org/abs/2507.02246}
  {arXiv:2507.02246 [astro-ph.HE]} \BibitemShut {NoStop}%
\bibitem [{\citenamefont {Farzan}\ and\ \citenamefont
  {Hostert}(2025)}]{Farzan:2025ydi}%
  \BibitemOpen
  \bibfield  {author} {\bibinfo {author} {\bibfnamefont {Y.}~\bibnamefont
  {Farzan}}\ and\ \bibinfo {author} {\bibfnamefont {M.}~\bibnamefont
  {Hostert}},\ }\href {\doibase 10.1007/JHEP10(2025)208} {\bibfield  {journal}
  {\bibinfo  {journal} {JHEP}\ }\textbf {\bibinfo {volume} {10}},\ \bibinfo
  {pages} {208} (\bibinfo {year} {2025})},\ \Eprint
  {http://arxiv.org/abs/2505.22711} {arXiv:2505.22711 [hep-ph]} \BibitemShut
  {NoStop}%
\bibitem [{\citenamefont {Fedynitch}\ \emph {et~al.}(2019)\citenamefont
  {Fedynitch}, \citenamefont {Riehn}, \citenamefont {Engel}, \citenamefont
  {Gaisser},\ and\ \citenamefont {Stanev}}]{Fedynitch:2018cbl}%
  \BibitemOpen
  \bibfield  {author} {\bibinfo {author} {\bibfnamefont {A.}~\bibnamefont
  {Fedynitch}}, \bibinfo {author} {\bibfnamefont {F.}~\bibnamefont {Riehn}},
  \bibinfo {author} {\bibfnamefont {R.}~\bibnamefont {Engel}}, \bibinfo
  {author} {\bibfnamefont {T.~K.}\ \bibnamefont {Gaisser}}, \ and\ \bibinfo
  {author} {\bibfnamefont {T.}~\bibnamefont {Stanev}},\ }\href {\doibase
  10.1103/PhysRevD.100.103018} {\bibfield  {journal} {\bibinfo  {journal}
  {Phys. Rev. D}\ }\textbf {\bibinfo {volume} {100}},\ \bibinfo {pages}
  {103018} (\bibinfo {year} {2019})},\ \Eprint
  {http://arxiv.org/abs/1806.04140} {arXiv:1806.04140 [hep-ph]} \BibitemShut
  {NoStop}%
\bibitem [{\citenamefont {Gaisser}(2012)}]{Gaisser:2011klf}%
  \BibitemOpen
  \bibfield  {author} {\bibinfo {author} {\bibfnamefont {T.~K.}\ \bibnamefont
  {Gaisser}},\ }\href {\doibase 10.1016/j.astropartphys.2012.02.010} {\bibfield
   {journal} {\bibinfo  {journal} {Astropart. Phys.}\ }\textbf {\bibinfo
  {volume} {35}},\ \bibinfo {pages} {801} (\bibinfo {year} {2012})},\ \Eprint
  {http://arxiv.org/abs/1111.6675} {arXiv:1111.6675 [astro-ph.HE]} \BibitemShut
  {NoStop}%
\bibitem [{\citenamefont {Aartsen}\ \emph
  {et~al.}(2018{\natexlab{c}})\citenamefont {Aartsen} \emph
  {et~al.}}]{IceCube:2018fhm}%
  \BibitemOpen
  \bibfield  {author} {\bibinfo {author} {\bibfnamefont {M.~G.}\ \bibnamefont
  {Aartsen}} \emph {et~al.} (\bibinfo {collaboration} {IceCube}),\ }\href
  {\doibase 10.1103/PhysRevD.98.062003} {\bibfield  {journal} {\bibinfo
  {journal} {Phys. Rev. D}\ }\textbf {\bibinfo {volume} {98}},\ \bibinfo
  {pages} {062003} (\bibinfo {year} {2018}{\natexlab{c}})},\ \Eprint
  {http://arxiv.org/abs/1807.01820} {arXiv:1807.01820 [astro-ph.HE]}
  \BibitemShut {NoStop}%
\bibitem [{\citenamefont {Abdul~Halim}\ \emph {et~al.}(2023)\citenamefont
  {Abdul~Halim} \emph {et~al.}}]{PierreAuger:2023pjg}%
  \BibitemOpen
  \bibfield  {author} {\bibinfo {author} {\bibfnamefont {A.}~\bibnamefont
  {Abdul~Halim}} \emph {et~al.} (\bibinfo {collaboration} {Pierre Auger}),\
  }\href {\doibase 10.22323/1.444.1488} {\bibfield  {journal} {\bibinfo
  {journal} {PoS}\ }\textbf {\bibinfo {volume} {ICRC2023}},\ \bibinfo {pages}
  {1488} (\bibinfo {year} {2023})}\BibitemShut {NoStop}%
\bibitem [{\citenamefont {Illana}\ \emph {et~al.}(2009)\citenamefont {Illana},
  \citenamefont {Masip},\ and\ \citenamefont {Meloni}}]{Illana:2009qv}%
  \BibitemOpen
  \bibfield  {author} {\bibinfo {author} {\bibfnamefont {J.~I.}\ \bibnamefont
  {Illana}}, \bibinfo {author} {\bibfnamefont {M.}~\bibnamefont {Masip}}, \
  and\ \bibinfo {author} {\bibfnamefont {D.}~\bibnamefont {Meloni}},\ }\href
  {\doibase 10.1088/1475-7516/2009/09/008} {\bibfield  {journal} {\bibinfo
  {journal} {JCAP}\ }\textbf {\bibinfo {volume} {09}},\ \bibinfo {pages} {008}
  (\bibinfo {year} {2009})},\ \Eprint {http://arxiv.org/abs/0907.1412}
  {arXiv:0907.1412 [hep-ph]} \BibitemShut {NoStop}%
\bibitem [{\citenamefont {Illana}\ \emph {et~al.}(2011)\citenamefont {Illana},
  \citenamefont {Lipari}, \citenamefont {Masip},\ and\ \citenamefont
  {Meloni}}]{Illana:2010gh}%
  \BibitemOpen
  \bibfield  {author} {\bibinfo {author} {\bibfnamefont {J.~I.}\ \bibnamefont
  {Illana}}, \bibinfo {author} {\bibfnamefont {P.}~\bibnamefont {Lipari}},
  \bibinfo {author} {\bibfnamefont {M.}~\bibnamefont {Masip}}, \ and\ \bibinfo
  {author} {\bibfnamefont {D.}~\bibnamefont {Meloni}},\ }\href {\doibase
  10.1016/j.astropartphys.2011.01.001} {\bibfield  {journal} {\bibinfo
  {journal} {Astropart. Phys.}\ }\textbf {\bibinfo {volume} {34}},\ \bibinfo
  {pages} {663} (\bibinfo {year} {2011})},\ \Eprint
  {http://arxiv.org/abs/1010.5084} {arXiv:1010.5084 [astro-ph.HE]} \BibitemShut
  {NoStop}%
\bibitem [{\citenamefont {Groom}\ \emph {et~al.}(2001)\citenamefont {Groom},
  \citenamefont {Mokhov},\ and\ \citenamefont {Striganov}}]{Groom:2001kq}%
  \BibitemOpen
  \bibfield  {author} {\bibinfo {author} {\bibfnamefont {D.~E.}\ \bibnamefont
  {Groom}}, \bibinfo {author} {\bibfnamefont {N.~V.}\ \bibnamefont {Mokhov}}, \
  and\ \bibinfo {author} {\bibfnamefont {S.~I.}\ \bibnamefont {Striganov}},\
  }\href {\doibase 10.1006/adnd.2001.0861} {\bibfield  {journal} {\bibinfo
  {journal} {Atom. Data Nucl. Data Tabl.}\ }\textbf {\bibinfo {volume} {78}},\
  \bibinfo {pages} {183} (\bibinfo {year} {2001})}\BibitemShut {NoStop}%
\bibitem [{\citenamefont {Group}()}]{mu-eloss-web}%
  \BibitemOpen
  \bibfield  {author} {\bibinfo {author} {\bibfnamefont {P.~D.}\ \bibnamefont
  {Group}},\ }\href@noop {} {\enquote {\bibinfo {title} {{Atomic and nuclear
  properties of water (liquid) (H2O)}},}\ }\bibinfo {howpublished} {Available
  at
  \url{https://pdg.lbl.gov/2023/AtomicNuclearProperties/HTML/water_liquid.html}
  (2025/04/16)}\BibitemShut {NoStop}%
\bibitem [{\citenamefont {Lipari}\ and\ \citenamefont
  {Stanev}(1991)}]{Lipari:1991ut}%
  \BibitemOpen
  \bibfield  {author} {\bibinfo {author} {\bibfnamefont {P.}~\bibnamefont
  {Lipari}}\ and\ \bibinfo {author} {\bibfnamefont {T.}~\bibnamefont
  {Stanev}},\ }\href {\doibase 10.1103/PhysRevD.44.3543} {\bibfield  {journal}
  {\bibinfo  {journal} {Phys. Rev. D}\ }\textbf {\bibinfo {volume} {44}},\
  \bibinfo {pages} {3543} (\bibinfo {year} {1991})}\BibitemShut {NoStop}%
\bibitem [{\citenamefont {Dutta}\ \emph {et~al.}(2001)\citenamefont {Dutta},
  \citenamefont {Reno}, \citenamefont {Sarcevic},\ and\ \citenamefont
  {Seckel}}]{Dutta:2000hh}%
  \BibitemOpen
  \bibfield  {author} {\bibinfo {author} {\bibfnamefont {S.~I.}\ \bibnamefont
  {Dutta}}, \bibinfo {author} {\bibfnamefont {M.~H.}\ \bibnamefont {Reno}},
  \bibinfo {author} {\bibfnamefont {I.}~\bibnamefont {Sarcevic}}, \ and\
  \bibinfo {author} {\bibfnamefont {D.}~\bibnamefont {Seckel}},\ }\href
  {\doibase 10.1103/PhysRevD.63.094020} {\bibfield  {journal} {\bibinfo
  {journal} {Phys. Rev. D}\ }\textbf {\bibinfo {volume} {63}},\ \bibinfo
  {pages} {094020} (\bibinfo {year} {2001})},\ \Eprint
  {http://arxiv.org/abs/hep-ph/0012350} {arXiv:hep-ph/0012350} \BibitemShut
  {NoStop}%
\bibitem [{\citenamefont {Lohmann}\ \emph {et~al.}(1985)\citenamefont
  {Lohmann}, \citenamefont {Kopp},\ and\ \citenamefont
  {Voss}}]{Lohmann:1985qg}%
  \BibitemOpen
  \bibfield  {author} {\bibinfo {author} {\bibfnamefont {W.}~\bibnamefont
  {Lohmann}}, \bibinfo {author} {\bibfnamefont {R.}~\bibnamefont {Kopp}}, \
  and\ \bibinfo {author} {\bibfnamefont {R.}~\bibnamefont {Voss}},\ }\href
  {\doibase 10.5170/CERN-1985-003} {\  (\bibinfo {year} {1985}),\
  10.5170/CERN-1985-003}\BibitemShut {NoStop}%
\bibitem [{\citenamefont {Abramowicz}\ and\ \citenamefont
  {Levy}(1997)}]{Abramowicz:1997ms}%
  \BibitemOpen
  \bibfield  {author} {\bibinfo {author} {\bibfnamefont {H.}~\bibnamefont
  {Abramowicz}}\ and\ \bibinfo {author} {\bibfnamefont {A.}~\bibnamefont
  {Levy}},\ }\href@noop {} {\  (\bibinfo {year} {1997})},\ \Eprint
  {http://arxiv.org/abs/hep-ph/9712415} {arXiv:hep-ph/9712415} \BibitemShut
  {NoStop}%
\bibitem [{\citenamefont {Abramowicz}\ \emph {et~al.}(1991)\citenamefont
  {Abramowicz}, \citenamefont {Levin}, \citenamefont {Levy},\ and\
  \citenamefont {Maor}}]{Abramowicz:1991xz}%
  \BibitemOpen
  \bibfield  {author} {\bibinfo {author} {\bibfnamefont {H.}~\bibnamefont
  {Abramowicz}}, \bibinfo {author} {\bibfnamefont {E.~M.}\ \bibnamefont
  {Levin}}, \bibinfo {author} {\bibfnamefont {A.}~\bibnamefont {Levy}}, \ and\
  \bibinfo {author} {\bibfnamefont {U.}~\bibnamefont {Maor}},\ }\href {\doibase
  10.1016/0370-2693(91)90202-2} {\bibfield  {journal} {\bibinfo  {journal}
  {Phys. Lett. B}\ }\textbf {\bibinfo {volume} {269}},\ \bibinfo {pages} {465}
  (\bibinfo {year} {1991})}\BibitemShut {NoStop}%
\bibitem [{\citenamefont {Gaisser}\ \emph {et~al.}(1995)\citenamefont
  {Gaisser}, \citenamefont {Halzen},\ and\ \citenamefont
  {Stanev}}]{Gaisser:1994yf}%
  \BibitemOpen
  \bibfield  {author} {\bibinfo {author} {\bibfnamefont {T.~K.}\ \bibnamefont
  {Gaisser}}, \bibinfo {author} {\bibfnamefont {F.}~\bibnamefont {Halzen}}, \
  and\ \bibinfo {author} {\bibfnamefont {T.}~\bibnamefont {Stanev}},\ }\href
  {\doibase 10.1016/0370-1573(95)00003-Y} {\bibfield  {journal} {\bibinfo
  {journal} {Phys. Rept.}\ }\textbf {\bibinfo {volume} {258}},\ \bibinfo
  {pages} {173} (\bibinfo {year} {1995})},\ \bibinfo {note} {[Erratum:
  Phys.Rept. 271, 355--356 (1996)]},\ \Eprint
  {http://arxiv.org/abs/hep-ph/9410384} {arXiv:hep-ph/9410384} \BibitemShut
  {NoStop}%
\bibitem [{\citenamefont {Halzen}\ and\ \citenamefont
  {Hooper}(2002)}]{Halzen:2002pg}%
  \BibitemOpen
  \bibfield  {author} {\bibinfo {author} {\bibfnamefont {F.}~\bibnamefont
  {Halzen}}\ and\ \bibinfo {author} {\bibfnamefont {D.}~\bibnamefont
  {Hooper}},\ }\href {\doibase 10.1088/0034-4885/65/7/201} {\bibfield
  {journal} {\bibinfo  {journal} {Rept. Prog. Phys.}\ }\textbf {\bibinfo
  {volume} {65}},\ \bibinfo {pages} {1025} (\bibinfo {year} {2002})},\ \Eprint
  {http://arxiv.org/abs/astro-ph/0204527} {arXiv:astro-ph/0204527} \BibitemShut
  {NoStop}%
\bibitem [{\citenamefont {Kotera}\ and\ \citenamefont
  {Olinto}(2011)}]{Kotera:2011cp}%
  \BibitemOpen
  \bibfield  {author} {\bibinfo {author} {\bibfnamefont {K.}~\bibnamefont
  {Kotera}}\ and\ \bibinfo {author} {\bibfnamefont {A.~V.}\ \bibnamefont
  {Olinto}},\ }\href {\doibase 10.1146/annurev-astro-081710-102620} {\bibfield
  {journal} {\bibinfo  {journal} {Ann. Rev. Astron. Astrophys.}\ }\textbf
  {\bibinfo {volume} {49}},\ \bibinfo {pages} {119} (\bibinfo {year} {2011})},\
  \Eprint {http://arxiv.org/abs/1101.4256} {arXiv:1101.4256 [astro-ph.HE]}
  \BibitemShut {NoStop}%
\bibitem [{\citenamefont {Anchordoqui}(2019)}]{Anchordoqui:2018qom}%
  \BibitemOpen
  \bibfield  {author} {\bibinfo {author} {\bibfnamefont {L.~A.}\ \bibnamefont
  {Anchordoqui}},\ }\href {\doibase 10.1016/j.physrep.2019.01.002} {\bibfield
  {journal} {\bibinfo  {journal} {Phys. Rept.}\ }\textbf {\bibinfo {volume}
  {801}},\ \bibinfo {pages} {1} (\bibinfo {year} {2019})},\ \Eprint
  {http://arxiv.org/abs/1807.09645} {arXiv:1807.09645 [astro-ph.HE]}
  \BibitemShut {NoStop}%
\bibitem [{\citenamefont {Coleman}\ \emph {et~al.}(2023)\citenamefont {Coleman}
  \emph {et~al.}}]{Coleman:2022abf}%
  \BibitemOpen
  \bibfield  {author} {\bibinfo {author} {\bibfnamefont {A.}~\bibnamefont
  {Coleman}} \emph {et~al.},\ }\href {\doibase
  10.1016/j.astropartphys.2023.102819} {\bibfield  {journal} {\bibinfo
  {journal} {Astropart. Phys.}\ }\textbf {\bibinfo {volume} {149}},\ \bibinfo
  {pages} {102819} (\bibinfo {year} {2023})},\ \Eprint
  {http://arxiv.org/abs/2205.05845} {arXiv:2205.05845 [astro-ph.HE]}
  \BibitemShut {NoStop}%
\bibitem [{\citenamefont {Stecker}\ \emph {et~al.}(1991)\citenamefont
  {Stecker}, \citenamefont {Done}, \citenamefont {Salamon},\ and\ \citenamefont
  {Sommers}}]{Stecker:1991vm}%
  \BibitemOpen
  \bibfield  {author} {\bibinfo {author} {\bibfnamefont {F.~W.}\ \bibnamefont
  {Stecker}}, \bibinfo {author} {\bibfnamefont {C.}~\bibnamefont {Done}},
  \bibinfo {author} {\bibfnamefont {M.~H.}\ \bibnamefont {Salamon}}, \ and\
  \bibinfo {author} {\bibfnamefont {P.}~\bibnamefont {Sommers}},\ }\href
  {\doibase 10.1103/PhysRevLett.66.2697} {\bibfield  {journal} {\bibinfo
  {journal} {Phys. Rev. Lett.}\ }\textbf {\bibinfo {volume} {66}},\ \bibinfo
  {pages} {2697} (\bibinfo {year} {1991})},\ \bibinfo {note} {[Erratum:
  Phys.Rev.Lett. 69, 2738 (1992)]}\BibitemShut {NoStop}%
\bibitem [{\citenamefont {Waxman}\ and\ \citenamefont
  {Bahcall}(1997)}]{Waxman:1997ti}%
  \BibitemOpen
  \bibfield  {author} {\bibinfo {author} {\bibfnamefont {E.}~\bibnamefont
  {Waxman}}\ and\ \bibinfo {author} {\bibfnamefont {J.~N.}\ \bibnamefont
  {Bahcall}},\ }\href {\doibase 10.1103/PhysRevLett.78.2292} {\bibfield
  {journal} {\bibinfo  {journal} {Phys. Rev. Lett.}\ }\textbf {\bibinfo
  {volume} {78}},\ \bibinfo {pages} {2292} (\bibinfo {year} {1997})},\ \Eprint
  {http://arxiv.org/abs/astro-ph/9701231} {arXiv:astro-ph/9701231} \BibitemShut
  {NoStop}%
\bibitem [{\citenamefont {Loeb}\ and\ \citenamefont
  {Waxman}(2006)}]{Loeb:2006tw}%
  \BibitemOpen
  \bibfield  {author} {\bibinfo {author} {\bibfnamefont {A.}~\bibnamefont
  {Loeb}}\ and\ \bibinfo {author} {\bibfnamefont {E.}~\bibnamefont {Waxman}},\
  }\href {\doibase 10.1088/1475-7516/2006/05/003} {\bibfield  {journal}
  {\bibinfo  {journal} {JCAP}\ }\textbf {\bibinfo {volume} {05}},\ \bibinfo
  {pages} {003} (\bibinfo {year} {2006})},\ \Eprint
  {http://arxiv.org/abs/astro-ph/0601695} {arXiv:astro-ph/0601695} \BibitemShut
  {NoStop}%
\bibitem [{\citenamefont {Greisen}(1966)}]{Greisen:1966jv}%
  \BibitemOpen
  \bibfield  {author} {\bibinfo {author} {\bibfnamefont {K.}~\bibnamefont
  {Greisen}},\ }\href {\doibase 10.1103/PhysRevLett.16.748} {\bibfield
  {journal} {\bibinfo  {journal} {Phys. Rev. Lett.}\ }\textbf {\bibinfo
  {volume} {16}},\ \bibinfo {pages} {748} (\bibinfo {year} {1966})}\BibitemShut
  {NoStop}%
\bibitem [{\citenamefont {Zatsepin}\ and\ \citenamefont
  {Kuzmin}(1966)}]{Zatsepin:1966jv}%
  \BibitemOpen
  \bibfield  {author} {\bibinfo {author} {\bibfnamefont {G.~T.}\ \bibnamefont
  {Zatsepin}}\ and\ \bibinfo {author} {\bibfnamefont {V.~A.}\ \bibnamefont
  {Kuzmin}},\ }\href@noop {} {\bibfield  {journal} {\bibinfo  {journal} {JETP
  Lett.}\ }\textbf {\bibinfo {volume} {4}},\ \bibinfo {pages} {78} (\bibinfo
  {year} {1966})}\BibitemShut {NoStop}%
\bibitem [{\citenamefont {Berezinsky}\ and\ \citenamefont
  {Zatsepin}(1969)}]{Berezinsky:1969erk}%
  \BibitemOpen
  \bibfield  {author} {\bibinfo {author} {\bibfnamefont {V.~S.}\ \bibnamefont
  {Berezinsky}}\ and\ \bibinfo {author} {\bibfnamefont {G.~T.}\ \bibnamefont
  {Zatsepin}},\ }\href {\doibase 10.1016/0370-2693(69)90341-4} {\bibfield
  {journal} {\bibinfo  {journal} {Phys. Lett. B}\ }\textbf {\bibinfo {volume}
  {28}},\ \bibinfo {pages} {423} (\bibinfo {year} {1969})}\BibitemShut
  {NoStop}%
\bibitem [{\citenamefont {Adriani}\ \emph
  {et~al.}(2025{\natexlab{a}})\citenamefont {Adriani} \emph
  {et~al.}}]{KM3NeT:2025vut}%
  \BibitemOpen
  \bibfield  {author} {\bibinfo {author} {\bibfnamefont {O.}~\bibnamefont
  {Adriani}} \emph {et~al.} (\bibinfo {collaboration} {KM3NeT}),\ }\href
  {\doibase 10.3847/2041-8213/adcc29} {\bibfield  {journal} {\bibinfo
  {journal} {Astrophys. J. Lett.}\ }\textbf {\bibinfo {volume} {984}},\
  \bibinfo {pages} {L41} (\bibinfo {year} {2025}{\natexlab{a}})},\ \Eprint
  {http://arxiv.org/abs/2502.08508} {arXiv:2502.08508 [astro-ph.HE]}
  \BibitemShut {NoStop}%
\bibitem [{\citenamefont {Muzio}\ \emph {et~al.}(2025)\citenamefont {Muzio},
  \citenamefont {Yuan},\ and\ \citenamefont {Lu}}]{Muzio:2025gbr}%
  \BibitemOpen
  \bibfield  {author} {\bibinfo {author} {\bibfnamefont {M.~S.}\ \bibnamefont
  {Muzio}}, \bibinfo {author} {\bibfnamefont {T.}~\bibnamefont {Yuan}}, \ and\
  \bibinfo {author} {\bibfnamefont {L.}~\bibnamefont {Lu}},\ }\href@noop {} {\
  (\bibinfo {year} {2025})},\ \Eprint {http://arxiv.org/abs/2502.06944}
  {arXiv:2502.06944 [astro-ph.HE]} \BibitemShut {NoStop}%
\bibitem [{\citenamefont {Kuznetsov}\ \emph {et~al.}(2026)\citenamefont
  {Kuznetsov}, \citenamefont {Petrov},\ and\ \citenamefont
  {Savchenko}}]{Kuznetsov:2025ehl}%
  \BibitemOpen
  \bibfield  {author} {\bibinfo {author} {\bibfnamefont {M.~Y.}\ \bibnamefont
  {Kuznetsov}}, \bibinfo {author} {\bibfnamefont {N.~A.}\ \bibnamefont
  {Petrov}}, \ and\ \bibinfo {author} {\bibfnamefont {Y.~S.}\ \bibnamefont
  {Savchenko}},\ }\href {\doibase 10.1134/S0021364025610061} {\bibfield
  {journal} {\bibinfo  {journal} {JETP Lett.}\ }\textbf {\bibinfo {volume}
  {123}},\ \bibinfo {pages} {287} (\bibinfo {year} {2026})},\ \Eprint
  {http://arxiv.org/abs/2509.09590} {arXiv:2509.09590 [astro-ph.HE]}
  \BibitemShut {NoStop}%
\bibitem [{\citenamefont {Alhebsi}\ \emph {et~al.}(2026)\citenamefont
  {Alhebsi}, \citenamefont {van Vliet}, \citenamefont {Ehlert},\ and\
  \citenamefont {Thoudam}}]{Alhebsi:2026bdk}%
  \BibitemOpen
  \bibfield  {author} {\bibinfo {author} {\bibfnamefont {A.}~\bibnamefont
  {Alhebsi}}, \bibinfo {author} {\bibfnamefont {A.}~\bibnamefont {van Vliet}},
  \bibinfo {author} {\bibfnamefont {D.}~\bibnamefont {Ehlert}}, \ and\ \bibinfo
  {author} {\bibfnamefont {S.}~\bibnamefont {Thoudam}},\ }\href {\doibase
  10.3847/1538-4357/ae6452} {\bibfield  {journal} {\bibinfo  {journal}
  {Astrophys. J.}\ }\textbf {\bibinfo {volume} {1003}},\ \bibinfo {pages} {235}
  (\bibinfo {year} {2026})},\ \Eprint {http://arxiv.org/abs/2603.13476}
  {arXiv:2603.13476 [astro-ph.HE]} \BibitemShut {NoStop}%
\bibitem [{\citenamefont {Abbasi}\ \emph {et~al.}(2025)\citenamefont {Abbasi}
  \emph {et~al.}}]{IceCubeCollaborationSS:2025jbi}%
  \BibitemOpen
  \bibfield  {author} {\bibinfo {author} {\bibfnamefont {R.}~\bibnamefont
  {Abbasi}} \emph {et~al.} (\bibinfo {collaboration} {(IceCube
  Collaboration){\textsection}, IceCube}),\ }\href {\doibase
  10.1103/PhysRevLett.135.031001} {\bibfield  {journal} {\bibinfo  {journal}
  {Phys. Rev. Lett.}\ }\textbf {\bibinfo {volume} {135}},\ \bibinfo {pages}
  {031001} (\bibinfo {year} {2025})},\ \Eprint
  {http://arxiv.org/abs/2502.01963} {arXiv:2502.01963 [astro-ph.HE]}
  \BibitemShut {NoStop}%
\bibitem [{\citenamefont {Li}\ \emph {et~al.}(2026)\citenamefont {Li},
  \citenamefont {Machado}, \citenamefont {Naredo-Tuero},\ and\ \citenamefont
  {Schwemberger}}]{Li:2025tqf}%
  \BibitemOpen
  \bibfield  {author} {\bibinfo {author} {\bibfnamefont {S.~W.}\ \bibnamefont
  {Li}}, \bibinfo {author} {\bibfnamefont {P.}~\bibnamefont {Machado}},
  \bibinfo {author} {\bibfnamefont {D.}~\bibnamefont {Naredo-Tuero}}, \ and\
  \bibinfo {author} {\bibfnamefont {T.}~\bibnamefont {Schwemberger}},\ }\href
  {\doibase 10.1016/j.physletb.2026.140293} {\bibfield  {journal} {\bibinfo
  {journal} {Phys. Lett. B}\ }\textbf {\bibinfo {volume} {875}},\ \bibinfo
  {pages} {140293} (\bibinfo {year} {2026})},\ \Eprint
  {http://arxiv.org/abs/2502.04508} {arXiv:2502.04508 [astro-ph.HE]}
  \BibitemShut {NoStop}%
\bibitem [{\citenamefont {Adriani}\ \emph
  {et~al.}(2025{\natexlab{b}})\citenamefont {Adriani} \emph
  {et~al.}}]{KM3NeT:2025ccp}%
  \BibitemOpen
  \bibfield  {author} {\bibinfo {author} {\bibfnamefont {O.}~\bibnamefont
  {Adriani}} \emph {et~al.} (\bibinfo {collaboration} {KM3NeT}),\ }\href
  {\doibase 10.1103/yypk-zmb8} {\bibfield  {journal} {\bibinfo  {journal}
  {Phys. Rev. X}\ }\textbf {\bibinfo {volume} {15}},\ \bibinfo {pages} {031016}
  (\bibinfo {year} {2025}{\natexlab{b}})},\ \Eprint
  {http://arxiv.org/abs/2502.08173} {arXiv:2502.08173 [astro-ph.HE]}
  \BibitemShut {NoStop}%
\bibitem [{\citenamefont {Yuan}\ and\ \citenamefont {Lu}(2025)}]{Yuan:2025isv}%
  \BibitemOpen
  \bibfield  {author} {\bibinfo {author} {\bibfnamefont {T.}~\bibnamefont
  {Yuan}}\ and\ \bibinfo {author} {\bibfnamefont {L.}~\bibnamefont {Lu}},\
  }\href {\doibase 10.53941/hihep.2025.100019} {\bibfield  {journal} {\bibinfo
  {journal} {HiHEP}\ }\textbf {\bibinfo {volume} {1}},\ \bibinfo {pages} {19}
  (\bibinfo {year} {2025})},\ \Eprint {http://arxiv.org/abs/2509.12628}
  {arXiv:2509.12628 [astro-ph.HE]} \BibitemShut {NoStop}%
\bibitem [{\citenamefont {Fang}\ \emph {et~al.}(2025)\citenamefont {Fang},
  \citenamefont {Halzen},\ and\ \citenamefont {Hooper}}]{Fang:2025nzg}%
  \BibitemOpen
  \bibfield  {author} {\bibinfo {author} {\bibfnamefont {K.}~\bibnamefont
  {Fang}}, \bibinfo {author} {\bibfnamefont {F.}~\bibnamefont {Halzen}}, \ and\
  \bibinfo {author} {\bibfnamefont {D.}~\bibnamefont {Hooper}},\ }\href
  {\doibase 10.3847/2041-8213/adbbec} {\bibfield  {journal} {\bibinfo
  {journal} {Astrophys. J. Lett.}\ }\textbf {\bibinfo {volume} {982}},\
  \bibinfo {pages} {L16} (\bibinfo {year} {2025})},\ \Eprint
  {http://arxiv.org/abs/2502.09545} {arXiv:2502.09545 [astro-ph.HE]}
  \BibitemShut {NoStop}%
\bibitem [{\citenamefont {Crnogor{\v{c}}evi{\'c}}\ \emph
  {et~al.}(2025)\citenamefont {Crnogor{\v{c}}evi{\'c}}, \citenamefont
  {Blanco},\ and\ \citenamefont {Linden}}]{Crnogorcevic:2025vou}%
  \BibitemOpen
  \bibfield  {author} {\bibinfo {author} {\bibfnamefont {M.}~\bibnamefont
  {Crnogor{\v{c}}evi{\'c}}}, \bibinfo {author} {\bibfnamefont {C.}~\bibnamefont
  {Blanco}}, \ and\ \bibinfo {author} {\bibfnamefont {T.}~\bibnamefont
  {Linden}},\ }\href {\doibase 10.1088/1475-7516/2025/10/009} {\bibfield
  {journal} {\bibinfo  {journal} {JCAP}\ }\textbf {\bibinfo {volume} {10}},\
  \bibinfo {pages} {009} (\bibinfo {year} {2025})},\ \Eprint
  {http://arxiv.org/abs/2503.16606} {arXiv:2503.16606 [astro-ph.HE]}
  \BibitemShut {NoStop}%
\bibitem [{\citenamefont {Neronov}\ \emph {et~al.}(2026)\citenamefont
  {Neronov}, \citenamefont {Oikonomou},\ and\ \citenamefont
  {Semikoz}}]{Neronov:2025jfj}%
  \BibitemOpen
  \bibfield  {author} {\bibinfo {author} {\bibfnamefont {A.}~\bibnamefont
  {Neronov}}, \bibinfo {author} {\bibfnamefont {F.}~\bibnamefont {Oikonomou}},
  \ and\ \bibinfo {author} {\bibfnamefont {D.}~\bibnamefont {Semikoz}},\ }\href
  {\doibase 10.1103/yj9s-jjjl} {\bibfield  {journal} {\bibinfo  {journal}
  {Phys. Rev. D}\ }\textbf {\bibinfo {volume} {113}},\ \bibinfo {pages}
  {083024} (\bibinfo {year} {2026})},\ \Eprint
  {http://arxiv.org/abs/2502.12986} {arXiv:2502.12986 [astro-ph.HE]}
  \BibitemShut {NoStop}%
\bibitem [{\citenamefont {Yuan}\ \emph {et~al.}(2026)\citenamefont {Yuan},
  \citenamefont {Pfeiffer}, \citenamefont {Winter}, \citenamefont {Zaballa},
  \citenamefont {Buson}, \citenamefont {Testagrossa},\ and\ \citenamefont
  {Azzollini}}]{Yuan:2025zwe}%
  \BibitemOpen
  \bibfield  {author} {\bibinfo {author} {\bibfnamefont {C.}~\bibnamefont
  {Yuan}}, \bibinfo {author} {\bibfnamefont {L.}~\bibnamefont {Pfeiffer}},
  \bibinfo {author} {\bibfnamefont {W.}~\bibnamefont {Winter}}, \bibinfo
  {author} {\bibfnamefont {J.~M.~S.}\ \bibnamefont {Zaballa}}, \bibinfo
  {author} {\bibfnamefont {S.}~\bibnamefont {Buson}}, \bibinfo {author}
  {\bibfnamefont {F.}~\bibnamefont {Testagrossa}}, \ and\ \bibinfo {author}
  {\bibfnamefont {A.}~\bibnamefont {Azzollini}},\ }\href {\doibase
  10.3847/2041-8213/ae3a73} {\bibfield  {journal} {\bibinfo  {journal}
  {Astrophys. J. Lett.}\ }\textbf {\bibinfo {volume} {998}},\ \bibinfo {pages}
  {L1} (\bibinfo {year} {2026})},\ \Eprint {http://arxiv.org/abs/2506.21111}
  {arXiv:2506.21111 [astro-ph.HE]} \BibitemShut {NoStop}%
\bibitem [{\citenamefont {Amelino-Camelia}\ \emph {et~al.}(2025)\citenamefont
  {Amelino-Camelia}, \citenamefont {D'Amico}, \citenamefont {Fabiano},
  \citenamefont {Frattulillo}, \citenamefont {Gubitosi}, \citenamefont {Moia},\
  and\ \citenamefont {Rosati}}]{Amelino-Camelia:2025lqn}%
  \BibitemOpen
  \bibfield  {author} {\bibinfo {author} {\bibfnamefont {G.}~\bibnamefont
  {Amelino-Camelia}}, \bibinfo {author} {\bibfnamefont {G.}~\bibnamefont
  {D'Amico}}, \bibinfo {author} {\bibfnamefont {G.}~\bibnamefont {Fabiano}},
  \bibinfo {author} {\bibfnamefont {D.}~\bibnamefont {Frattulillo}}, \bibinfo
  {author} {\bibfnamefont {G.}~\bibnamefont {Gubitosi}}, \bibinfo {author}
  {\bibfnamefont {A.}~\bibnamefont {Moia}}, \ and\ \bibinfo {author}
  {\bibfnamefont {G.}~\bibnamefont {Rosati}},\ }\href {\doibase
  10.1016/j.physletb.2025.139764} {\bibfield  {journal} {\bibinfo  {journal}
  {Phys. Lett. B}\ }\textbf {\bibinfo {volume} {868}},\ \bibinfo {pages}
  {139764} (\bibinfo {year} {2025})},\ \Eprint
  {http://arxiv.org/abs/2502.13093} {arXiv:2502.13093 [astro-ph.HE]}
  \BibitemShut {NoStop}%
\bibitem [{\citenamefont {Adriani}\ \emph
  {et~al.}(2026{\natexlab{a}})\citenamefont {Adriani} \emph
  {et~al.}}]{KM3NeT:2025bxl}%
  \BibitemOpen
  \bibfield  {author} {\bibinfo {author} {\bibfnamefont {O.}~\bibnamefont
  {Adriani}} \emph {et~al.} (\bibinfo {collaboration} {KM3NeT, MessMapp Group,
  Fermi-LAT, Owens Valley Radio Observatory 40-m Telescope Group, SVOM}),\
  }\href {\doibase 10.3847/1538-3881/ae7a4e} {\bibfield  {journal} {\bibinfo
  {journal} {Astron. J.}\ }\textbf {\bibinfo {volume} {172}},\ \bibinfo {pages}
  {139} (\bibinfo {year} {2026}{\natexlab{a}})},\ \Eprint
  {http://arxiv.org/abs/2502.08484} {arXiv:2502.08484 [astro-ph.HE]}
  \BibitemShut {NoStop}%
\bibitem [{\citenamefont {Padovani}\ \emph {et~al.}(2019)\citenamefont
  {Padovani}, \citenamefont {Oikonomou}, \citenamefont {Petropoulou},
  \citenamefont {Giommi},\ and\ \citenamefont {Resconi}}]{Padovani:2019xcv}%
  \BibitemOpen
  \bibfield  {author} {\bibinfo {author} {\bibfnamefont {P.}~\bibnamefont
  {Padovani}}, \bibinfo {author} {\bibfnamefont {F.}~\bibnamefont {Oikonomou}},
  \bibinfo {author} {\bibfnamefont {M.}~\bibnamefont {Petropoulou}}, \bibinfo
  {author} {\bibfnamefont {P.}~\bibnamefont {Giommi}}, \ and\ \bibinfo {author}
  {\bibfnamefont {E.}~\bibnamefont {Resconi}},\ }\href {\doibase
  10.1093/mnrasl/slz011} {\bibfield  {journal} {\bibinfo  {journal} {Mon. Not.
  Roy. Astron. Soc.}\ }\textbf {\bibinfo {volume} {484}},\ \bibinfo {pages}
  {L104} (\bibinfo {year} {2019})},\ \Eprint {http://arxiv.org/abs/1901.06998}
  {arXiv:1901.06998 [astro-ph.HE]} \BibitemShut {NoStop}%
\bibitem [{\citenamefont {Petropoulou}\ \emph {et~al.}(2020)\citenamefont
  {Petropoulou} \emph {et~al.}}]{Petropoulou:2019zqp}%
  \BibitemOpen
  \bibfield  {author} {\bibinfo {author} {\bibfnamefont {M.}~\bibnamefont
  {Petropoulou}} \emph {et~al.},\ }\href {\doibase 10.3847/1538-4357/ab76d0}
  {\bibfield  {journal} {\bibinfo  {journal} {Astrophys. J.}\ }\textbf
  {\bibinfo {volume} {891}},\ \bibinfo {pages} {115} (\bibinfo {year}
  {2020})},\ \Eprint {http://arxiv.org/abs/1911.04010} {arXiv:1911.04010
  [astro-ph.HE]} \BibitemShut {NoStop}%
\bibitem [{\citenamefont {Sherman}\ \emph {et~al.}(2026)\citenamefont
  {Sherman}, \citenamefont {Mirabal}, \citenamefont {Guevel}, \citenamefont
  {Fang}, \citenamefont {Murase},\ and\ \citenamefont
  {Hays}}]{Sherman:2025gir}%
  \BibitemOpen
  \bibfield  {author} {\bibinfo {author} {\bibfnamefont {A.}~\bibnamefont
  {Sherman}}, \bibinfo {author} {\bibfnamefont {N.}~\bibnamefont {Mirabal}},
  \bibinfo {author} {\bibfnamefont {D.}~\bibnamefont {Guevel}}, \bibinfo
  {author} {\bibfnamefont {K.}~\bibnamefont {Fang}}, \bibinfo {author}
  {\bibfnamefont {K.}~\bibnamefont {Murase}}, \ and\ \bibinfo {author}
  {\bibfnamefont {E.}~\bibnamefont {Hays}},\ }\href {\doibase
  10.3847/1538-4357/ae75e3} {\bibfield  {journal} {\bibinfo  {journal}
  {Astrophys. J.}\ }\textbf {\bibinfo {volume} {1005}},\ \bibinfo {pages} {194}
  (\bibinfo {year} {2026})},\ \Eprint {http://arxiv.org/abs/2510.09101}
  {arXiv:2510.09101 [astro-ph.HE]} \BibitemShut {NoStop}%
\bibitem [{\citenamefont {Kivokurtseva}\ and\ \citenamefont
  {Troitsky}(2026)}]{Kivokurtseva:2025sui}%
  \BibitemOpen
  \bibfield  {author} {\bibinfo {author} {\bibfnamefont {P.}~\bibnamefont
  {Kivokurtseva}}\ and\ \bibinfo {author} {\bibfnamefont {S.}~\bibnamefont
  {Troitsky}},\ }\href {\doibase 10.1088/1475-7516/2026/08/026} {\bibfield
  {journal} {\bibinfo  {journal} {JCAP}\ }\textbf {\bibinfo {volume} {08}},\
  \bibinfo {pages} {026} (\bibinfo {year} {2026})},\ \Eprint
  {http://arxiv.org/abs/2509.10352} {arXiv:2509.10352 [astro-ph.HE]}
  \BibitemShut {NoStop}%
\bibitem [{\citenamefont {Borah}\ \emph {et~al.}(2025)\citenamefont {Borah},
  \citenamefont {Das}, \citenamefont {Okada},\ and\ \citenamefont
  {Sarmah}}]{Borah:2025igh}%
  \BibitemOpen
  \bibfield  {author} {\bibinfo {author} {\bibfnamefont {D.}~\bibnamefont
  {Borah}}, \bibinfo {author} {\bibfnamefont {N.}~\bibnamefont {Das}}, \bibinfo
  {author} {\bibfnamefont {N.}~\bibnamefont {Okada}}, \ and\ \bibinfo {author}
  {\bibfnamefont {P.}~\bibnamefont {Sarmah}},\ }\href {\doibase
  10.1103/shsw-mct6} {\bibfield  {journal} {\bibinfo  {journal} {Phys. Rev. D}\
  }\textbf {\bibinfo {volume} {111}},\ \bibinfo {pages} {123022} (\bibinfo
  {year} {2025})},\ \Eprint {http://arxiv.org/abs/2503.00097} {arXiv:2503.00097
  [hep-ph]} \BibitemShut {NoStop}%
\bibitem [{\citenamefont {Kohri}\ \emph {et~al.}(2025)\citenamefont {Kohri},
  \citenamefont {Paul},\ and\ \citenamefont {Sahu}}]{Kohri:2025bsn}%
  \BibitemOpen
  \bibfield  {author} {\bibinfo {author} {\bibfnamefont {K.}~\bibnamefont
  {Kohri}}, \bibinfo {author} {\bibfnamefont {P.~K.}\ \bibnamefont {Paul}}, \
  and\ \bibinfo {author} {\bibfnamefont {N.}~\bibnamefont {Sahu}},\ }\href
  {\doibase 10.1103/vvqq-1z2t} {\bibfield  {journal} {\bibinfo  {journal}
  {Phys. Rev. D}\ }\textbf {\bibinfo {volume} {112}},\ \bibinfo {pages}
  {L031703} (\bibinfo {year} {2025})},\ \Eprint
  {http://arxiv.org/abs/2503.04464} {arXiv:2503.04464 [hep-ph]} \BibitemShut
  {NoStop}%
\bibitem [{\citenamefont {Jho}\ \emph {et~al.}(2025)\citenamefont {Jho},
  \citenamefont {Park},\ and\ \citenamefont {Shin}}]{Jho:2025gaf}%
  \BibitemOpen
  \bibfield  {author} {\bibinfo {author} {\bibfnamefont {Y.}~\bibnamefont
  {Jho}}, \bibinfo {author} {\bibfnamefont {S.~C.}\ \bibnamefont {Park}}, \
  and\ \bibinfo {author} {\bibfnamefont {C.~S.}\ \bibnamefont {Shin}},\ }\href
  {\doibase 10.1103/7jgm-f852} {\bibfield  {journal} {\bibinfo  {journal}
  {Phys. Rev. D}\ }\textbf {\bibinfo {volume} {112}},\ \bibinfo {pages}
  {115004} (\bibinfo {year} {2025})},\ \Eprint
  {http://arxiv.org/abs/2503.18737} {arXiv:2503.18737 [hep-ph]} \BibitemShut
  {NoStop}%
\bibitem [{\citenamefont {Barman}\ \emph {et~al.}(2025)\citenamefont {Barman},
  \citenamefont {Das},\ and\ \citenamefont {Sarmah}}]{Barman:2025hoz}%
  \BibitemOpen
  \bibfield  {author} {\bibinfo {author} {\bibfnamefont {B.}~\bibnamefont
  {Barman}}, \bibinfo {author} {\bibfnamefont {A.}~\bibnamefont {Das}}, \ and\
  \bibinfo {author} {\bibfnamefont {P.}~\bibnamefont {Sarmah}},\ }\href
  {\doibase 10.1103/t58k-sjq5} {\bibfield  {journal} {\bibinfo  {journal}
  {Phys. Rev. D}\ }\textbf {\bibinfo {volume} {112}},\ \bibinfo {pages}
  {075014} (\bibinfo {year} {2025})},\ \Eprint
  {http://arxiv.org/abs/2504.01447} {arXiv:2504.01447 [hep-ph]} \BibitemShut
  {NoStop}%
\bibitem [{\citenamefont {Murase}\ \emph {et~al.}(2025)\citenamefont {Murase},
  \citenamefont {Narita},\ and\ \citenamefont {Yin}}]{Murase:2025uwv}%
  \BibitemOpen
  \bibfield  {author} {\bibinfo {author} {\bibfnamefont {K.}~\bibnamefont
  {Murase}}, \bibinfo {author} {\bibfnamefont {Y.}~\bibnamefont {Narita}}, \
  and\ \bibinfo {author} {\bibfnamefont {W.}~\bibnamefont {Yin}},\ }\href
  {\doibase 10.1088/1475-7516/2025/10/109} {\bibfield  {journal} {\bibinfo
  {journal} {JCAP}\ }\textbf {\bibinfo {volume} {10}},\ \bibinfo {pages} {109}
  (\bibinfo {year} {2025})},\ \Eprint {http://arxiv.org/abs/2504.15272}
  {arXiv:2504.15272 [hep-ph]} \BibitemShut {NoStop}%
\bibitem [{\citenamefont {Khan}\ \emph {et~al.}(2025)\citenamefont {Khan},
  \citenamefont {Kim},\ and\ \citenamefont {Ko}}]{Khan:2025gxs}%
  \BibitemOpen
  \bibfield  {author} {\bibinfo {author} {\bibfnamefont {S.}~\bibnamefont
  {Khan}}, \bibinfo {author} {\bibfnamefont {J.}~\bibnamefont {Kim}}, \ and\
  \bibinfo {author} {\bibfnamefont {P.}~\bibnamefont {Ko}},\ }\href {\doibase
  10.1088/1475-7516/2025/11/033} {\bibfield  {journal} {\bibinfo  {journal}
  {JCAP}\ }\textbf {\bibinfo {volume} {11}},\ \bibinfo {pages} {033} (\bibinfo
  {year} {2025})},\ \Eprint {http://arxiv.org/abs/2504.16040} {arXiv:2504.16040
  [hep-ph]} \BibitemShut {NoStop}%
\bibitem [{\citenamefont {Aloisio}\ \emph {et~al.}(2026)\citenamefont
  {Aloisio}, \citenamefont {Ambrosone},\ and\ \citenamefont
  {Evoli}}]{Aloisio:2025nts}%
  \BibitemOpen
  \bibfield  {author} {\bibinfo {author} {\bibfnamefont {R.}~\bibnamefont
  {Aloisio}}, \bibinfo {author} {\bibfnamefont {A.}~\bibnamefont {Ambrosone}},
  \ and\ \bibinfo {author} {\bibfnamefont {C.}~\bibnamefont {Evoli}},\ }\href
  {\doibase 10.1103/rc2p-53yg} {\bibfield  {journal} {\bibinfo  {journal}
  {Phys. Rev. D}\ }\textbf {\bibinfo {volume} {113}},\ \bibinfo {pages}
  {043024} (\bibinfo {year} {2026})},\ \Eprint
  {http://arxiv.org/abs/2508.08779} {arXiv:2508.08779 [astro-ph.HE]}
  \BibitemShut {NoStop}%
\bibitem [{\citenamefont {Barger}(2026)}]{Barger:2026zkd}%
  \BibitemOpen
  \bibfield  {author} {\bibinfo {author} {\bibfnamefont {V.}~\bibnamefont
  {Barger}},\ }\href {\doibase 10.1103/51px-xpdh} {\bibfield  {journal}
  {\bibinfo  {journal} {APS Open Sci.}\ }\textbf {\bibinfo {volume} {1}},\
  \bibinfo {pages} {000180} (\bibinfo {year} {2026})},\ \Eprint
  {http://arxiv.org/abs/2607.10760} {arXiv:2607.10760 [hep-ph]} \BibitemShut
  {NoStop}%
\bibitem [{\citenamefont {Adriani}\ \emph
  {et~al.}(2026{\natexlab{b}})\citenamefont {Adriani} \emph
  {et~al.}}]{KM3NeT:2026flu}%
  \BibitemOpen
  \bibfield  {author} {\bibinfo {author} {\bibfnamefont {O.}~\bibnamefont
  {Adriani}} \emph {et~al.} (\bibinfo {collaboration} {KM3NeT}),\ }\href@noop
  {} {\  (\bibinfo {year} {2026}{\natexlab{b}})},\ \Eprint
  {http://arxiv.org/abs/2606.09986} {arXiv:2606.09986 [hep-ex]} \BibitemShut
  {NoStop}%
\bibitem [{\citenamefont {Grimbaum~Yamamoto}\ and\ \citenamefont
  {Hambye}(2025)}]{GrimbaumYamamoto:2025nai}%
  \BibitemOpen
  \bibfield  {author} {\bibinfo {author} {\bibfnamefont {N.}~\bibnamefont
  {Grimbaum~Yamamoto}}\ and\ \bibinfo {author} {\bibfnamefont {T.}~\bibnamefont
  {Hambye}},\ }\href {\doibase 10.1088/1475-7516/2025/10/092} {\bibfield
  {journal} {\bibinfo  {journal} {JCAP}\ }\textbf {\bibinfo {volume} {10}},\
  \bibinfo {pages} {092} (\bibinfo {year} {2025})},\ \Eprint
  {http://arxiv.org/abs/2507.02063} {arXiv:2507.02063 [hep-ph]} \BibitemShut
  {NoStop}%
\bibitem [{\citenamefont {Grimbaum~Yamamoto}\ and\ \citenamefont
  {Hambye}(2026)}]{GrimbaumYamamoto:2026kam}%
  \BibitemOpen
  \bibfield  {author} {\bibinfo {author} {\bibfnamefont {N.}~\bibnamefont
  {Grimbaum~Yamamoto}}\ and\ \bibinfo {author} {\bibfnamefont {T.}~\bibnamefont
  {Hambye}},\ }\href@noop {} {\  (\bibinfo {year} {2026})},\ \Eprint
  {http://arxiv.org/abs/2604.18677} {arXiv:2604.18677 [astro-ph.HE]}
  \BibitemShut {NoStop}%
\bibitem [{\citenamefont {Klipfel}\ and\ \citenamefont
  {Kaiser}(2025)}]{Klipfel:2025jql}%
  \BibitemOpen
  \bibfield  {author} {\bibinfo {author} {\bibfnamefont {A.~P.}\ \bibnamefont
  {Klipfel}}\ and\ \bibinfo {author} {\bibfnamefont {D.~I.}\ \bibnamefont
  {Kaiser}},\ }\href {\doibase 10.1103/vnm4-7wdc} {\bibfield  {journal}
  {\bibinfo  {journal} {Phys. Rev. Lett.}\ }\textbf {\bibinfo {volume} {135}},\
  \bibinfo {pages} {121003} (\bibinfo {year} {2025})},\ \Eprint
  {http://arxiv.org/abs/2503.19227} {arXiv:2503.19227 [hep-ph]} \BibitemShut
  {NoStop}%
\bibitem [{\citenamefont {Airoldi}\ \emph {et~al.}(2026)\citenamefont
  {Airoldi}, \citenamefont {Alves}, \citenamefont {Perez-Gonzalez},
  \citenamefont {Salla},\ and\ \citenamefont {Funchal}}]{Airoldi:2025opo}%
  \BibitemOpen
  \bibfield  {author} {\bibinfo {author} {\bibfnamefont {L.~F.~T.}\
  \bibnamefont {Airoldi}}, \bibinfo {author} {\bibfnamefont {G.~F.~S.}\
  \bibnamefont {Alves}}, \bibinfo {author} {\bibfnamefont {Y.~F.}\ \bibnamefont
  {Perez-Gonzalez}}, \bibinfo {author} {\bibfnamefont {G.~M.}\ \bibnamefont
  {Salla}}, \ and\ \bibinfo {author} {\bibfnamefont {R.~Z.}\ \bibnamefont
  {Funchal}},\ }\href {\doibase 10.1103/w9dp-dfkx} {\bibfield  {journal}
  {\bibinfo  {journal} {Phys. Rev. Lett.}\ }\textbf {\bibinfo {volume} {136}},\
  \bibinfo {pages} {041002} (\bibinfo {year} {2026})},\ \Eprint
  {http://arxiv.org/abs/2505.24666} {arXiv:2505.24666 [hep-ph]} \BibitemShut
  {NoStop}%
\bibitem [{\citenamefont {Klipfel}\ and\ \citenamefont
  {Kaiser}(2026)}]{Klipfel:2026jrx}%
  \BibitemOpen
  \bibfield  {author} {\bibinfo {author} {\bibfnamefont {A.~P.}\ \bibnamefont
  {Klipfel}}\ and\ \bibinfo {author} {\bibfnamefont {D.~I.}\ \bibnamefont
  {Kaiser}},\ }\href {\doibase 10.3390/universe12080245} {\bibfield  {journal}
  {\bibinfo  {journal} {Universe}\ }\textbf {\bibinfo {volume} {12}},\ \bibinfo
  {pages} {245} (\bibinfo {year} {2026})},\ \Eprint
  {http://arxiv.org/abs/2605.26204} {arXiv:2605.26204 [hep-ph]} \BibitemShut
  {NoStop}%
\bibitem [{\citenamefont {Anchordoqui}\ \emph {et~al.}(2025)\citenamefont
  {Anchordoqui}, \citenamefont {Halzen},\ and\ \citenamefont
  {L\"ust}}]{Anchordoqui:2025xug}%
  \BibitemOpen
  \bibfield  {author} {\bibinfo {author} {\bibfnamefont {L.~A.}\ \bibnamefont
  {Anchordoqui}}, \bibinfo {author} {\bibfnamefont {F.}~\bibnamefont {Halzen}},
  \ and\ \bibinfo {author} {\bibfnamefont {D.}~\bibnamefont {L\"ust}},\ }\href
  {\doibase 10.1103/5kt2-5pvj} {\bibfield  {journal} {\bibinfo  {journal}
  {Phys. Rev. D}\ }\textbf {\bibinfo {volume} {112}},\ \bibinfo {pages}
  {083034} (\bibinfo {year} {2025})},\ \Eprint
  {http://arxiv.org/abs/2505.23414} {arXiv:2505.23414 [hep-ph]} \BibitemShut
  {NoStop}%
\bibitem [{\citenamefont {Anchordoqui}\ \emph {et~al.}(2026)\citenamefont
  {Anchordoqui}, \citenamefont {Bedroya},\ and\ \citenamefont
  {L{\"u}st}}]{Anchordoqui:2025opy}%
  \BibitemOpen
  \bibfield  {author} {\bibinfo {author} {\bibfnamefont {L.~A.}\ \bibnamefont
  {Anchordoqui}}, \bibinfo {author} {\bibfnamefont {A.}~\bibnamefont
  {Bedroya}}, \ and\ \bibinfo {author} {\bibfnamefont {D.}~\bibnamefont
  {L{\"u}st}},\ }\href {\doibase 10.1103/g12h-93th} {\bibfield  {journal}
  {\bibinfo  {journal} {Phys. Rev. D}\ }\textbf {\bibinfo {volume} {114}},\
  \bibinfo {pages} {063551} (\bibinfo {year} {2026})},\ \Eprint
  {http://arxiv.org/abs/2506.14874} {arXiv:2506.14874 [hep-ph]} \BibitemShut
  {NoStop}%
\bibitem [{\citenamefont {Anchordoqui}\ and\ \citenamefont
  {Vafa}(2026)}]{Anchordoqui:2026ret}%
  \BibitemOpen
  \bibfield  {author} {\bibinfo {author} {\bibfnamefont {L.~A.}\ \bibnamefont
  {Anchordoqui}}\ and\ \bibinfo {author} {\bibfnamefont {C.}~\bibnamefont
  {Vafa}},\ }\href@noop {} {\  (\bibinfo {year} {2026})},\ \Eprint
  {http://arxiv.org/abs/2610.15209} {arXiv:2610.15209 [hep-ph]} \BibitemShut
  {NoStop}%
\bibitem [{\citenamefont {Brdar}\ and\ \citenamefont
  {Chattopadhyay}(2026)}]{Brdar:2025azm}%
  \BibitemOpen
  \bibfield  {author} {\bibinfo {author} {\bibfnamefont {V.}~\bibnamefont
  {Brdar}}\ and\ \bibinfo {author} {\bibfnamefont {D.~S.}\ \bibnamefont
  {Chattopadhyay}},\ }\href {\doibase 10.1103/xcnt-trs2} {\bibfield  {journal}
  {\bibinfo  {journal} {Phys. Rev. Lett.}\ }\textbf {\bibinfo {volume} {136}},\
  \bibinfo {pages} {081001} (\bibinfo {year} {2026})},\ \Eprint
  {http://arxiv.org/abs/2502.21299} {arXiv:2502.21299 [hep-ph]} \BibitemShut
  {NoStop}%
\bibitem [{\citenamefont {Dev}\ \emph {et~al.}(2025)\citenamefont {Dev},
  \citenamefont {Dutta}, \citenamefont {Karthikeyan}, \citenamefont {Maitra},
  \citenamefont {Strigari},\ and\ \citenamefont {Verma}}]{Dev:2025czz}%
  \BibitemOpen
  \bibfield  {author} {\bibinfo {author} {\bibfnamefont {P.~S.~B.}\
  \bibnamefont {Dev}}, \bibinfo {author} {\bibfnamefont {B.}~\bibnamefont
  {Dutta}}, \bibinfo {author} {\bibfnamefont {A.}~\bibnamefont {Karthikeyan}},
  \bibinfo {author} {\bibfnamefont {W.}~\bibnamefont {Maitra}}, \bibinfo
  {author} {\bibfnamefont {L.~E.}\ \bibnamefont {Strigari}}, \ and\ \bibinfo
  {author} {\bibfnamefont {A.}~\bibnamefont {Verma}},\ }\href@noop {} {\
  (\bibinfo {year} {2025})},\ \Eprint {http://arxiv.org/abs/2505.22754}
  {arXiv:2505.22754 [hep-ph]} \BibitemShut {NoStop}%
\bibitem [{\citenamefont {Reno}\ \emph {et~al.}(2021)\citenamefont {Reno},
  \citenamefont {Venters},\ and\ \citenamefont {Krizmanic}}]{Reno:2021veo}%
  \BibitemOpen
  \bibfield  {author} {\bibinfo {author} {\bibfnamefont {M.~H.}\ \bibnamefont
  {Reno}}, \bibinfo {author} {\bibfnamefont {T.}~\bibnamefont {Venters}}, \
  and\ \bibinfo {author} {\bibfnamefont {J.}~\bibnamefont {Krizmanic}}
  (\bibinfo {collaboration} {POEMMA, JEM-EUSO}),\ }\href {\doibase
  10.22323/1.395.1201} {\bibfield  {journal} {\bibinfo  {journal} {PoS}\
  }\textbf {\bibinfo {volume} {ICRC2021}},\ \bibinfo {pages} {1201} (\bibinfo
  {year} {2021})}\BibitemShut {NoStop}%
\bibitem [{\citenamefont {Dutta}\ \emph {et~al.}(2005)\citenamefont {Dutta},
  \citenamefont {Huang},\ and\ \citenamefont {Reno}}]{Dutta:2005yt}%
  \BibitemOpen
  \bibfield  {author} {\bibinfo {author} {\bibfnamefont {S.~I.}\ \bibnamefont
  {Dutta}}, \bibinfo {author} {\bibfnamefont {Y.}~\bibnamefont {Huang}}, \ and\
  \bibinfo {author} {\bibfnamefont {M.~H.}\ \bibnamefont {Reno}},\ }\href
  {\doibase 10.1103/PhysRevD.72.013005} {\bibfield  {journal} {\bibinfo
  {journal} {Phys. Rev. D}\ }\textbf {\bibinfo {volume} {72}},\ \bibinfo
  {pages} {013005} (\bibinfo {year} {2005})},\ \Eprint
  {http://arxiv.org/abs/hep-ph/0504208} {arXiv:hep-ph/0504208} \BibitemShut
  {NoStop}%
\bibitem [{\citenamefont {Dziewonski}\ and\ \citenamefont
  {Anderson}(1981)}]{PREM}%
  \BibitemOpen
  \bibfield  {author} {\bibinfo {author} {\bibfnamefont {A.~M.}\ \bibnamefont
  {Dziewonski}}\ and\ \bibinfo {author} {\bibfnamefont {D.~L.}\ \bibnamefont
  {Anderson}},\ }\href {\doibase https://doi.org/10.1016/0031-9201(81)90046-7}
  {\bibfield  {journal} {\bibinfo  {journal} {Physics of the Earth and
  Planetary Interiors}\ }\textbf {\bibinfo {volume} {25}},\ \bibinfo {pages}
  {297} (\bibinfo {year} {1981})}\BibitemShut {NoStop}%
\bibitem [{\citenamefont {Reno}\ \emph {et~al.}(2019)\citenamefont {Reno},
  \citenamefont {Krizmanic},\ and\ \citenamefont {Venters}}]{Reno:2019jtr}%
  \BibitemOpen
  \bibfield  {author} {\bibinfo {author} {\bibfnamefont {M.~H.}\ \bibnamefont
  {Reno}}, \bibinfo {author} {\bibfnamefont {J.~F.}\ \bibnamefont {Krizmanic}},
  \ and\ \bibinfo {author} {\bibfnamefont {T.~M.}\ \bibnamefont {Venters}},\
  }\href {\doibase 10.1103/PhysRevD.100.063010} {\bibfield  {journal} {\bibinfo
   {journal} {Phys. Rev. D}\ }\textbf {\bibinfo {volume} {100}},\ \bibinfo
  {pages} {063010} (\bibinfo {year} {2019})},\ \Eprint
  {http://arxiv.org/abs/1902.11287} {arXiv:1902.11287 [astro-ph.HE]}
  \BibitemShut {NoStop}%
\bibitem [{\citenamefont {Reno}\ and\ \citenamefont
  {Krizmanic}(2025)}]{Reno:2025tpw}%
  \BibitemOpen
  \bibfield  {author} {\bibinfo {author} {\bibfnamefont {M.~H.}\ \bibnamefont
  {Reno}}\ and\ \bibinfo {author} {\bibfnamefont {J.~F.}\ \bibnamefont
  {Krizmanic}},\ }\href {\doibase 10.22323/1.501.1082} {\bibfield  {journal}
  {\bibinfo  {journal} {PoS}\ }\textbf {\bibinfo {volume} {ICRC2025}},\
  \bibinfo {pages} {1082} (\bibinfo {year} {2025})},\ \Eprint
  {http://arxiv.org/abs/2509.15469} {arXiv:2509.15469 [astro-ph.HE]}
  \BibitemShut {NoStop}%
\bibitem [{\citenamefont {Garg}\ \emph {et~al.}(2023)\citenamefont {Garg} \emph
  {et~al.}}]{Garg:2022ugd}%
  \BibitemOpen
  \bibfield  {author} {\bibinfo {author} {\bibfnamefont {D.}~\bibnamefont
  {Garg}} \emph {et~al.},\ }\href {\doibase 10.1088/1475-7516/2023/01/041}
  {\bibfield  {journal} {\bibinfo  {journal} {JCAP}\ }\textbf {\bibinfo
  {volume} {01}},\ \bibinfo {pages} {041} (\bibinfo {year} {2023})},\ \Eprint
  {http://arxiv.org/abs/2209.15581} {arXiv:2209.15581 [astro-ph.HE]}
  \BibitemShut {NoStop}%
\bibitem [{\citenamefont {Alvarez-Mu{\~n}iz}\ \emph {et~al.}(2018)\citenamefont
  {Alvarez-Mu{\~n}iz}, \citenamefont {Carvalho}, \citenamefont {Cummings},
  \citenamefont {Payet}, \citenamefont {Romero-Wolf}, \citenamefont
  {Schoorlemmer},\ and\ \citenamefont {Zas}}]{Alvarez-Muniz:2018owm}%
  \BibitemOpen
  \bibfield  {author} {\bibinfo {author} {\bibfnamefont {J.}~\bibnamefont
  {Alvarez-Mu{\~n}iz}}, \bibinfo {author} {\bibfnamefont {W.~R.}\ \bibnamefont
  {Carvalho}}, \bibinfo {author} {\bibfnamefont {A.~L.}\ \bibnamefont
  {Cummings}}, \bibinfo {author} {\bibfnamefont {K.}~\bibnamefont {Payet}},
  \bibinfo {author} {\bibfnamefont {A.}~\bibnamefont {Romero-Wolf}}, \bibinfo
  {author} {\bibfnamefont {H.}~\bibnamefont {Schoorlemmer}}, \ and\ \bibinfo
  {author} {\bibfnamefont {E.}~\bibnamefont {Zas}},\ }\href {\doibase
  10.1103/PhysRevD.97.023021} {\bibfield  {journal} {\bibinfo  {journal} {Phys.
  Rev. D}\ }\textbf {\bibinfo {volume} {97}},\ \bibinfo {pages} {023021}
  (\bibinfo {year} {2018})},\ \bibinfo {note} {[Erratum: Phys.Rev.D 99, 069902
  (2019)]},\ \Eprint {http://arxiv.org/abs/1707.00334} {arXiv:1707.00334
  [astro-ph.HE]} \BibitemShut {NoStop}%
\bibitem [{\citenamefont {Heibges}\ \emph
  {et~al.}(2026{\natexlab{b}})\citenamefont {Heibges}, \citenamefont
  {Gu{\'e}pin}, \citenamefont {Kupari}, \citenamefont {Wistrand}, \citenamefont
  {Lahm}, \citenamefont {Eser}, \citenamefont {Reno}, \citenamefont {Venters},\
  and\ \citenamefont {Wiencke}}]{Heibges:2025nuts}%
  \BibitemOpen
  \bibfield  {author} {\bibinfo {author} {\bibfnamefont {T.}~\bibnamefont
  {Heibges}}, \bibinfo {author} {\bibfnamefont {C.}~\bibnamefont {Gu{\'e}pin}},
  \bibinfo {author} {\bibfnamefont {L.}~\bibnamefont {Kupari}}, \bibinfo
  {author} {\bibfnamefont {H.}~\bibnamefont {Wistrand}}, \bibinfo {author}
  {\bibfnamefont {R.}~\bibnamefont {Lahm}}, \bibinfo {author} {\bibfnamefont
  {J.}~\bibnamefont {Eser}}, \bibinfo {author} {\bibfnamefont {M.~H.}\
  \bibnamefont {Reno}}, \bibinfo {author} {\bibfnamefont {T.~M.}\ \bibnamefont
  {Venters}}, \ and\ \bibinfo {author} {\bibfnamefont {L.}~\bibnamefont
  {Wiencke}},\ }\href {\doibase 10.1016/j.cpc.2026.110310} {\bibfield
  {journal} {\bibinfo  {journal} {Comput. Phys. Commun.}\ }\textbf {\bibinfo
  {volume} {328}},\ \bibinfo {pages} {110310} (\bibinfo {year}
  {2026}{\natexlab{b}})},\ \Eprint {http://arxiv.org/abs/2509.13844}
  {arXiv:2509.13844 [astro-ph.IM]} \BibitemShut {NoStop}%
\bibitem [{\citenamefont {{Emde}}\ \emph {et~al.}(2016)\citenamefont {{Emde}}
  \emph {et~al.}}]{2016GMD.....9.1647E}%
  \BibitemOpen
  \bibfield  {author} {\bibinfo {author} {\bibfnamefont {C.}~\bibnamefont
  {{Emde}}} \emph {et~al.},\ }\href {\doibase 10.5194/gmd-9-1647-2016}
  {\bibfield  {journal} {\bibinfo  {journal} {Geoscientific Model Development}\
  }\textbf {\bibinfo {volume} {9}},\ \bibinfo {pages} {1647} (\bibinfo {year}
  {2016})}\BibitemShut {NoStop}%
\bibitem [{\citenamefont {Venters}\ \emph
  {et~al.}(2021{\natexlab{a}})\citenamefont {Venters}, \citenamefont {Reno},\
  and\ \citenamefont {Krizmanic}}]{Venters:2021xos}%
  \BibitemOpen
  \bibfield  {author} {\bibinfo {author} {\bibfnamefont {T.~M.}\ \bibnamefont
  {Venters}}, \bibinfo {author} {\bibfnamefont {M.~H.}\ \bibnamefont {Reno}}, \
  and\ \bibinfo {author} {\bibfnamefont {J.~F.}\ \bibnamefont {Krizmanic}}
  (\bibinfo {collaboration} {POEMMA, JEM-EUSO}),\ }\href {\doibase
  10.22323/1.395.0977} {\bibfield  {journal} {\bibinfo  {journal} {PoS}\
  }\textbf {\bibinfo {volume} {ICRC2021}},\ \bibinfo {pages} {977} (\bibinfo
  {year} {2021}{\natexlab{a}})},\ \Eprint {http://arxiv.org/abs/2206.02730}
  {arXiv:2206.02730 [astro-ph.HE]} \BibitemShut {NoStop}%
\bibitem [{\citenamefont {Venters}\ \emph
  {et~al.}(2021{\natexlab{b}})\citenamefont {Venters}, \citenamefont {Reno},\
  and\ \citenamefont {Krizmanic}}]{Venters:2021RV}%
  \BibitemOpen
  \bibfield  {author} {\bibinfo {author} {\bibfnamefont {T.}~\bibnamefont
  {Venters}}, \bibinfo {author} {\bibfnamefont {M.~H.}\ \bibnamefont {Reno}}, \
  and\ \bibinfo {author} {\bibfnamefont {J.}~\bibnamefont {Krizmanic}},\ }\href
  {\doibase 10.22323/1.395.0977} {\bibfield  {journal} {\bibinfo  {journal}
  {PoS}\ }\textbf {\bibinfo {volume} {ICRC2021}},\ \bibinfo {pages} {977}
  (\bibinfo {year} {2021}{\natexlab{b}})}\BibitemShut {NoStop}%
\bibitem [{\citenamefont {Heibges}\ \emph {et~al.}(2023)\citenamefont
  {Heibges}, \citenamefont {Posligua}, \citenamefont {Wistrand}, \citenamefont
  {Gu{\'e}pin}, \citenamefont {Reno},\ and\ \citenamefont
  {Venters}}]{Heibges:2023yhn}%
  \BibitemOpen
  \bibfield  {author} {\bibinfo {author} {\bibfnamefont {T.}~\bibnamefont
  {Heibges}}, \bibinfo {author} {\bibfnamefont {J.}~\bibnamefont {Posligua}},
  \bibinfo {author} {\bibfnamefont {H.}~\bibnamefont {Wistrand}}, \bibinfo
  {author} {\bibfnamefont {C.}~\bibnamefont {Gu{\'e}pin}}, \bibinfo {author}
  {\bibfnamefont {M.~H.}\ \bibnamefont {Reno}}, \ and\ \bibinfo {author}
  {\bibfnamefont {T.~M.}\ \bibnamefont {Venters}} (\bibinfo {collaboration}
  {JEM-EUSO}),\ }\href {\doibase 10.22323/1.444.1134} {\bibfield  {journal}
  {\bibinfo  {journal} {PoS}\ }\textbf {\bibinfo {volume} {ICRC2023}},\
  \bibinfo {pages} {1134} (\bibinfo {year} {2023})},\ \Eprint
  {http://arxiv.org/abs/2310.12310} {arXiv:2310.12310 [astro-ph.HE]}
  \BibitemShut {NoStop}%
\bibitem [{\citenamefont {Abe}\ \emph {et~al.}(2024)\citenamefont {Abe} \emph
  {et~al.}}]{JEM-EUSO:2023fyg}%
  \BibitemOpen
  \bibfield  {author} {\bibinfo {author} {\bibfnamefont {S.}~\bibnamefont
  {Abe}} \emph {et~al.} (\bibinfo {collaboration} {JEM-EUSO}),\ }\href
  {\doibase 10.1088/1748-0221/19/01/P01007} {\bibfield  {journal} {\bibinfo
  {journal} {JINST}\ }\textbf {\bibinfo {volume} {19}},\ \bibinfo {pages}
  {P01007} (\bibinfo {year} {2024})},\ \Eprint
  {http://arxiv.org/abs/2309.02577} {arXiv:2309.02577 [astro-ph.IM]}
  \BibitemShut {NoStop}%
\bibitem [{\citenamefont {Anchordoqui}\ \emph {et~al.}(2002)\citenamefont
  {Anchordoqui}, \citenamefont {Feng}, \citenamefont {Goldberg},\ and\
  \citenamefont {Shapere}}]{Anchordoqui:2002vb}%
  \BibitemOpen
  \bibfield  {author} {\bibinfo {author} {\bibfnamefont {L.~A.}\ \bibnamefont
  {Anchordoqui}}, \bibinfo {author} {\bibfnamefont {J.~L.}\ \bibnamefont
  {Feng}}, \bibinfo {author} {\bibfnamefont {H.}~\bibnamefont {Goldberg}}, \
  and\ \bibinfo {author} {\bibfnamefont {A.~D.}\ \bibnamefont {Shapere}},\
  }\href {\doibase 10.1103/PhysRevD.66.103002} {\bibfield  {journal} {\bibinfo
  {journal} {Phys. Rev. D}\ }\textbf {\bibinfo {volume} {66}},\ \bibinfo
  {pages} {103002} (\bibinfo {year} {2002})},\ \Eprint
  {http://arxiv.org/abs/hep-ph/0207139} {arXiv:hep-ph/0207139} \BibitemShut
  {NoStop}%
\bibitem [{\citenamefont {Feldman}\ and\ \citenamefont
  {Cousins}(1998)}]{Feldman:1997qc}%
  \BibitemOpen
  \bibfield  {author} {\bibinfo {author} {\bibfnamefont {G.~J.}\ \bibnamefont
  {Feldman}}\ and\ \bibinfo {author} {\bibfnamefont {R.~D.}\ \bibnamefont
  {Cousins}},\ }\href {\doibase 10.1103/PhysRevD.57.3873} {\bibfield  {journal}
  {\bibinfo  {journal} {Phys. Rev. D}\ }\textbf {\bibinfo {volume} {57}},\
  \bibinfo {pages} {3873} (\bibinfo {year} {1998})},\ \Eprint
  {http://arxiv.org/abs/physics/9711021} {arXiv:physics/9711021} \BibitemShut
  {NoStop}%
\bibitem [{\citenamefont {{Gelaro}}\ \emph {et~al.}(2017)\citenamefont
  {{Gelaro}}, \citenamefont {{McCarty}}, \citenamefont {{Su{\'a}rez}},
  \citenamefont {{Todling}}, \citenamefont {{Molod}}, \citenamefont {{Takacs}},
  \citenamefont {{Randles}}, \citenamefont {{Darmenov}}, \citenamefont
  {{Bosilovich}}, \citenamefont {{Reichle}}, \citenamefont {{Wargan}},
  \citenamefont {{Coy}}, \citenamefont {{Cullather}}, \citenamefont {{Draper}},
  \citenamefont {{Akella}}, \citenamefont {{Buchard}}, \citenamefont
  {{Conaty}}, \citenamefont {{da Silva}}, \citenamefont {{Gu}}, \citenamefont
  {{Kim}}, \citenamefont {{Koster}}, \citenamefont {{Lucchesi}}, \citenamefont
  {{Merkova}}, \citenamefont {{Nielsen}}, \citenamefont {{Partyka}},
  \citenamefont {{Pawson}}, \citenamefont {{Putman}}, \citenamefont
  {{Rienecker}}, \citenamefont {{Schubert}}, \citenamefont {{Sienkiewicz}},\
  and\ \citenamefont {{Zhao}}}]{2017JCli...30.5419G}%
  \BibitemOpen
  \bibfield  {author} {\bibinfo {author} {\bibfnamefont {R.}~\bibnamefont
  {{Gelaro}}}, \bibinfo {author} {\bibfnamefont {W.}~\bibnamefont {{McCarty}}},
  \bibinfo {author} {\bibfnamefont {M.~J.}\ \bibnamefont {{Su{\'a}rez}}},
  \bibinfo {author} {\bibfnamefont {R.}~\bibnamefont {{Todling}}}, \bibinfo
  {author} {\bibfnamefont {A.}~\bibnamefont {{Molod}}}, \bibinfo {author}
  {\bibfnamefont {L.}~\bibnamefont {{Takacs}}}, \bibinfo {author}
  {\bibfnamefont {C.~A.}\ \bibnamefont {{Randles}}}, \bibinfo {author}
  {\bibfnamefont {A.}~\bibnamefont {{Darmenov}}}, \bibinfo {author}
  {\bibfnamefont {M.~G.}\ \bibnamefont {{Bosilovich}}}, \bibinfo {author}
  {\bibfnamefont {R.}~\bibnamefont {{Reichle}}}, \bibinfo {author}
  {\bibfnamefont {K.}~\bibnamefont {{Wargan}}}, \bibinfo {author}
  {\bibfnamefont {L.}~\bibnamefont {{Coy}}}, \bibinfo {author} {\bibfnamefont
  {R.}~\bibnamefont {{Cullather}}}, \bibinfo {author} {\bibfnamefont
  {C.}~\bibnamefont {{Draper}}}, \bibinfo {author} {\bibfnamefont
  {S.}~\bibnamefont {{Akella}}}, \bibinfo {author} {\bibfnamefont
  {V.}~\bibnamefont {{Buchard}}}, \bibinfo {author} {\bibfnamefont
  {A.}~\bibnamefont {{Conaty}}}, \bibinfo {author} {\bibfnamefont {A.~M.}\
  \bibnamefont {{da Silva}}}, \bibinfo {author} {\bibfnamefont
  {W.}~\bibnamefont {{Gu}}}, \bibinfo {author} {\bibfnamefont {G.-K.}\
  \bibnamefont {{Kim}}}, \bibinfo {author} {\bibfnamefont {R.}~\bibnamefont
  {{Koster}}}, \bibinfo {author} {\bibfnamefont {R.}~\bibnamefont
  {{Lucchesi}}}, \bibinfo {author} {\bibfnamefont {D.}~\bibnamefont
  {{Merkova}}}, \bibinfo {author} {\bibfnamefont {J.~E.}\ \bibnamefont
  {{Nielsen}}}, \bibinfo {author} {\bibfnamefont {G.}~\bibnamefont
  {{Partyka}}}, \bibinfo {author} {\bibfnamefont {S.}~\bibnamefont {{Pawson}}},
  \bibinfo {author} {\bibfnamefont {W.}~\bibnamefont {{Putman}}}, \bibinfo
  {author} {\bibfnamefont {M.}~\bibnamefont {{Rienecker}}}, \bibinfo {author}
  {\bibfnamefont {S.~D.}\ \bibnamefont {{Schubert}}}, \bibinfo {author}
  {\bibfnamefont {M.}~\bibnamefont {{Sienkiewicz}}}, \ and\ \bibinfo {author}
  {\bibfnamefont {B.}~\bibnamefont {{Zhao}}},\ }\href {\doibase
  10.1175/JCLI-D-16-0758.1} {\bibfield  {journal} {\bibinfo  {journal} {Journal
  of Climate}\ }\textbf {\bibinfo {volume} {30}},\ \bibinfo {pages} {5419}
  (\bibinfo {year} {2017})}\BibitemShut {NoStop}%
\bibitem [{\citenamefont {Gorham}\ \emph {et~al.}(2016)\citenamefont {Gorham}
  \emph {et~al.}}]{ANITA:2016vrp}%
  \BibitemOpen
  \bibfield  {author} {\bibinfo {author} {\bibfnamefont {P.~W.}\ \bibnamefont
  {Gorham}} \emph {et~al.} (\bibinfo {collaboration} {ANITA}),\ }\href
  {\doibase 10.1103/PhysRevLett.117.071101} {\bibfield  {journal} {\bibinfo
  {journal} {Phys. Rev. Lett.}\ }\textbf {\bibinfo {volume} {117}},\ \bibinfo
  {pages} {071101} (\bibinfo {year} {2016})},\ \Eprint
  {http://arxiv.org/abs/1603.05218} {arXiv:1603.05218 [astro-ph.HE]}
  \BibitemShut {NoStop}%
\bibitem [{\citenamefont {Gorham}\ \emph {et~al.}(2018)\citenamefont {Gorham}
  \emph {et~al.}}]{ANITA:2018sgj}%
  \BibitemOpen
  \bibfield  {author} {\bibinfo {author} {\bibfnamefont {P.~W.}\ \bibnamefont
  {Gorham}} \emph {et~al.} (\bibinfo {collaboration} {ANITA}),\ }\href
  {\doibase 10.1103/PhysRevLett.121.161102} {\bibfield  {journal} {\bibinfo
  {journal} {Phys. Rev. Lett.}\ }\textbf {\bibinfo {volume} {121}},\ \bibinfo
  {pages} {161102} (\bibinfo {year} {2018})},\ \Eprint
  {http://arxiv.org/abs/1803.05088} {arXiv:1803.05088 [astro-ph.HE]}
  \BibitemShut {NoStop}%
\bibitem [{\citenamefont {Chipman}\ \emph {et~al.}(2019)\citenamefont
  {Chipman}, \citenamefont {Diesing}, \citenamefont {Reno},\ and\ \citenamefont
  {Sarcevic}}]{Chipman:2019vjm}%
  \BibitemOpen
  \bibfield  {author} {\bibinfo {author} {\bibfnamefont {S.}~\bibnamefont
  {Chipman}}, \bibinfo {author} {\bibfnamefont {R.}~\bibnamefont {Diesing}},
  \bibinfo {author} {\bibfnamefont {M.~H.}\ \bibnamefont {Reno}}, \ and\
  \bibinfo {author} {\bibfnamefont {I.}~\bibnamefont {Sarcevic}},\ }\href
  {\doibase 10.1103/PhysRevD.100.063011} {\bibfield  {journal} {\bibinfo
  {journal} {Phys. Rev. D}\ }\textbf {\bibinfo {volume} {100}},\ \bibinfo
  {pages} {063011} (\bibinfo {year} {2019})},\ \Eprint
  {http://arxiv.org/abs/1906.11736} {arXiv:1906.11736 [astro-ph.HE]}
  \BibitemShut {NoStop}%
\bibitem [{\citenamefont {Marsh}(2016)}]{Marsh:2015xka}%
  \BibitemOpen
  \bibfield  {author} {\bibinfo {author} {\bibfnamefont {D.~J.~E.}\
  \bibnamefont {Marsh}},\ }\href {\doibase 10.1016/j.physrep.2016.06.005}
  {\bibfield  {journal} {\bibinfo  {journal} {Phys. Rept.}\ }\textbf {\bibinfo
  {volume} {643}},\ \bibinfo {pages} {1} (\bibinfo {year} {2016})},\ \Eprint
  {http://arxiv.org/abs/1510.07633} {arXiv:1510.07633 [astro-ph.CO]}
  \BibitemShut {NoStop}%
\bibitem [{\citenamefont {Calibbi}\ \emph {et~al.}(2023)\citenamefont
  {Calibbi}, \citenamefont {Huang}, \citenamefont {Qin}, \citenamefont {Yang},\
  and\ \citenamefont {Yin}}]{Calibbi:2022izs}%
  \BibitemOpen
  \bibfield  {author} {\bibinfo {author} {\bibfnamefont {L.}~\bibnamefont
  {Calibbi}}, \bibinfo {author} {\bibfnamefont {Z.}~\bibnamefont {Huang}},
  \bibinfo {author} {\bibfnamefont {S.}~\bibnamefont {Qin}}, \bibinfo {author}
  {\bibfnamefont {Y.}~\bibnamefont {Yang}}, \ and\ \bibinfo {author}
  {\bibfnamefont {X.}~\bibnamefont {Yin}},\ }\href {\doibase
  10.1103/PhysRevD.108.015002} {\bibfield  {journal} {\bibinfo  {journal}
  {Phys. Rev. D}\ }\textbf {\bibinfo {volume} {108}},\ \bibinfo {pages}
  {015002} (\bibinfo {year} {2023})},\ \Eprint
  {http://arxiv.org/abs/2212.02818} {arXiv:2212.02818 [hep-ph]} \BibitemShut
  {NoStop}%
\bibitem [{\citenamefont {Nomura}\ and\ \citenamefont
  {Thaler}(2009)}]{Nomura:2008ru}%
  \BibitemOpen
  \bibfield  {author} {\bibinfo {author} {\bibfnamefont {Y.}~\bibnamefont
  {Nomura}}\ and\ \bibinfo {author} {\bibfnamefont {J.}~\bibnamefont
  {Thaler}},\ }\href {\doibase 10.1103/PhysRevD.79.075008} {\bibfield
  {journal} {\bibinfo  {journal} {Phys. Rev. D}\ }\textbf {\bibinfo {volume}
  {79}},\ \bibinfo {pages} {075008} (\bibinfo {year} {2009})},\ \Eprint
  {http://arxiv.org/abs/0810.5397} {arXiv:0810.5397 [hep-ph]} \BibitemShut
  {NoStop}%
\end{thebibliography}
%\end{document}

\providecommand{\noopsort}[1]{}\providecommand{\singleletter}[1]{#1}%

\end{document}